\documentclass[aps,prd,preprintnumbers,nofootinbib,superscriptaddress,10pt]{revtex4-2}
\usepackage[latin1]{inputenc}
\usepackage{slashed,xspace,nicefrac}
\usepackage{amsmath,amssymb,amsfonts}
\usepackage{textcomp}
\usepackage{enumitem}
\usepackage{threeparttable}
\usepackage{appendix}
\usepackage[table,xcdraw]{xcolor}
\usepackage{color} 
\usepackage{tikz}
\usepackage[normalem]{ulem}
\usepackage[unicode]{hyperref}
\usepackage{verbatim}
\usepackage{placeins}

\newcommand{\be}{\begin{equation}}
\newcommand{\ee}{\end{equation}}
\newcommand{\bea}{\begin{eqnarray}}
\newcommand{\eea}{\end{eqnarray}}

\newcommand{\hf}{\nicefrac{1}{2}\xspace}
\newcommand{\nn}{\nonumber\\}

\newcommand{\tev}{\mbox{TeV}\xspace}

\newcommand{\gd}{\ensuremath{g_{\mathrm{D}}}\xspace}
\usepackage{graphicx,here}

\newcommand{\ovl}{\overline}

\DeclareGraphicsExtensions{.pdf,.png,.jpg}
\graphicspath{{./}{./figs/}{./plots/}}
\begin{document}
\begin{flushleft} 
KCL-PH-TH/2025-{\bf 45}
\end{flushleft}

\title{Magnetic-monopole resummation justifies perturbatively calculated collider production cross sections}

\author{Jean Alexandre}
\email{jean.alexandre@kcl.ac.uk}
\affiliation{Theoretical Particle Physics and Cosmology group, Department of Physics, King's College London, London WC2R 2LS, UK}

\author{Nick E.\ Mavromatos}
\email{nikolaos.mavromatos@cern.ch}
\affiliation{Department of Theoretical Physics and IFIC, University of Valencia and CSIC, 46100, Valencia, Spain}
\affiliation{Theoretical Particle Physics and Cosmology group, Department of Physics, King's College London, London WC2R 2LS, UK}
\affiliation{Physics Division, School of Applied Mathematical and Physical Sciences, National Technical University of Athens, 15780 Zografou Campus,
Athens, Greece}

\author{Vasiliki A.\ Mitsou}
\email{vasiliki.mitsou@ific.uv.es}
\affiliation{Instituto de F\'isica Corpuscular (IFIC), CSIC -- Universitat de Val\`encia,
C/ Catedr\'atico Jos\'e Beltr\'an 2, 46980 Paterna (Valencia), Spain}

\author{Emanuela Musumeci}
\email{emanuela.musumeci@cern.ch}
\affiliation{Department of Physics and Astronomy, University of Alabama, Tuscaloosa, Alabama, USA}

\begin{abstract}
A one-loop resummation scheme, inspired by Dyson--Schwinger (DS) formalism of strongly coupled quantum field theories, is applied  to spin-\hf magnetic monopoles (MMs), in the context of an effective field theory (EFT), invariant under the gauge group $\rm U(1)_{em}\otimes U(1)^\prime$, where $\mathrm U(1)^\prime$ is a dual strongly coupled Abelian interaction, associated with a `dark photon'. The latter leads to a MM appearing as a limiting case of a `dyon', with a tiny `electric charge', compensated by a huge wave-function renormalization of the MM, thereby leading to a finite renormalized MM-photon coupling. An ultra-violet fixed point structure is found in the resummed theory, which is purely non-perturbative due to different boundary conditions of the resummation equations, compared to the weak coupling (perturbative) case. The renormalized coupling of the MM to the electromagnetic photon in the fixed-point theory is identified with the magnetic charge, compatible with the Dirac quantization condition. This provides for the first time a formal justification of the use of tree-level Drell--Yan and photon-fusion MM production processes in collider searches, and of the corresponding cross sections and MM mass bounds thereof. The latter provide a means to constrain the resummed-EFT parameters experimentally. The DS resummation applies here primarily to elementary MMs. However, this approach may also be applied to the last stage (collapse) of the formation of composite MM pairs at colliders, in case they behave as quantum excitations, with their core radius comparable to the Compton wavelength, thereby avoiding the extreme suppression of their production. 
\end{abstract}

\maketitle

\section{Introduction}

Experimental searches for magnetic monopoles (MM) at colliders have been revived in recent years~\cite{ATLAS:2012bda,MoEDAL:2014ttp,ATLAS:2015tyu,MoEDAL:2016jlb,MoEDAL:2016lxh,MoEDAL:2017vhz,MoEDAL:2019ort,ATLAS:2019wkg,MoEDAL:2020pyb,MoEDAL:2021mpi,MoEDAL:2021vix,MoEDAL:2023ost,ATLAS:2023esy,MoEDAL:2024wbc,ATLAS:2024nzp}, partly due to theoretical developments (for a review, see~\cite{Mavromatos:2020gwk} and references therein) pointing towards the existence of solitonic solutions of field equations of MM type, with masses at least of order of a few TeV, which makes such objects in principle detectable at current colliders, such as the Large Hadron Collider (LHC).  In all searches so far, with the exception of the (non-perturbative) Schwinger-mechanism for the production of MM from the vacuum in the presence of strong magnetic fields~\cite{Affleck:1981ag,Gould:2018efv,Gould:2017fve,Gould:2017zwi,Gould:2018ovk,Bruce:2018yzs,Gould:2019myj,Gould:2021bre,MoEDAL:2021vix,ATLAS:2024nzp,MoEDAL:2024wbc}, tree-level Drell--Yan (DY) and Photon-Fusion (PF) processes are used~\cite{Dougall:2007tt,Baines:2018ltl,Song:2021vpo}. The tree-level DY or PF production processes are typically perturbatively unreliable, due to the large MM couplings involved. This calls for a non-perturbative treatment of the MM production at colliders. 

One such treatment has been proposed~\cite{Alexandre:2019iub}, based on a continuous field theory which generalizes --- at a quantum level --- a classical effective theory of MM proposed by Zwanziger~\cite{Zwanziger:1970hk}, which constitutes a \emph{local} description of electric and magnetic charges. The latter is assumed to be of Dirac type~\cite{Dirac:1931kp,Dirac:1948um}, i.e.\ considers the MM as a new structureless elementary particle, without involving the line-singularities of the Dirac string. Instead, it involves a Lorentz-Invariance-Violating (LIV) fixed-in-space unit vector, which represents the position of the string. In \cite{Alexandre:2019iub}, the authors have made use of the conjecture of \cite{Terning:2018udc}, based on toy models of perturbative MM in the dark sector of a theory, according to which the effects of the Dirac string can  most affect the phases of scattering amplitudes, and, if the Dirac charge quantization condition (DQC) is satisfied, they drop completely out of the formalism. Hence, physical quantities such as cross sections, remain unaffected by the
presence of the string, which prompted the authors of \cite{Alexandre:2019iub}
to formally ignore 
the LIV terms attributed to the Dirac string in the two-gauge potential model of \cite{Zwanziger:1970hk}.  This allows for a proper Lorentz-invariant quantum-field theoretic formulation of the model. In \cite{Alexandre:2019iub}, the MM is assumed to be a Dirac-fermion field, and one of the gauge potentials is assumed to be strongly coupled, thereby necessitating a Dyson--Schwinger (DS) resummation treatment. 
The magnetic charge of the formalism is proportional to the non-perturbative, due to the strong coupling, wavefunction renormalization of the fermion monopole, and obeys the Dirac charge quantization. 

We demonstrate here the appearance of an ultraviolet (UV) fixed point of the renormalization group of the resummed theory, at which the pertinent cross sections, describing the production of elementary MM of Dirac type, are calculated. This feature has not been discussed in the analysis of \cite{Alexandre:2019iub}.
Such non-trivial UV fixed points are known to characterize one-loop resummed effective theories of high-electric-charge objects (HECOs),
studied in \cite{Alexandre:2023qjo,Alexandre:2024pbs}, which are strongly coupled field theories of quantum electrodynamics (QED) type. In ref.~\cite{Alexandre:2019iub} and here, we employ one-loop resummation. We shall use the UV-fixed-point theory to estimate the spin-\hf MM-production cross section primarily for elementary MMs. However, we shall also provide arguments, that, under some circumstances, the theory can apply to composite MMs. 

The structure of the article is the following:  
In Section~\ref{sec:resumMmodel}, we extend the one-loop resummation of the effective model of \cite{Alexandre:2019iub} for spin-\hf MMs by determining a non-trivial UV fixed point, at which the magnetic charge of the theory, satisfying the DQC, is defined. 
In Section~\ref{sec:elvscompres}, we review  the known problems in the production of composite MMs at colliders due to a huge suppression~ as a result of their high-degree of compositeness~\cite{Drukier:1981fq}. This makes the resummation methods presented in this work suitable for elementary MMs, for which there is no such suppression. Nevertheless, in Section~\ref{sec:comporesum} we point out some features of the resummed fixed-point theory which could provide compensation for the aforementioned suppression, under some circumstances. In Section~\ref{sec:mass-limits}, we constrain the resummation parameters by means of the mass bounds set by LHC experiments.
Finally, our conclusions and outlook are given in Section~\ref{sec:concl}. Some technical, but important, issues of the effective theory, not discussed previously, are presented in two Appendices. In Appendix \ref{sec:app}, we discuss the  connection of the effective Lagrangian of ref.~\cite{Alexandre:2019iub} with Zwanzinger's classical approach~\cite{Zwanziger:1970hk}; in particular, we provide formal arguments in support of the assumption of \cite{Alexandre:2019iub} that the two gauge potentials satisfy the constraint of \cite{Zwanziger:1970hk} \emph{only} at the classical level. This is based on the  independence of the UV fixed-point on gauge backgrounds.
In Appendix~\ref{sec:k1unitarity}, we discuss the boundaries in the transmutation-mass-scale space, for which unitarity for elementary MM is valid. Supplementary exclusion plots of the MM dressed mass for various magnetic charges are provided in Appendix~\ref{sec:exclusions}.

\section{The (one-loop) Resummed Model and its  UltraViolet fixed-point structure}\label{sec:resumMmodel}

In ref.~\cite{Alexandre:2019iub}, an effective field theoretic model for the description of the scattering dynamics of magnetic monopoles of spin~\hf and (bare) mass $M$, with charged fermionic (spin~\hf) matter $\psi$, of (bare) mass $m$, has been considered. The model is based on a two-pontential formalism for MM, developed by Zwanziger~\cite{Zwanziger:1970hk}.\footnote{An earlier approach, using also a two-gauge-potential formalism, has been developed in \cite{Cabibbo:1962td}, which however uses non-local field equations, in contrast to the approach of \cite{Zwanziger:1970hk}, where the formalism employs local Lagrangians for a MM field in the presence of its Dirac string.} To keep the correct number of degrees of freedom, expected in electromagnetism, in the presence of MM, Zwanziger assumed a constraint between the classical gauge potentials. In \cite{Farakos:2024ggp}, which is a variant of the effective field theory (EFT) of  \cite{Alexandre:2019iub} for scalar MM, this constraint was implemented in a (Euclidean) path integral at a full quantum level. It was demonstrated there that  
the quantum fluctuations of the gauge potentials remain unconstrained.  In  Appendix \ref{sec:app} of the current manuscript, 
we adapt the proof of this 
to our context of fermionic monopoles, for completeness, 
thus verifying formally the basic assumption underlying the model of \cite{Alexandre:2019iub}.

\subsection{The model and its basic resummation features}\label{sec:resumU1U1}

The bare Lagrangian for the model is given by
\be\label{Lagrangian}
L=-\frac{1}{4}F_A^{\mu\nu}F^A_{\mu\nu}-\frac{1}{4} F_B^{\mu\nu}~ F^B_{\mu\nu}+\ovl\psi(i\gamma^\mu D^A_\mu-m)\psi+\ovl\chi(i\gamma^\mu D^{A+B}_\mu-M)\chi~,
\ee
where $A_\mu$ is a vector and $B_\mu$ is an \emph{axial} vector (pseudovector).
The field $A_\mu$ is the physical photon, corresponding to the carrier of the U(1)$_{\rm em}$ weak interaction of electromagnetism, while $B_\mu$ represents a `dark photon', pertaining to a dual U(1)$^\prime$ strongly-coupled Abelian gauge group.\footnote{Our EFT resembles, somewhat the effective field theories of anyon superconductivity \cite{Dorey:1990sz,Dorey:1991kp,Kovner:1990zz}, which introduce an effective statistical U(1) gauge field to represent the new attraction among the quasiparticle degrees of freedom (holons), arising effectively from their deviation 
from integer or half-integer spins, and 
hence from Bose or fermion statistics. In that case the microscopic theory is just electromagnetism, but the EFT describing the complicated environments of holon and spinons is different.}  
The corresponding field strength tensors are given by
\be
F_A^{\mu\nu}=\partial^\mu A^\nu-\partial^\nu A^\mu~~~,~~~~F_B^{\mu\nu}= \partial^\mu B^\nu -\partial^\nu B^\mu~,
\ee
and the covariant derivatives are:
\be
D^A_\mu=\partial_\mu-i\,q_e A_\mu~~~~\mbox{and}~~~~D^{A+B}_\mu=\partial_\mu-ie_A A_\mu-ie_B B_\mu~,
\ee
where $q_e$ is the coupling (electric charge) the charged fermion $\psi$ to the elecromagnetic gauge field (photon) $A_\mu$ and $e_A,e_B$ are the couplings of the monopole to $A_\mu$ and $B_\mu$ gauge fields, respectively.
These covariant derivatives ensure that the Lagrangian is invariant under the gauge transformation
\bea\label{gauge}
A_\mu&\to& A_\mu+\partial_\mu\theta_A\\
B_\mu&\to& B_\mu+\partial_\mu\theta_B\nn
\psi&\to& \exp(ie\theta_A)~\psi\nn
\chi&\to& \exp(ie_A\theta_A+ie_B\theta_B)~\chi~.\nonumber
\eea
Few remarks are in order at this point, which describe the basic features of the model, for details of which we refer the reader to \cite{Alexandre:2019iub}:
\begin{description}

 \item[Discrete symmetries] The axial nature of the gauge field $B_\mu$ is required for consistency of the field equations, under improper Lorentz transformations, including spatial reflections and reversal in time. As a consequence, the Lagrangian (\ref{Lagrangian}) breaks parity, P, and time reversal symmetry, T, but preserves CPT, where C denotes charge conjugation. It should be noted that such an explicit parity and time reversal symmetry breaking, but CPT conservation, is a generic feature of theories with magnetic charges~\cite{cptmag}. We also note that the action of the CP operator on MM states transforms them to their antiparticles~\cite{Ramsey:1958gvj,Weinberg:1965rz}.
 
 \item[Coupling hierarchy] The (bare) electric charges $q_e$ and $e_A$ are both assumed to be perturbative: $q_e\ll1$ and $e_A\ll1$. Our analysis can be of course extended trivially to incorporate any other charged matter, 
 in which case the coupling $e$ will be replaced by the corresponding  electric charge $q_e$. 
 \item[Magnetic charge]  In the classical theory \eqref{Lagrangian}, the coupling $e_B$ of the spin-\hf monopole to the dual photon can be identified with the magnetic charge, $g_m$, given that its equation of motion yields the MM current~\cite{Zwanziger:1970hk}. In the quantum theory, $e_B$ should not be identified immediately with the magnetic charge $g_m$. The latter is defined~\cite{Alexandre:2019iub} via the scattering process of MM with matter fermions in the context of the renormalized effective theory \eqref{Lagrangian},
 and  the identification of the relevant cross sections in the non-relativistic limit with the corresponding ones in the quantum-mechanical approach of \cite{Schwinger:1976fr,Milton:2006cp,Epele:2012jn}, which guarantee electric-magnetic duality. In the current work we shall demonstrate that the bare $e_B$ gets renormalized at the quantum level of the UV-fixed point theory stemming from resummation, with its dressed value being multiplied by  a non-trivial wavefunction renormalization factor of the strongly coupled U(1)$^\prime$ gauge group. 

 In view of the coupling $e_A$ in \eqref{Lagrangian}, which provides a direct interaction of the MM with the electromagnetic photon, one might naively think that the $\chi$ field is a dyon in the Schwinger sense~\cite{Schwinger:1969ib,Schwinger:1975ww}, i.e.\ an object carrying both electric $q_e$ and magnetic $g_m$ charge. In the original formulation of the dyon, Lorentz invariance, which was broken by the presence of the Dirac string, is restored but at the cost of  introducing a non-local Hamiltonian formulation of the dyon field. The two-potential effective formalism of \cite{Alexandre:2019iub} could thus be seen as an attempt to describe the interactions of a dyon in a local Lagrangian formalism \eqref{Lagrangian}, as is the case of the MM of Zwanziger~\cite{Zwanziger:1970hk}. As we shall demonstrate in the current work, however, the resummed quantum theory based on \eqref{Lagrangian}, at its non-trivial UV renormalization fixed point, is such that, \emph{the dressed (renormalized) $e_A$ coupling of the photon to the dyon is actually the magnetic charge,} $g_m$, obeying DQC. Hence, the (quantum) resummed Lagrangian \eqref{Lagrangian} describes the effective interactions of MM, and \emph{not} dyons, with photons, of the type used in experimental searches~\cite{Mavromatos:2020gwk}. This is a crucial point to bear in mind when confronting the predictions of this effective field theory model with experimental results in Section~\ref{sec:mass-limits}.

\item[Charge quantization] In the case of Schwinger's dyons
there is a  generalization of DQC~\cite{Dirac:1931kp,Dirac:1948um}. 
This can be obtained by considering the scattering of dyon configurations with charges $q_e^n$, $g_m^\ell$, with $n, \ell$ positive integers, and reads~\cite{Schwinger:1969ib,Schwinger:1975ww}:
\begin{equation}\label{schquant}
\left(q_e^n\, g_m^\ell  - q_e^\ell \,g_m^n \right)/4\pi = {\texttt Z}_{n\ell} \in \mathbb Z~, 
\end{equation}
where ${\mathbb  Z}$ denotes the set of integers. 

In the case of interest, that is interaction of ordinary charged matter with MM, we may consider one of the dyons, as having only electric charge i.e.\ 
$q_e^n \ne 0$ and $g^n_m =0$ in \eqref{schquant}. For ordinary positrons (electrons) $q_e=e > 0 \, (q_e = -e <0)$, and then \eqref{schquant} reduces to  the standard DQC~\cite{Dirac:1931kp,Dirac:1948um} \emph{but} with magnetic charge an \emph{even multiple} of the fundamental Dirac unit \gd:
\be\label{schquant2}
g_m = \frac{n}{\alpha} = 2n \, \gd \, , \quad n \in \mathbb Z\,, \quad 
\gd \equiv \frac{1}{2\alpha}\, e\,,
\ee
with $e >0$ the positron charge, and $\alpha = e^2$ the fine-structure constant in the system of natural units we are working on,  $\hbar=c= 4\pi \epsilon_0 = 1$, where $\epsilon_0$ is the vacuum permittivity. 
This is the situation characterizing all known field-theoretic composite-MM solitonic solutions of interest nowadays, such as 't Hooft-Polyakov~\cite{tHooft:1974kcl,Polyakov:1974ek} MM in unified gauge theories, Cho-Maison electroweak monopole~\cite{Cho:1996qd} and its finite-energy and non-linear -electrodynamics extensions~\cite{Cho:2013vba,Ellis:2016glu,Arunasalam:2017eyu,Mavromatos:2018kcd}, etc. 

\end{description}

The bare fermion propagators are 
\be
S_\psi=i\frac{\slashed p+m}{p^2-m^2}\, , \quad S_\chi=i\frac{\slashed p+M}{p^2-M^2}\, ,
\ee
and we consider the approximation where the dressed fermion propagators are
\be
G_\psi\simeq S_\psi \ , \quad
G_\chi=i\frac{Z\slashed p+\widetilde M}{Z^2p^2-\widetilde M^2} \, ,
\ee
where the approximate  symbol ``$\simeq$" indicates that a tree-level propagator for the electron suffices, due to the assumed perturbatively weak electric charge~\cite{Alexandre:2019iub}, 
\begin{align}\label{ebea}
e\,, \, |e_A| \ll |e_B| \,, \quad {\rm with} \quad  0 < e \ll 1, \,\, |e_A| \ll 1\,,
\end{align}
that is, both $e, \, e_A$ are treated as perturbative,
as in the standard QED case, but $e_B$ is a strong coupling. The hierarchy \eqref{ebea} will be justified later on in the article, specifically in Section~\ref{sec:comporesum}, where we shall also discuss  concrete models of composite MMs with a specific hierarchy between $e$ and $e_A$.

An important comment for the phenomenology of the model  concerns an estimate of the strong coupling $e_B$. In the context of the Zwanziger model~\cite{Zwanziger:1970hk}, the dual gauge potential $B_\mu$
is connected to the magnetic current, via the equations of motion stemming from the tree-level (bare) Lagrangian \eqref{Lagrangian}, and therefore a natural identification of the bare $e_B$ with the magnetic charge $g_m$, satisfying the DQC \eqref{schquant2}:
\begin{align}\label{eBgm}
    |e_B| = |g_m| = 2n \, \gd  =\frac{n}{\alpha} \, e \, \, \gg \, \,  |e_A|, \, \, e\,.  
\end{align}
We shall make use of this relation when we discuss the collider phenomenology of the resummed model in Section~\ref{sec:mass-limits}.

The quantity 
$Z$ is the monopole wave-function renormalization and $\widetilde M$ is the dressed monopole mass. If we assume that $Z$ and 
the dressed vertices $\Gamma^{B\chi}_\mu$ and $\Gamma^{A\chi}_\mu$ are momentum-independent, the gauge-invariance Ward identities given in 
\cite{Alexandre:2019iub}
 imply 
\be\label{vertices}
\Gamma_\mu^{B\chi}=e_B Z\gamma_\mu~~~~\mbox{and}~~~~\Gamma_\mu^{A\chi}=e_A Z\gamma_\mu~.
\ee
In a generic covariant gauge parametrized by $\lambda>0$, the gauge propagators assume the form~\cite{Itzykson:1980rh}
\bea\label{emU1prop}
\Delta^A_{\mu\nu}&\simeq&D_{\mu\nu}=\frac{-i}{q^2}\left(\eta_{\mu\nu}+\frac{1-\lambda}{\lambda}\frac{q_\mu q_\nu}{q^2}\right) \  ,
\eea
\bea\label{magU1prop}
\Delta^B_{\mu\nu}&=&\frac{-i}{(1+\omega)q^2}\left(\eta_{\mu\nu}+\frac{1+\omega-\lambda}{\lambda}\frac{q_\mu q_\nu}{q^2}\right)~,
\eea
where $\omega$ is the quantum correction responsible for the dual photon transverse polarization tensor, defined by
\be
\Delta^{B~-1}_{\mu\nu}-D^{-1}_{\mu\nu}=i\omega\left(q^2\eta_{\mu\nu}-q_\mu q_\nu\right)~,
\ee
with $D$ the corresponding bare propagator, given in \eqref{emU1prop}; in the approach of \cite{Alexandre:2019iub} adopted here, 
$\omega$ is also assumed to be momentum independent.

Above, we treated the electromagnetic gauge field as perturbative, according to the standard rules of perturbative QED, ignoring resummations due to the weakness of the pertinent coupling (indicated, once again,  by the approximate symbol ``$\simeq$" in the pertinent gauge field propagator \eqref{emU1prop}).

The DS approach of \cite{Alexandre:2019iub} 
entails a certain set of coupled equations, stemming from resumming appropriate subsets of one-loop diagrams of dressed propagators and vertices. 
The equations involve masses and wave-function renormalizations, which are  functions of the 
dimensional-regularization transmutation mass scale $k$, and the derivatives with respect to $k$. The couplings in \eqref{Lagrangian} are assumed independent of $k$.
The pertinent one-loop-resummed, SD-like, equations are derived in detail in \cite{Alexandre:2019iub}. It is important to stress that the resummation scheme we adopt here and in \cite{Alexandre:2019iub} is based on different boundary conditions than in the perturbative cases, with weak gauge couplings for the $U(1)\otimes U(1) $ gauge group, that underlies the physics of the model \eqref{Lagrangian}. This is crucial, and the reason as to why (some of) the resummed quantities do not have a smooth limit when the corresponding strong coupling is set to zero. 

In this approach we used the boundary condition 
\begin{align}\label{Zto0}
Z(k_0=2M)=0\,, 
\end{align}
where the scale 
\begin{align}\label{k0}
k_0=2M\, ,
\end{align}
with $M$ the bare monopole mass, 
corresponds to a particular value of the dimensional transmutation-mass running scale $k$. 

It is important to remark at this point that the boundary condition \eqref{Zto0} does not admit a perturbative limit, nor does it respect unitarity. The latter point implies that the resummed EFT of \cite{Alexandre:2019iub} applies to composite MM and does not connect smoothly to perturbative field theories. Later on in the current article we shall discuss unitarity regions in $k$-space, at which $Z(k) >1$, as appropriate for elementary MMs. 

The boundary condition \eqref{Zto0} leads to
\be\label{Z}
Z(k)=\frac{e_A^2+e_B^2}{8\pi^2\lambda}\ln\left(\frac{k M_0}{k_0M_r(k)}\right)~,
\ee
where $M_r(k)$ is defined as 
\be\label{runningmass}
M_r(k)\equiv\frac{\widetilde M}{Z(k)}~,
\ee
and represents the physical (non-perturbatively renormalized) running mass, which equals the pole of the dressed propagator, taking $Z$ into account. 
In what follows, we use the shorthand notation
\begin{align}\label{m)mequal}
M_0 \equiv M_r(k_0)~,
\end{align}
to denote the value of the renormalized (running) mass at the scale $k_0$. 

Regarding the form \eqref{Z}, we would like to stress again, in conjunction to what we have mentioned above about the different boundary conditions from the perturbative case, that the boundary condition of vanishing of
the wavefunction renormalization $Z$ in the limit where the couplings go to zero, is \emph{in contrast} to the perturbative QED-like case, where $Z \to 1$ when the perturbative couplings vanish.

For the given bare mass $M$, $M_r(k)$  is given by~\cite{Alexandre:2019iub}
\be\label{M}
\frac{M}{M_r(k)}=Z(k)-\frac{\kappa^2}{8\pi^2\lambda}\ln\left(\frac{k}{M_r(k)}\right)-C~,
\ee
where $C$ is a constant of integration and 
\be
\kappa^2\equiv e_A^2+e_B^2+3\lambda\left(e_A^2+\frac{e_B^2}{1+\omega_0}\right)~.
\ee
The quantum correction to the transverse polarization tensor (associated with the wave-function renormalization) of the strong-U(1) gauge bosons $B_\mu$ of the model \eqref{Lagrangian} turns out to be independent of $k$
for $k$-independent gauge couplings $e_A, e_B$~\cite{Alexandre:2019iub}:
\be\label{omega0}
\omega=\frac{e_B^2}{6\pi^2Z}\ln\left(\frac{kM_0}{k_0M_r(k)}\right) \stackrel{\eqref{Z}}{=}\frac{4\lambda e_B^2}{3(e_A^2+e_B^2)} \equiv \omega_0~.
\ee

Taking into account that 
$M_r$ increases with $k$ since, from \eqref{M}
one has:
\be\label{dMdk}
k\partial_k M_r=\frac{(\kappa^2-e_A^2-e_B^2)M_r}{\kappa^2-e_A^2-e_B^2+8\pi^2\lambda M/M_r}
=\frac{3[e_A^2+e_B^2/(1+\omega_0)]M_r}{3[e_A^2+e_B^2/(1+\omega_0)]+8\pi^2 M/M_r}~>0~,
\, \forall\, \, \,   k \in [0, +\infty)\,,
\ee
and for $M > 0$ and $\omega_0 > 0$ we easily see that:
\be
\frac{k \partial_k M_r}{M_r} \, < \, 1 ~.
\ee
Consequently, 
as discussed in detail in \cite{Alexandre:2019iub}, we obtain 
that 
$Z$ is a \emph{monotonically increasing} function of the transmutation scale $k$: 
\be\label{dZdk}
k\partial_k Z(k)=\frac{e_A^2+e_B^2}{8\pi^2\lambda}\left(1-\frac{k\partial_k M_r}{M_r}\right)>0~.
\ee

We next proceed in determining the non-trivial UV fixed point, which occurs in the limit $k \to \infty$.\footnote{We remark for completion that the point $k \simeq k_0=2M$ is \emph{not} a fixed point, as becomes evident from \eqref{dMdk}, \eqref{dZdk}.}

\subsection{The UV non-perturbative fixed point and the effective Lagrangian}\label{sec:UVfp}

From \eqref{M} we can see that the ratio $k/M_r(k)$ must be finite in the limit $k\to\infty$, and we note
\be\label{eta}
\lim_{k\to\infty}\frac{k}{M_r(k)}\equiv\eta~.
\ee
From \eqref{Z} we can see that $Z(k)$ then goes to the UV fixed point expression
\be\label{Zast}
Z^\ast\equiv \lim_{k\to\infty}Z(k)=\frac{e_A^2+e_B^2}{8\pi^2\lambda}\ln\left(\frac{\eta M_0}{k_0}\right)~.
\ee
From \eqref{M} we can also see that
\bea
&&\mbox{for}~k\to\infty~~~~C=Z^\ast-\frac{\kappa^2}{8\pi^2\lambda}\ln(\eta)\\
&&\mbox{for}~k\to k_0~~~~~C=-\frac{M}{M_0}-\frac{\kappa^2}{8\pi^2\lambda}\ln\left(\frac{k_0}{M_0}\right)~,\nonumber
\eea
such that 
\be\label{MoverM0}
\frac{M}{M_0}=\frac{3}{8\pi^2}\left(e_A^2+\frac{e_B^2}{1+\omega_0}\right)\ln\left(\frac{\eta M_0}{k_0}\right)~.
\ee

As discussed in \cite{Alexandre:2019iub}, we consider the Feynman gauge $\lambda=1$ as a convenient gauge to adopt, following related literature on resummation techniques, e.g.\ pinch techniques, of DS situations in strongly coupled gauge quantum field theories~\cite{pinch1,pinch2,pinch3,pinch4}, including Abelian ones. 

We now augment the model by imposing the boundary condition
\begin{align}\label{bc}
M_0 = \varepsilon M\,, \quad \varepsilon > 0\,,
\end{align}
with $\varepsilon$ \emph{independent} of $M$. 
This can be considered as a \emph{boundary condition} for the resummation model, since it does not ``run'' with $k$. In the next subsection \ref{sec:modmagcharge}, we shall see that only a specific range of $\varepsilon \ll 1$ is consistent. It is important to stress that in case $\varepsilon \ll 1$, the parameter is kept finite, $\varepsilon \ne 0$, never allowed to reach exactly zero. 

For the sake of generality and completeness, we first keep $e_A, e_B$ generic, without any specific hierarchy, but we remind the reader that the gauge coupling do not run with $k$~\cite{Alexandre:2019iub}. From \eqref{omega0}, \eqref{Zast} and \eqref{MoverM0}, we obtain:
\bea
Z^\ast&=&\frac{e_A^2+e_B^2}{8\pi^2}\ln\left(\frac{\eta\, \varepsilon}{2}\right)\\
\frac{1}{\varepsilon} &=&\frac{3}{8\pi^2}\frac{3(e_A^4+e_B^4)+10e_A^2e_B^2}{3e_A^2+7e_B^2}\ln\left(\frac{\eta\,\varepsilon}{2}\right)~.\nonumber
\eea
Eliminating the variable $\eta$ from the above system of equations, we then determine the UV fixed point, using \eqref{k0},\eqref{Z}: 
\begin{align}\label{fpfull}
Z^\star &= \frac{1}{3\, \varepsilon} \, \frac{3 + 7\zeta^4 + 10\zeta^2}{3 + 3 \zeta^4 + 10 \zeta^2}\,, \nn 
\omega^\star & = \omega_0 = \frac{4}{3}\, \frac{\zeta^2}{1 + \zeta^2}\,, \nn 
\lim_{k\to\infty}\frac{M_r(k)}{\varepsilon\, k}
& = \frac{1}{2} \exp\left(-\frac{8\pi^2}{3\, \varepsilon\,e_B^2} \, \, \frac{\zeta^2}{1 + \zeta^2}\, \frac{3 + 7\zeta^4 + 10\zeta^2}{3 + 3 \zeta^4 + 10 \zeta^2}\right)\, , \qquad \zeta \equiv \frac{e_B}{e_A}\,.
\end{align}

The presence of $\varepsilon$
accompanying $k$ on the left-hand side of the last relation in \eqref{fpfull} calls for a careful discussion of the definition of the UV limit $k \to \infty$, which needs regularization. If one naively replaced $k$ by a large but finite energy scale $\widetilde \Lambda$ when regularizing the UV limit, then the fixed-point-value of the MM mass $M^\star$ would be proportional to $\varepsilon \widetilde \Lambda $. The quantity $\varepsilon \widetilde \Lambda$ 
should be viewed as 
providing a physical, subplanckian cutoff energy scale:
\be\label{Lphys}
\varepsilon \widetilde \Lambda \equiv \Lambda  \lesssim M_{\rm Pl}\,,
\ee
with $M_{\rm Pl} = 2.435 \times 10^{18}~{\rm GeV}$ the reduced Planck-mass scale. 
This is important, since, otherwise, for very small
$0 < \varepsilon \ll 1$, which, as we shall discuss in Section~\ref{sec:comporesum}, are relevant to composite MM, the MM mass would appear to be unnaturally small for $\widetilde \Lambda \lesssim M_{\rm Pl}$. In such a case, our effective theory could not accommodate electroweak MMs, whose mass is expected to be in the range of $M_W/\alpha = \mathcal O(10~{\rm TeV})$, with $M_W = \mathcal O(100~{\rm GeV})$ the electroweak symmetry breaking scale~\cite{Mavromatos:2020gwk}, unless, of course, one used transplanckian UV cutoffs $\widetilde \Lambda \gg M_{\rm Pl}$. The latter, however, would be inconsistent with an effective theory operating at much lower energy scales than the Planck scale, at which quantum gravity effects are expected to decouple. Such problems are avoided if one uses \eqref{Lphys} as the physical cutoff.  

A formally elegant way to implement the above considerations, which we adopt here, is to first rescale $k$ in  \eqref{fpfull} as:
\be\label{rescalek}
k \, \to \, \frac{k}{\varepsilon}
\ee
before taking the limit $k \to \infty$, which, since $\varepsilon >0$ is kept fixed and finite, can be consistently implemented. The rescaling \eqref{rescalek} is innocuous for the value of $Z^\star$.
Then, in order to regularize the 
running mass \eqref{eta} in this UV limit, we 
follow 
the analysis in the HECO case~\cite{Alexandre:2023qjo}, and replace the $k \to \infty$ limit in \eqref{eta} and \eqref{fpfull} by a \emph{regularized} UV fixed-point, defined as the limit of the rescaled $k$ \eqref{rescalek}: 
\begin{align}\label{kLambda}
    k \,\to \, \Lambda\,  < \, \infty\,.
\end{align}
Here, 
$\Lambda  < \infty$ defines the physical (subplanckian) energy scale above which new physics is expected in the pertinent experimental searches. In this way all the physical masses, momenta and energies in the effective theory, \emph{both} in the standard-model and the magnetic monopole sectors, remain below $\Lambda$, and are thus subplanckian.

Then, on account of \eqref{fpfull}, the MM physical mass $M^\star$ at the UV fixed point reads:\footnote{In terms of the rescaled variable $k$ \eqref{rescalek}, the argument of the running mass at the UV fixed-point  looks transplanckian, but this has no physical meaning, as we have already mentioned, given that the true (physical) UV cutoff of the effective theory is at $\Lambda=\Lambda_\text{phys}$.} 
\begin{align}\label{physmass}
M^\star \equiv \lim_{k \to \Lambda}M_r(k/\/\varepsilon)  \simeq 
\frac{1}{2} \, \Lambda \, \exp\Big(-\frac{8\pi^2}{3\, \varepsilon\,e_B^2} \, \, \frac{\zeta^2}{1 + \zeta^2}\, \frac{3 + 7\zeta^4 + 10\zeta^2}{3 + 3 \zeta^4 + 10 \zeta^2}\Big)\, , \qquad \zeta \equiv \frac{e_B}{e_A}\,,
\end{align}
where we remind the reader that $1 \gg \varepsilon > 0 $ remains finite. In fact, the strong dual coupling $e_B$ should be such that the quantity $\varepsilon \, e_B^2 >0$ in the argument of the exponential in \eqref{physmass} takes on values that allow for MM masses of order 
of the electroweak MM, as mentioned previously. 

The fixed point in the case of the MM effective model of \cite{Alexandre:2019iub}, based on the two-gauge potential formalism, depends on the values of the 
two couplings $e_A$, $e_B$ appearing in \eqref{Lagrangian}. However, things simplify, upon taking into account \eqref{ebea}.  
Indeed. in that limit, 
the UV fixed point \eqref{fpfull} is well approximated by:
\begin{align}\label{fp}
Z^\star &\simeq \frac{7}{9\, \varepsilon}\,,\nn 
\omega^\star &=\omega_0 \simeq \frac{4}{3}\,, \nn 
M^\star &\simeq \frac{1}{2} \Lambda \, \exp\left(-\frac{56\pi^2}{9\, \varepsilon\, e_B^2}\right)~.
\end{align}
The form of $M^\star$ as a function of the strong coupling $e_B$ resembles the situation of the HECOs~\cite{Alexandre:2023qjo}, for large wavefunction renormalization $Z^\star \gg 1$ (cf.\ the first of \eqref{fp}.

For the current LHC searches, which 
the current analysis refers to, the UV  scale $\Lambda$, for instance, should be at least larger than the maximally available center-of-mass collision energy, e.g.\ currently 14~TeV in order of magnitude:
\begin{align}\label{Lambdafix}
    \Lambda \, \gtrsim \, 14~\rm TeV\,.
\end{align}
However, as we shall discuss in Section~\ref{sec:massresum}, the actual value of the physical cutoff depends on microscopic details of the (UV complete) theory.
MM mass bounds, which, in view of the current resummation approach, can be reliably extracted from experimental collider searches, can actually constrain $\Lambda$ in concrete frameworks. 

At the fixed point \eqref{fpfull}-\eqref{physmass} (or \eqref{fp}), 
the following effective Lagrangian, valid for energy scales $E < \Lambda$, describes the dynamics of the interaction of MM with fermionic matter of electric charge $q_e$. From \eqref{Lagrangian}, and the above analysis, we observe that 
the UV-fixed-point \eqref{fp} Lagrangian, based on the running effective parameters, reads
(in the Feynman gauge $\lambda=1$)
\begin{align}\label{renlag}
{\cal L}_\text{eff} &= \frac{1}{2}A_\mu\,\eta^{\mu\nu}\Box\, A_\nu 
+ 
\frac{1}{2}B_\mu\Big((1+\omega^\star)\eta^{\mu\nu}\Box-\omega^\star \,\partial^\mu\partial^\nu\Big)B_\nu 
+
\overline\psi\Big(i\slashed\partial+ q_e\, \slashed{A}-m\Big)\psi  \nonumber \\
&+ \overline\chi\Big( i\, Z^\star\, \slashed\partial+  Z^\star\, e_A \slashed{A} + Z^\star\, e_B\slashed{B}-\widetilde M^\star\Big)\chi~
,\end{align}
where $\widetilde M$ denotes the renormalized mass at the UV fixed point. 
The running MM mass  and fine structure constant of the dual coupling $e_B$ at the UV fixed point, are,  respectively (for the rescaled $k$ (\eqref{rescalek}) ~\cite{Alexandre:2019iub}:
\be
M^\star \equiv \lim_{k\to \Lambda}M_r (k/\varepsilon)  = \frac{\widetilde M^\star }{Z^\star}~, \quad
\lim_{k\to \Lambda}\alpha_B(k/\varepsilon)=\frac{e_B^2/(4\pi)}{1+\omega^\star}~.
\ee
We then perform the simultaneous rescalings 
\be\label{resc}
B_\mu\to B_\mu/\sqrt{1+\omega^\star}~~~~\mbox{and}~~~~\chi\to\chi/\sqrt{Z}~, 
\ee
to obtain the canonically normalized Lagrangian
\begin{align}\label{efflag}
{\cal L}_\text{eff} &\to 
\frac{1}{2}A_\mu\,\eta^{\mu\nu}\Box\, A_\nu 
+  \frac{1}{2}B_\mu\left(\eta^{\mu\nu}\Box -\frac{\omega^\star}{1+\omega^\star}\partial^\mu\partial^\nu\right) B_\nu +
\overline\psi\Big(i\slashed\partial+ q_e\, \slashed{A}-m\Big)\psi \nn 
&+\overline\chi\left( i\slashed\partial+ e_A\, \slashed{A} + \frac{e_B}{\sqrt{1+\omega^\star}}\slashed B- M^\star\right)\chi ~.
\end{align}
Thus we observe that, for a fixed $k$, the effective  Lagrangian \eqref{efflag},  corresponding to the above-described DS resummation, is nothing other than a QED Lagrangian, but with a ``shifted'' covariant gauge parameter
\bea\label{shiftcg}
\lambda_\text{eff} = \frac{1}{1 + \omega^\star}= 1 - \frac{\omega^\star}{1+\omega^\star} 
\eea
for the strongly-coupled U(1) group, but normal Feynman gauge fixing for the electromagnetic U(1) and fermion matter sector. 

As discussed in the HECO case~\cite{Alexandre:2023qjo}, which also applies here, one observes from 
\eqref{shiftcg} that the effective action \eqref{efflag} is \emph{gauge fixed}. Hence,  the usual arguments on implementing Ward identities that would eliminate the wave-function renormalization factor $Z$ from the Feynman rule for the MM-fermion-gauge-boson vertex do not apply. According to the arguments given in \cite{Alexandre:2023qjo}, then, the Feynman rule to be applied on the pertinent vertices should be such that 
\begin{align}\label{gfshift}
e_{A,B} &\to Z\, e_{A,B}\,,
\end{align}
but the matter $\psi$-sector coupling remains the standard QED one, $q_e$. 

Thus, the final effective Lagrangian that produces the correct rules to be used in the computation of the pertinent MM-production cross sections, is:
\begin{align}\label{efflag2}
{\cal L}_\text{eff} &=
\frac{1}{2}A_\mu\,\eta^{\mu\nu}\Box\, A_\nu 
+   \frac{1}{2}B_\mu\left(\eta^{\mu\nu}\Box -\frac{\omega^\star}{1+\omega^\star}\partial^\mu\partial^\nu\right) B_\nu +
\overline\psi\Big(i\slashed\partial+ q_e\, \slashed{A}-m\Big)\psi \nn 
&+\overline\chi\left( i\slashed\partial+ Z^\star\,e_A\, \slashed{A} + \frac{Z^\star\,e_B}{\sqrt{1+\omega^\star}}\slashed B-M^\star \right)\chi ~.
\end{align}

\begin{figure}[t]
 \centering
\includegraphics[clip,width=0.25\linewidth]{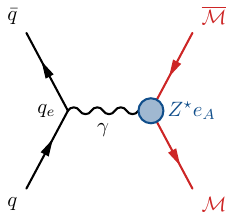} \hspace*{1cm}
\includegraphics[clip,width=0.26\linewidth]{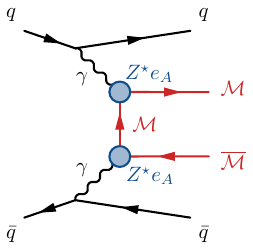} 
\hspace*{1cm}
\includegraphics[clip,width=0.27\linewidth]{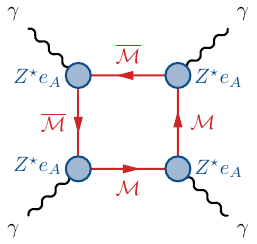} 
\caption{Drell--Yan (left), photon fusion (center) and one-loop (right) processes for the production of structureless MMs, $\mathcal{M}$, of spin-\hf at colliders. Wavy lines denote electrodynamical photons, $\gamma$. Straight black lines with arrows denote the 
scattered (anti)quarks, while red continuous arrow lines denote the (anti)monopoles. The shaded blob denotes the dressed coupling $Z^\star e_A$ due to quantum corrections induced by the strongly
coupled dual photon, which, in the EFT of \cite{Alexandre:2019iub}, couples only to monopoles. The fixed-point wavefunction renormalization  $Z^\star$ is given in \eqref{fp}. }  
\label{fig:DYPF}
\end{figure}

To avoid overcounting of the resummed loops, we need to restrict ourselves to tree-level DY or PF processes for monopole production (cf.\ Fig.~\ref{fig:DYPF}), of the type discussed in Refs.~\cite{MoEDAL:2021mpi,ATLAS:2023esy}, 
but with the important replacement of the standard tree-level Feynman rules by the above, effective ones, stemming from \eqref{efflag2}. 
For these physical processes,
at the one-loop resummation approximation, adopted here, 
and in \cite{Alexandre:2019iub,Alexandre:2023qjo}, the strong-coupling sector of the MM lagrangian plays no role, the only remnant being the renormalization \eqref{gfshift} of $e_A$ by $Z$. 

Thus, the pertinent Feynman rules, used in the  calculation of the pertinent cross sections in DY or PF processes, stem from the effective Lagrangian:
\begin{align}\label{efflag3}
{\cal L}_{\chi,\psi,A}^{{\rm DY/PF~eff}} =
\frac{1}{2}A_\mu\,\eta^{\mu\nu}\Box\, A_\nu 
+   
\overline\psi\Big(i\slashed\partial+ q_e\, \slashed{A}-m\Big)\psi +\overline\chi\Big( i\slashed\partial+ Z^\star e_A\, \slashed{A} - M^\star \Big)\chi ~,
\end{align}
with the various fixed-point (starred) quantities 
to be determined later on.
We stress at this point that, when evaluating the PF production cross section, one should use the Feynman rule for the ``tree-level'' MM propagator, as implied by the UV fixed-point  tree-level effective Lagrangian \eqref{efflag3}, in order to describe the internal MM line in the pertinent graph of Fig.~\ref{fig:DYPF}. 

We note at this point that the dressed coupling $Z^\ast e_A$ is also valid for the ``box-diagram'' of Fig.~\ref{fig:DYPF} (right), where virtual MMs in the loop may enhance the light-by-light scattering measurements at colliders~\cite{Ginzburg:1998vb,Ginzburg:1999ej,Epele:2012jn}. Such diagrams have constrained MMs indirectly, in a way complementary to direct searches, at the LHC~\cite{Ellis:2016glu,Ellis:2017edi,Mitsou:2026gvp} and the prospects at future $e^+e^-$ colliders are promising~\cite{Ellis:2022uxv}. 

We stress once again that, within the context of the effective theory of \cite{Alexandre:2019iub}, the parameters $\varepsilon$, $e_A$ and $e_B$ are, at this stage treated, as phenomenological, provided the hierarchy \eqref{ebea} is respected. 
In what follows, we shall discuss further, physically-motivated, relations between the couplings.
 As we shall discuss in the following section, $\varepsilon$ can be constrained, and expressed in terms of $e_A$ and $e$, by requiring that the DQC \eqref{schquant2} be satisfied, so that \eqref{Lagrangian} describes MMs and not dyons. 

\subsection{Definition of the magnetic charge and boundary conditions}\label{sec:modmagcharge}

In the presence of a UV non-trivial fixed point, 
with well defined wave-function renormalizations $Z$ and $\omega$, 
and a finite ratio 
\eqref{eta} of the 
MM-running-mass to UV-cutoff-$\Lambda$,
we define the magnetic charge of the effective fixed-point theory, which should satisfy the DQC \eqref{schquant2}, by means of the fixed-point value of $g_\text{eff}((k \to\Lambda)/\varepsilon)$ only (we remind the reader that we use the rescaled $k$ \eqref{rescalek}):
\begin{align}\label{fpgeff}
g_m = g_\text{eff}^\star \equiv \lim_{k \to \Lambda}g_\text{eff}(k/\varepsilon) = Z^\star \, e_A = \frac{7}{9\, \varepsilon} e_A = \frac{n}{\alpha}\, e\,.
\end{align}

This is an important point of our UV-fixed-point effective field theory approach to MM, within which MM is considered as a limiting case of `dyons' with a tiny electric bare charge, $e_A \propto \varepsilon$. Then, it follows from \eqref{fpgeff}, that the renormalized $e_a$ charge in the effective theory at the UV fixed point \eqref{fp}, $Z^\star\, e_A$, is finite, leading to a standard magnetic charge, obeying the DQC \eqref{schquant2}. This justifies the use of the tree-level cross sections of the DY and PF processes (cf.\ Fig.~\ref{fig:DYPF}) in experimental MM searches at colliders.\footnote{In this respect our `dyon' EFT to MM presented in \cite{Alexandre:2019iub} and here, has a broken CP symmetry, 
in agreement with generic dyon models~\cite{Terning:2020dzg}, otherwise the amplitude of the single DY process
would vanish~\cite{Ignatiev:1997pm,Terning:2020dzg}.}

With this interpretation, which is also consistent with the physical process of the formation of composite MM, as we shall discuss in section \ref{sec:elvscompres} (cf.\ Fig.~\ref{fig:compform}), the dual gauge coupling $e_B$ is assumed large compared to the other gauge couplings, $e_B \gg e_A, e$. Furthermore, the relative hierarchy between the couplings $e$ and $e_A$ is left unspecified, provided they are both perturbative, and thus much smaller than $e_B$. However, in the framework of the Zwanziger model~\cite{Zwanziger:1970hk}, which motivates the construction \eqref{Lagrangian}~\cite{Alexandre:2019iub}, one is forced to impose the restriction \eqref{eBgm} on $e_B$, i.e.\  $e_B=g_m$. 

It should be stressed that our approach is different from that of \cite{Calucci:1982wy}, where the renormalization of electric (``$e$'') and magnetic (``$m$'') charges is such that the corresponding product of wavefunction renormalizations equals unity, $Z_{e}\, Z_m =1$, so Dirac's condition is \emph{not} renormalized. In our approach, we introduce the $\beta$-independent effective magnetic charge \eqref{fpgeff}, and we do not renormalize the electron (or, more general, matter) charge, which is assumed to be approximately bare, given that the QED part of our action is in its strong perturbative regime at the energy scales we are interested in. For us, the gauge couplings do \emph{not} depend on the transmutation mass scale $k$, hence the non-renormalized Dirac condition is the one defined by means of  \eqref{fpgeff}.

We also note, for completeness, that the definition \eqref{fpgeff} and the associated consistency with the DQC \eqref{schquant3}, are compatible with the analysis of \cite{Alexandre:2019iub} in the non-fixed-point region $k \to k_0$. There, it was shown that, for $e_B \gg e_A$, of interest to us here, the non-fixed-point (running) effective coupling of photons with the MM, $g_\text{eff}(k) = Z(k)\, e_A \stackrel{k \to k_0}\simeq \frac{k-k_0}{k_0} \, \frac{7}{9\, \varepsilon} e_A  =  \beta (k \to k_0) \,  g_m $, where the last equality stems from \eqref{fpgeff}, and the quantity $0 < \beta (k \to k_0) = \frac{k-k_0}{k_0} \ll 1 $ has been argued~\cite{Alexandre:2019iub} to play the r\^ole of the Lorentz-invariant MM ``velocity'' $\beta$ appearing in the non-relativistic dyon--dyon scattering~\cite{Schwinger:1976fr,Milton:2006cp,Epele:2012jn}, which defines an effective  $\beta$-dependent magnetic charge $\beta g_m$, with $g_m$ obeying the DQC~\eqref{schquant2}. In our fixed-point theory, presented here, which is argued to describe the collider production of MM via DY and PF processes, we have, on account of \eqref{fpgeff}, that $\beta (k \to \Lambda) \equiv  \beta^\star = 1$, and, thus, there is no $\beta$-dependent magnetic coupling entering the pertinent MM production processes. Thus, we believe that our approach offers a clarification on this point. 

Let us now examine the consistency of the boundary conditions of the form \eqref{bc}. The correct relation between $M_0$ and $M$, should be the one that respects the DQC for the magnetic charge \eqref{schquant2}, which, in view of the new definition of the magnetic charge at the UV fixed point of the theory, is equivalent to the last equality in \eqref{fpgeff}. This implies:
\be\label{epsvseA}
 \varepsilon = \frac{7}{9} \frac{e_A}{g_m} = \frac{7}{18\,n} \frac{e_A}{\gd}\,.
\ee
where the fundamental Dirac charge has been defined in \eqref{schquant2}. 

Because of the assumed hierarchy 
\eqref{ebea},
and the perturbativity condition $|e_A| \ll \gd$, we observe from \eqref{epsvseA} that 
one must have 
\begin{align}\label{MepsM0}
M_0 = \varepsilon  M\,, \quad 0 < \varepsilon \ll 1\,,
\end{align} 
as the only type of allowed boundary conditions \eqref{bc} for consistency with the DQC.\footnote{In this sense, the initial conclusion of \cite{Alexandre:2019iub} on the incompatibility of the $M_0 \gg M$ case with the DQC remains valid, but under the lens of the new definition \eqref{fpgeff} of the magnetic charge.}

We stress that on imposing  \eqref{fpgeff}, the effective resummed fixed-point Lagrangian \eqref{efflag3} becomes equivalent to the one in the tree-level DY and PF processes shown in Fig.~\ref{fig:DYPF}) used in experimental searches~\cite{CDF:2005cvf,MoEDAL:2014ttp,ATLAS:2015tyu,MoEDAL:2016jlb,MoEDAL:2016lxh,MoEDAL:2017vhz,MoEDAL:2019ort,ATLAS:2019wkg,MoEDAL:2020pyb,MoEDAL:2021mpi,MoEDAL:2021vix,MoEDAL:2023ost,ATLAS:2023esy,MoEDAL:2024wbc,ATLAS:2024nzp}. This provides support to the MM production cross sections assumed, and consequent MM mass bounds set, in collider experiments~\cite{Mavromatos:2020gwk,Mitsou:2026zcf}.
This is an important aspect of this EFT, which is thus distinct from the case of (spin-\hf) HECOs~\cite{Alexandre:2023qjo}, where the DQC is absent, because there the only gauge coupling of the theory is the QED charge. Once the latter is renormalized by the non-perturbative wavefunction renormalization of the HECO field at the UV fixed point, it leads to an enhancement of the corresponding cross sections, calculated using the resummed theory~\cite{Alexandre:2023qjo,Alexandre:2024pbs}, as compared to the tree-level ones used in experimental searches.

\subsection{Magnetic monopole mass in the UV-fixed-point resummation approach}\label{sec:massresum}

Let us now  discuss in detail the behavior of the effective UV-fixed-point theory with this type of boundary condition. To determine the UV fixed point for such boundary conditions, 
we follow the analysis leading to \eqref{fp}, but now taking into account \eqref{MepsM0}. From \eqref{Zast} and \eqref{MoverM0}, by elimination of the variable $\eta$ then, 
and assuming again \eqref{ebea}, we obtain at the UV fixed point $k \to \Lambda$ (cf.\ \eqref{fp}):
\begin{align}\label{Zastnew}
Z^\star = \frac{7}{9\, \varepsilon} \, \gg \, 1\,.
\end{align}
This fixed point wave-function renormalization is larger than unity, and therefore, it is consistent with Unitarity even for elementary monopoles.\footnote{The situation in this respect resembles that of HECO resummation ~\cite{Alexandre:2023qjo}, however there the boundary condition for the wave-function renormalization at $k=k_0$ was 1 and \emph{not} zero, in contrast to our MM case here. The HECO resummation has led to $Z > 1$ behavior in the entire range of running of the transmutation-mass scale $k$.} As we lower the transmutation mass scale from the fixed point $k \to \Lambda$ to $k_0$, the wavefunction renormalization decreases smoothly, due to \eqref{dZdk}, crossing unity at a given $k=k_1$, determined from the appropriate running $\partial_k Z(k)$. For elementary (non composite) monopoles, going below the scale $k_1$ would not be allowed, as this would imply violations of Unitarity. However, by
considering composite monopoles, as we do here, we avoid such inconsistencies, and one can make sense of the theory in the regime $k \simeq k_0$, where the production of scattering of monopoles off matter acquires the $\beta$-dependent form suggested in \cite{Schwinger:1976fr,Milton:2006cp}.  In Appendix \ref{sec:k1unitarity} we estimate $k_1$ in some representative cases, and find it to lie near $k_0$.
 
\begin{figure}[ht]
    \centering
\includegraphics[width=0.65\linewidth]{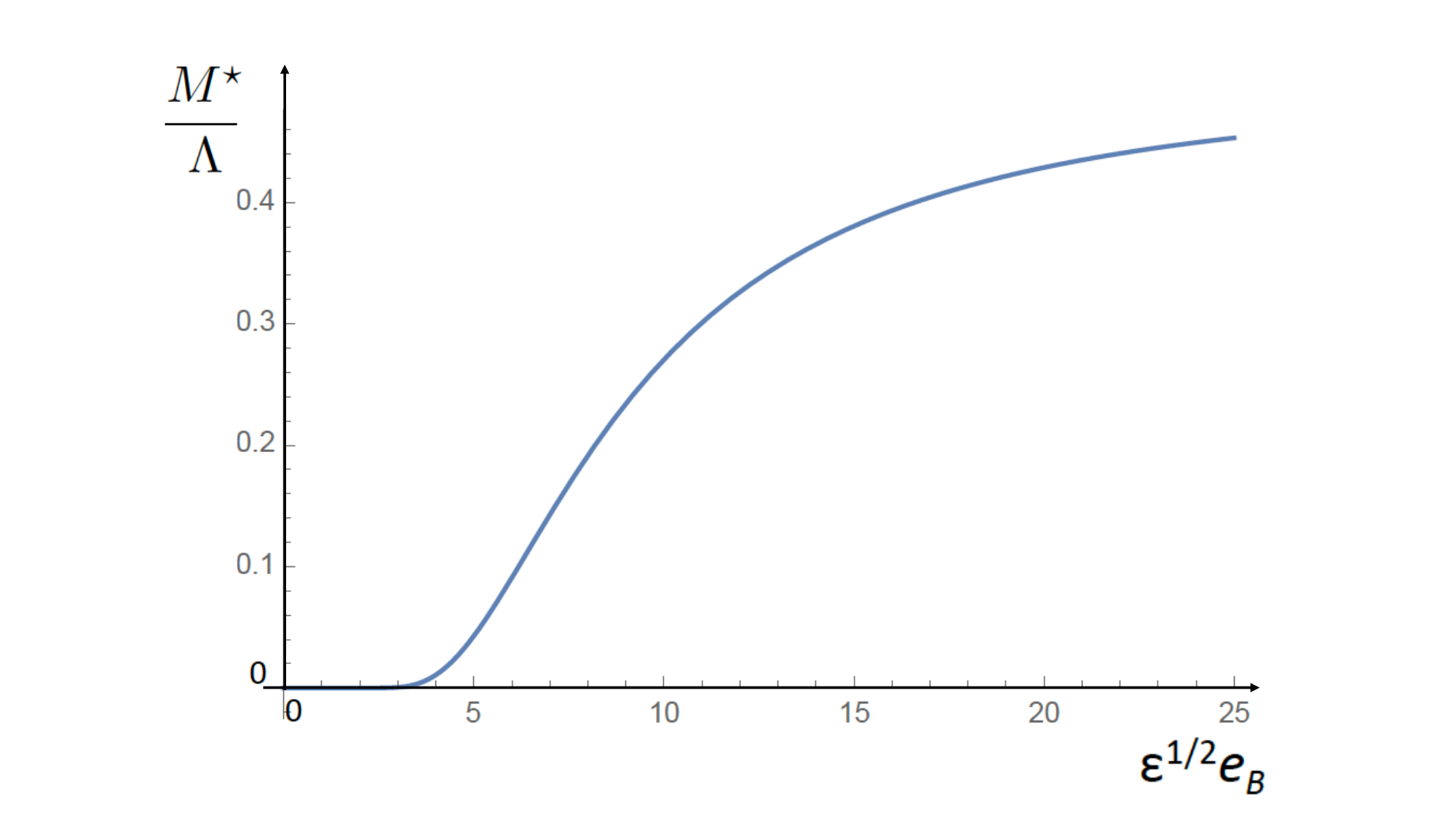}
    \caption{The MM mass \eqref{MMmasseps} (in units of $\Lambda$, as a function of $\varepsilon^{1/2}\, e_B$, where $\varepsilon$ is given in \eqref{epsvseA}, and is proportional to $e_A/n$. Qualitatively, the dependence on the strong coupling is similar to the case of fermion HECOs~\cite{Alexandre:2023qjo}, when their mass is plotted against the high electric charge in that case.}
    \label{fig:MMmass}
\end{figure}

One can then go on and calculate the MM mass at the UV fixed point \eqref{fp}, for the case $e_B \gg e_A$ we concentrate upon here. We give it below again for convenience of the reader:
\begin{align}\label{MMmasseps}
M^\star \stackrel{(e_B \gg e_A)}{\simeq} 
\frac{\Lambda}{2}\, \exp\left(-\frac{8\,Z^\star\pi^2}{e_B^2} \right) =
\frac{\Lambda}{2}\, \exp\left(-\frac{56\pi^2}{9\, \varepsilon\, e_B^2} \right)   \stackrel{\eqref{epsvseA}}{=} 
\frac{\Lambda}{2}\, \exp\left(-\frac{8\pi^2}{\frac{e_A}{g_m}\, e_B^2} \right) =
\frac{\Lambda}{2}\, \exp\left(-\frac{16\pi^2}{\frac{e_A}{n \,\gd}\, e_B^2} \right)\,, \quad n \in \mathbb Z\,.
\end{align}
This MM mass formula works for both elementary and composite MMs, as we shall discuss below. 
The MM mass is a monotonically increasing function of $\sqrt{\varepsilon}\, e_B$, as shown in Fig.~\ref{fig:MMmass}), and is proportional to the UV physical cutoff $\Lambda$, 
which in our approach is a phenomenological subplanckian energy scale. The mass \eqref{MMmasseps} is always smaller than $\Lambda/2$, consistently with the role of $\Lambda$ as an UV cutoff of an effective theory, which incorporates the MM. The cutoff $\Lambda$ has been assumed to
satisfy \eqref{Lambdafix}. As already mentioned, this allows comfortably for the incorporation into the formalism of electroweak MM with mass of order $M_W/\alpha =\mathcal O(10~{\rm TeV})$. 

\begin{figure}[htbp]
  \centering
    \includegraphics[width=0.49\linewidth]{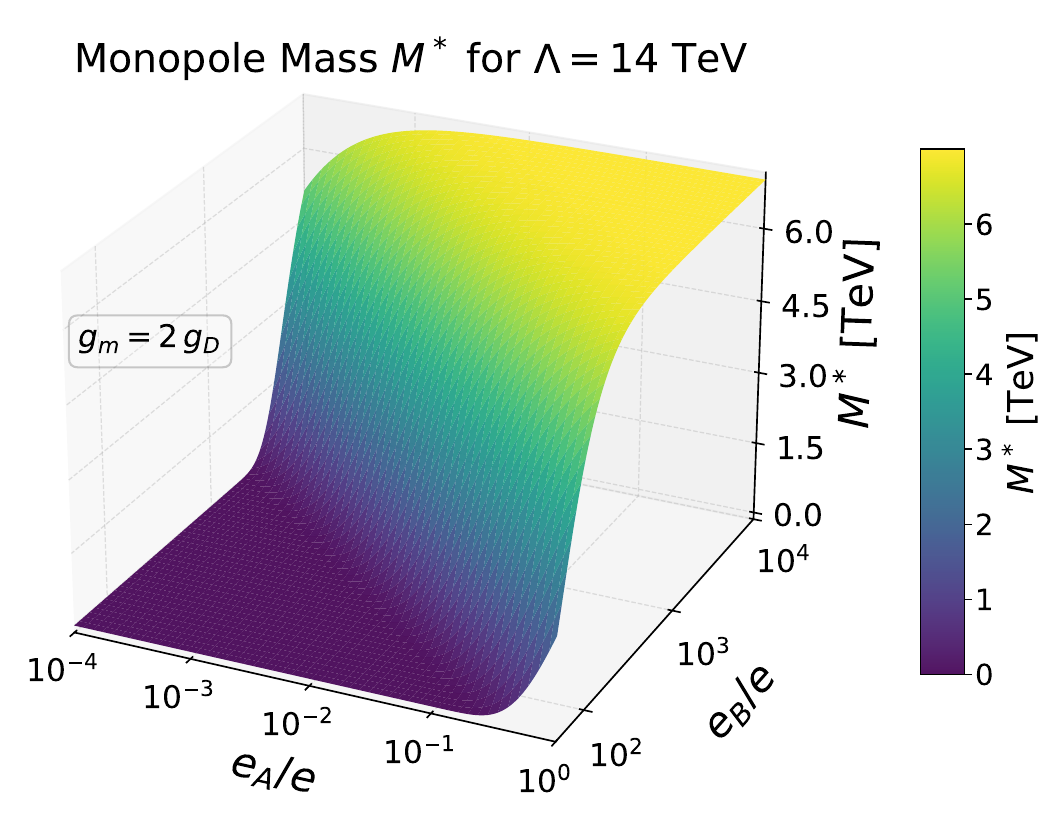}
  \hfill
    \includegraphics[width=0.49\linewidth]{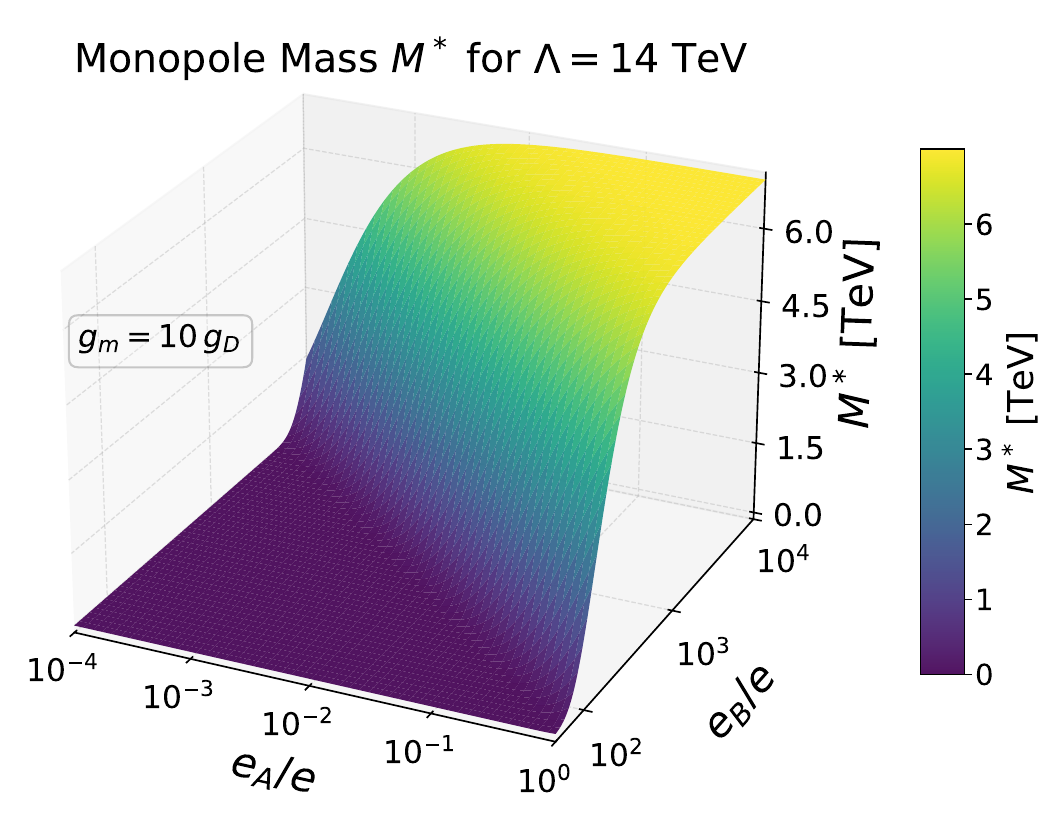}
  \caption{Monopole mass $M^*$ \eqref{MMmasseps} versus parameters $e_A$ and $e_B$ for $\Lambda = 14~\tev$ and for different magnetic charges: $g_m = 2\gd$ (left) and $g_m = 10\gd$ (right).}
  \label{fig:monopole_lambda14_alln}
\end{figure}

The dependence of the MM mass \eqref{MMmasseps} on the two parameters $e_A$ and $e_B$, taken as free parameters yet respecting the hierarchy \eqref{ebea}, is graphically presented in Fig.~\ref{fig:monopole_lambda14_alln} for MM of charge 2\gd and 10\gd. The mass-scale parameter is assumed equal to the minimum value \eqref{Lambdafix} allowed at the LHC energies: $\Lambda = 14~\tev$. As expected, the mass $M^\star$ increases with $e_A$ and $e_B$ approaching a plateau just below $\Lambda/2$. The ranges of $e_A$ and $e_B$ values have been chosen so that the hierarchy $e_B \gg e_A$ is respected.

\begin{figure}[htbp]
  \centering
    \includegraphics[width=0.49\linewidth]{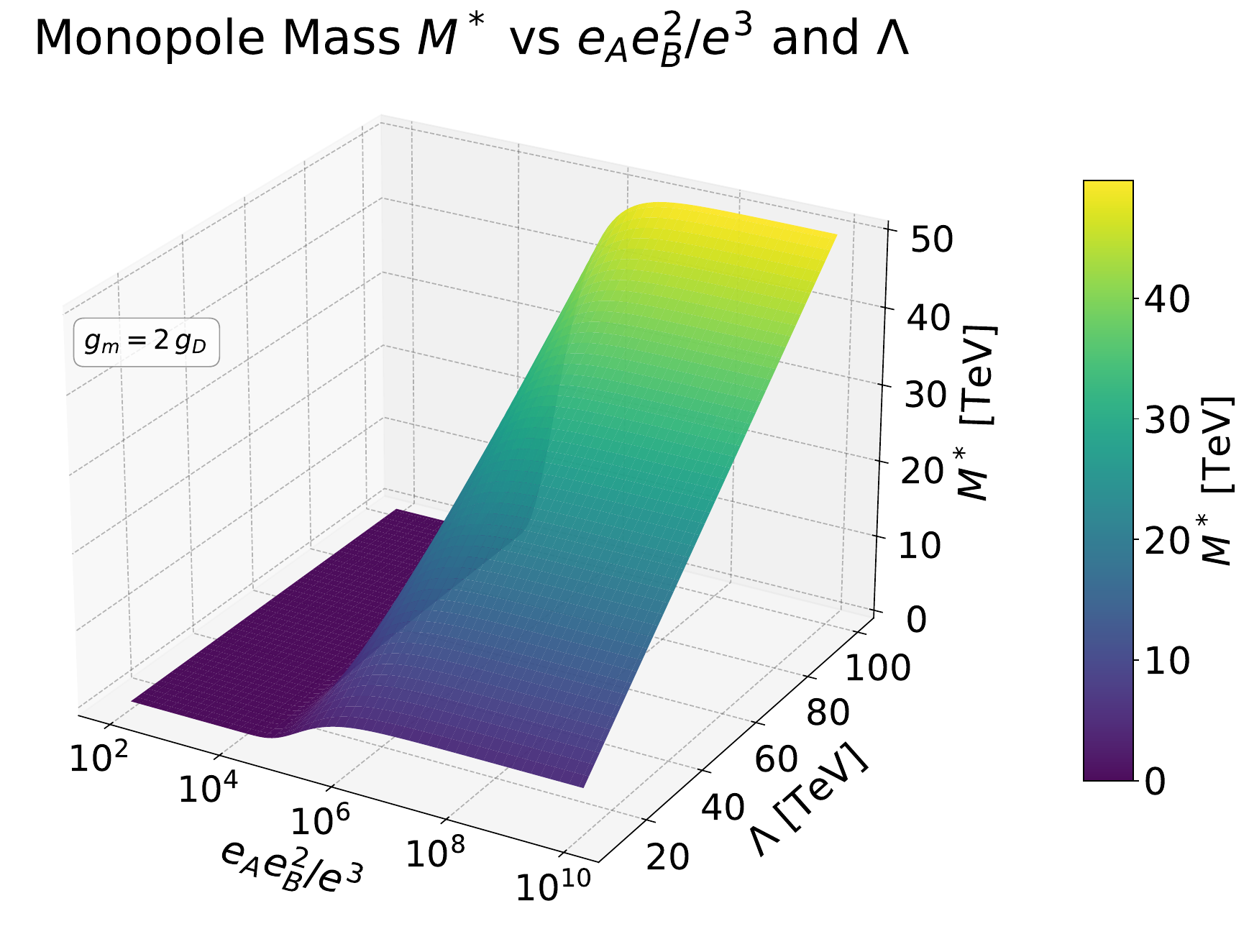}
  \hfill
    \includegraphics[width=0.49\linewidth]{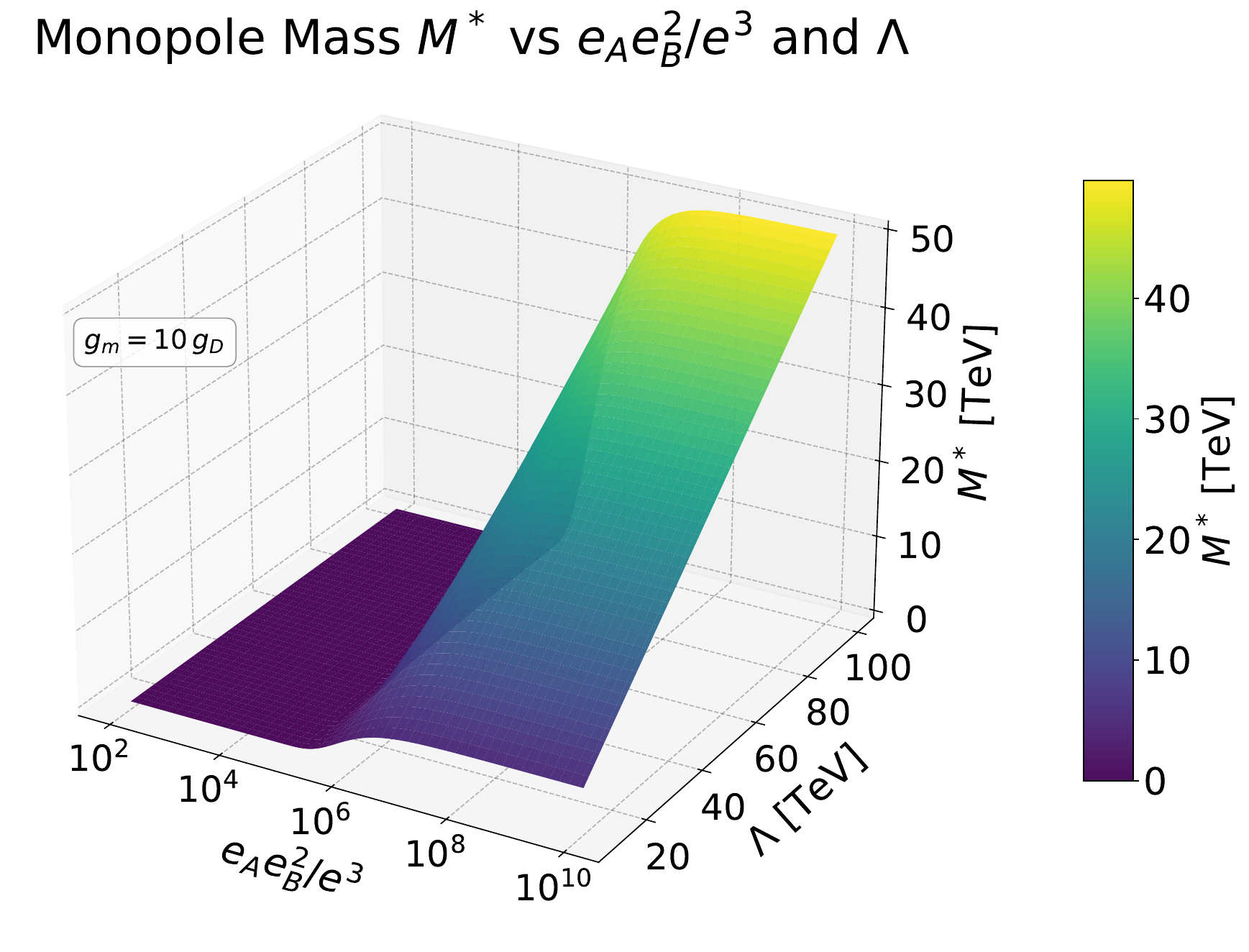}
  \caption{Monopole mass $M^*$ \eqref{MMmasseps} as a function of the mass scale $\Lambda$ and the product $e_Ae_B^2$ for different magnetic charges: $g_m = 2\gd$ (left) and $g_m = 10\gd$ (right).}
  \label{fig:3d_massvslambdaeaeb2_alln}
\end{figure}

Taking advantage of the modular form \eqref{MMmasseps} with which the MM mass depends on $e_A$ and $e_B$, a more general depiction of $M^\star$ as a function of the product $e_A e_B^2$ and the mass scale $\Lambda$ is presented in Fig.~\ref{fig:3d_massvslambdaeaeb2_alln}, again for $g_m = 2\gd$ and $g_m = 10\gd$. The linear increase with $\Lambda$ is evident, as well as the growth with $e_A e_B^2$. The values of $e_A$ and $e_B$ are expressed as multiples of the positron charge $e$, to facilitate comparisons with QED and maintain the result independent of the electromagnetism units used.

\subsubsection{Specific cases of phenomenological interest motivated by the Zwanziger model }\label{sec:zwanz}

The reader is reminded at this point that in the Zwanziger model one has the relation \eqref{eBgm}, which, in view of our definition of the magnetic charge $g_m$ \eqref{fpgeff}, 
would read 
\begin{align}\label{eAR}
e_B = g_m = Z^\star e_A \equiv e_A^R\,,
\end{align}
consistent with the coupling hierarchy \eqref{eBgm}. Notice, then, that, at the level of the resummed UV-fixed-point Lagrangian \eqref{efflag2}, the equations of motion with respect the strongly-coupled gauge potential $B_\mu$ yields a magnetic current but with a modified (non-perturbatively renormalized) magnetic coupling 
\begin{align}\label{eBR}
e_B^R = \frac{Z^\star}{\sqrt{1 + \omega^\star}} \, g_m \stackrel{\eqref{eAR}}{=} \, \frac{Z^\star}{\sqrt{1 + \omega^\star}} \, e_A^R\,, 
\end{align}
which is still discretized in view of the DQC \eqref{schquant2}. It should be stressed that the prefactor in front of the $g_m$ in the above expression is much larger than  unity (cf.\ \eqref{fp}) for $\varepsilon \ll 1$, thus respecting  
the coupling hierarchy \eqref{ebea}, which our analysis is based upon, at a full quantum (renormalized coupling) level.
On account of \eqref{eBgm}, then, the MM mass \eqref{MMmasseps}
becomes:
\begin{align}\label{massLambda1}
 M^\star \simeq \frac{\Lambda}{2}\, \exp \left(-\frac{e}{e_A}\,\frac{2\pi}{n}\right)\,, \quad n \in \mathbb Z\,,   
\end{align}
where we used the DQC \eqref{schquant2}. 

A convenient, and phenomenologically motivated choice of $e_A$ is the one in which
case \eqref{massLambda1} reduces to
\begin{align}\label{eeA}
    |e_A| \sim e \,.
\end{align}
which, combined with \eqref{eBgm}, implies for the MM mass \eqref{massLambda1}:
\begin{align}\label{massLambda2}
 M^\star \simeq \frac{\Lambda}{2}\, \exp \left(-\frac{2\pi}{n}\right)\,, \quad n \in \mathbb Z\,,   
\end{align}
where we used the DQC \eqref{schquant3}. In that case, bounds on MM masses $M^\star$, imposed by collider searches, such as the LHC (by ATLAS~\cite{ATLAS:2023esy} and MoEDAL~\cite{MoEDAL:2023ost} Experiments, will impose lower bounds on the effective cutoff $\Lambda$. For $n^\prime =2$, which corresponds to the lowest MM number that characterize miroscopic models in the current literature~\cite{Mavromatos:2020gwk}, which also admit structureless (Dirac-type) MM in their solutions~(e.g.\ those in \cite{Mavromatos:2018kcd}), we do have $M^\star \sim 1.7 \times 10^{-6} \, \Lambda$. In that case, we observe that for the theory to admit electroweak-scale MM of masses od order $\mathcal{O}(10~{\rm TeV})$, we need a $\Lambda \sim 5.7 \times 10^{9}$~GeV. On the other hand for $\Lambda \sim M_{\rm Pl}$, which is naturally expected when embed the models in UV complete theories of quantum gravity, with quantum-gravitational effects expected to dominate above the reduced Planck energy scale $M_{\rm Pl}$, we arrive at MM masses $M^\star \sim 4 \times 10^{12}$~GeV.  

In the next section \ref{sec:elvscompres}, we shall consider the application of resummation in the case of production of composite MMs,  within the context of the model of \cite{Alexandre:2019iub}, as augmented with the UV fixed-point discussion in the present article, and examine potential (but stringent) conditions, under which the entropy-aergument in \cite{Drukier:1981fq} for the suppression of the production cross sections at colliders, might be avoided.

\section{Elementary vs.\ composite magnetic Monopoles and Resummation}\label{sec:elvscompres}

Before proceeding to the construction of our UV-fixed-point resummed theory, we discuss an important difference between elementary and composite MM, as far as their production at colliders is concerned, which is not often discussed in the pertinent literature. This has to do with an extreme suppression of the production of composite MM, which consist of a large number of coherent quanta of SM charged particles, like $W^\pm$-boson and charged Higgs excitations $h^\pm$, whose collective coupling to photons define the magnetic charge, according to the argumentation of \cite{Drukier:1981fq} (the fact that the composite MMs contain an unbroken Higgs phase at their centre, implies that the charged higgs goldstone bosons of the SM group will play a role in the formation of the MM through appropriate diagrams~\cite{Drukier:1981fq}). 

\subsection{An ``entropy-mismatch'' argument for the suppression of the composite-MM production at colliders}\label{sec:entropy}

\begin{figure}[ht]
 \centering
\includegraphics[clip,width=0.80\linewidth]{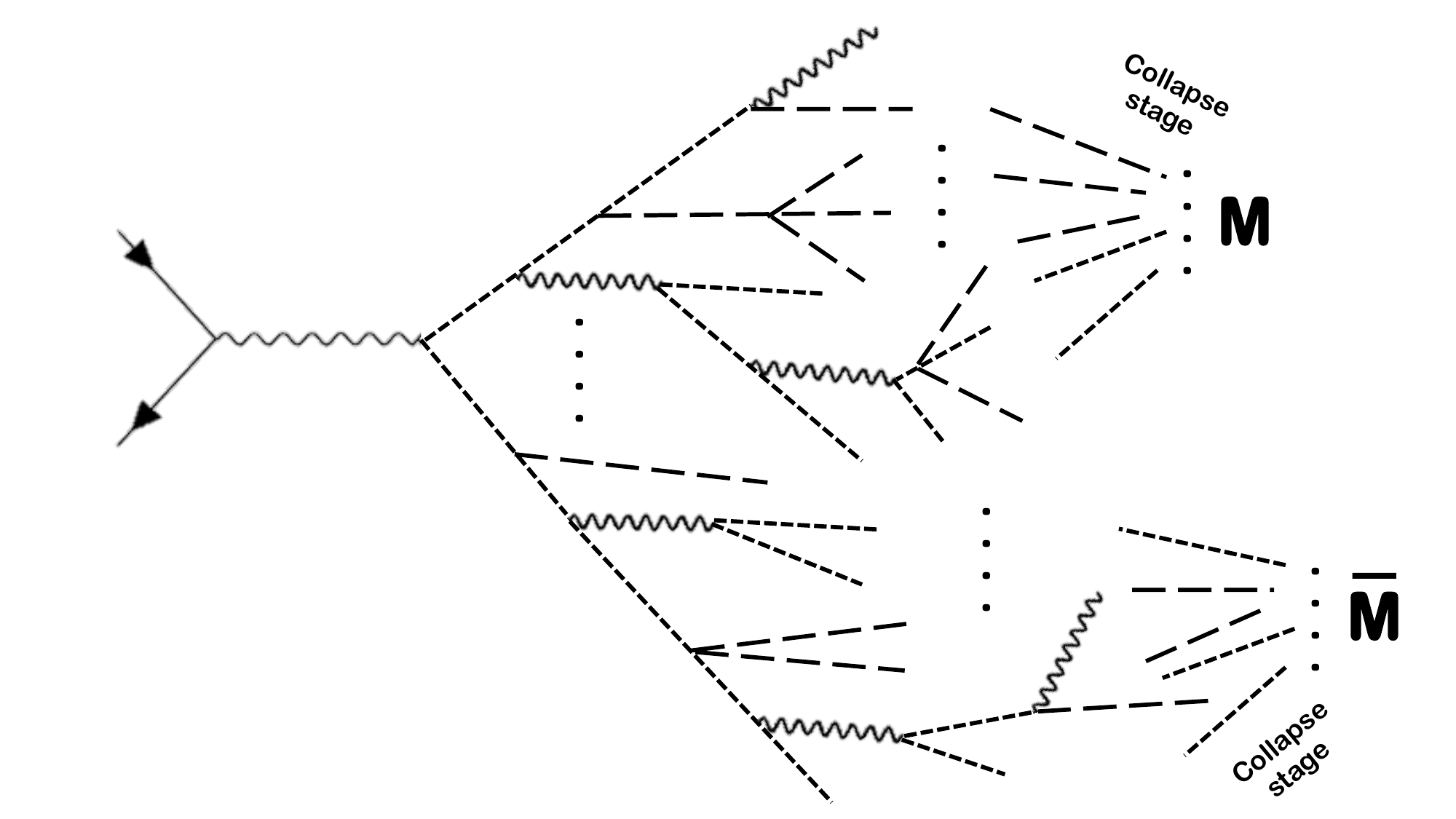}
\caption{Tree-level Feynman Diagram relevant for the formation of a composite electroweak magnetic-monopole ($\mathbf M$)-antimonopole ($\mathbf{\overline{M}}$) pair from elementary SM particles, according to \cite{Drukier:1981fq}. The diagram can represent the DY process or the appropriate part of the PF process. Continuous lines with arrows denote SM fermions (quarks or leptons, depending on the collider). Wavy lines denote $\gamma/Z^0$ combinations, short-dash lines $W^\pm$ SU(2) gauge bosons, while long-dash lines denote charged Higgs excitations $h^\pm$, which play a role in the presence of MMs, since the gauge symmetry is unbroken near the MM centre. The ``Collapse stage'' denotes the region at which $2/\alpha$ coherent-mode quanta of $W^\pm$ and $h^\pm$ collapse to form $\mathbf{M}$  or $\mathbf{\overline{M}}$.}
\label{fig:compform}
\end{figure}

As argued in \cite{Drukier:1981fq}, the main reason for suppression of the production cross section of MM which are composite of $W^\pm$ and $h^\pm$, is not the final stage of formation of MM (through collapse processes of $W^\pm, h^\pm$ quanta) in the collision of SM particles, say quarks in DY, or photons in PF processes, 
but rather the many successive intermediate stages that the system has to undergo before the collapse to a monopole-antimonopole pair (cf.\ Fig.~\ref{fig:compform}). In view of the fact that the energy scale involved (virtual momenta in exchange tree-level graphs) is much larger than the characteriztic electroweak scale (of order of the mass of the $W$-bosons, for the cases of the electroweak MM, say of Cho-Maison type~\cite{Cho:1996qd}, and its finite-energy variants~\cite{Cho:2013vba,Ellis:2016glu,Arunasalam:2017eyu,Mavromatos:2018kcd}), these processes are all tree level, as a consequence of the asymptotic freedom  of the non-Abelian SU(2) gauge group at such high energy regime. 
The so formed MM contains a large $\sim \frac{2}{\alpha}$ number of $W^\pm, h^\pm$ quanta, where $\alpha = e^2/(4\pi)$ (in units $\hbar=c=1$ we work throughout)  is the fine structure constant of QED. Based on this fact, the authors of \cite{Drukier:1981fq} argued that the corresponding diagram which determines the formation of the MM-anti-MM pair is of $\frac{2}{\alpha}$-th order in perturbation theory of the electroweak SU(2) coupling $g = e/\sin\theta_{\rm w}$ (with $\theta_{\rm w}$ the weak mixing angle.  Vertices $W^+\,W^-\,\gamma/Z^0$, with $\gamma$ denoting photons, and $Z^0$ the neutral SU(2) bosons, 
or $W^+\,W^- h^+\, h^-$, Higgs self couplings etc., appear in this complicated diagram, which involves tree-level exchanges (cf.\ Fig.~\ref{fig:compform}). In each of them, the emission of relevant quanta contributes 
an SU(2)-coupling-factor $g \propto e = (4\pi \alpha)^{1/2}$  
(we ignore the r\^ole of the weak mixing angle, which is not important for our purposes here). 
Thus, the entire tree diagram is characterized by an overall  suppression factor 
in the corresponding  amplitude 
of the form 
\be\label{collcoupl}
(\sqrt{4\pi\alpha})^{2/\alpha} \sim e^{-2/\alpha}\,, 
\ee
where we took into account~\cite{Drukier:1981fq} that, in order of magnitude, one has $(4\pi \alpha)^{1/2} \sim e^{-1}$.

This yields~\cite{Drukier:1981fq} for the cross section a
suppression factor 
$\mathcal F \sim e^{-4/\alpha} 
\sim 10^{-250}$,  implying that composite-MM production through elementary-particle collisions at current colliders 
is negligible.
This is essentially an argument on an ``entropy mismatch'' between initial and final states in the process of MM production at colliders. There is an  incoming configuration of particles, highly localized in spacetime (effectively ``point-like'': quarks, photons etc.), which has very low entropy as compared with the macroscopic coherent superposition of charged elementary constituents ($W^\pm, h^\pm$) which form the produced MMs, and which are characterized collectively by a huge entropy.

We stress, however, that the above argument applies only to composite MM, consisting of coherent states of $W^\pm$ and $h^\pm$ quanta, that is the aforementioned electroweak MM models, or even the electroweak MM model of \cite{Hung:2020vuo,Ellis:2020bpy}, based on a particle physics model for neutrino masses~\cite{Hung:2006ap}, with an extra Higgs-like real triplet.
The above discussion could be interpreted, therefore, that there would be only two ways for an unsuppressed production of 
such composite MM:  either through appropriate phase transitions in the early Universe~\cite{Drukier:1981fq}, in which simple gauge groups are spontaneously broken, or 
via the Schwinger mechanism~\cite{Affleck:1981ag,Gould:2017fve,Gould:2018efv,Gould:2017zwi,Gould:2018ovk,Gould:2018efv,Gould:2019myj,Gould:2021bre}, mentioned above, which yields the only reliable experimental mass bounds for composite MMs at colliders~\cite{MoEDAL:2021vix}. 

The aforementioned  suppression does \emph{not} characterize the cases of elementary (structureless)  Dirac-type MMs. It is important to notice at this stage that elementary MM can exist as appropriate limiting cases of composite-MM models, as happens, for instance, for the model of \cite{Mavromatos:2018kcd}, where a Dirac type monopole arises in a specific limiting case of the pertinent parameter space. 
The effective two-gauge-potential resummed field theory of \cite{Alexandre:2019iub} does apply to such elementary MMs, if they have spin \hf  (appropriate extensions of the model of \cite{Alexandre:2019iub} for scalar or vector MMs are required, and are straightforward to construct). The theory is essential for a quantification of the production of elementary MM at current colliders, 
via the calculation of the pertinent cross sections, after appropriate resummation in the spirit of DS.\footnote{The EFT of \cite{Alexandre:2019iub} does not characterize the Schwinger vacuum production~\cite{Affleck:1981ag,Gould:2017fve,Gould:2018efv,Gould:2017zwi,Gould:2018ovk,Gould:2018efv,Gould:2019myj,Gould:2021bre}, which is a truly non-perturbative quantum process.} On the other hand, as already mentioned, the approach of \cite{Drukier:1981fq} gave a physical microscopic meaning to the magnetic charge, that of a collective coupling of $2/\alpha$ coherent quanta of $W^\pm, h^\pm$ involved in the formation of a MM,  to a photon field.  Indeed, the new collective coupling assumes the value $\frac{2}{\alpha}\, e$, which is nothing other than the magnetic charge of all these composite MM. 

This concept of ``emergent'' magnetic charge has been used, in a different context, in the EFT of \cite{Alexandre:2019iub}, and we also make use of it here, specifically in section \ref{sec:modmagcharge} (cf.\ \eqref{fpgeff}), when we apply the resummed effective theory at the non-trivial UV fixed point to estimate the pertinent production cross sections. In the next subsection \ref{sec:comporesum}, we shall also argue in favour of the fact that resummation in the presence of composite MMs, formed by the collapse of the $W^\pm-h^\pm$ quanta to a monopole-antimnopole pair, could provide, under certain circumstances, an amplification factor which avoids the suppression arguments of \cite{Drukier:1981fq}. As we shall explain, it is the huge value of the wave-function renormalization of the MM at the UV fixed point that is responsible for this feature. This  formally would place the composite MM case on equal footing with that of  the elementary ones, as far as their production at colliders is concerned. However, we stress that, at present, connecting the EFT developed here and in \cite{Alexandre:2019iub} to microscopic models containing solitonic composite MM configurations in their spectra is a highly non-trivial open task. 

\subsection{Resummation and composite MM production at colliders}\label{sec:comporesum}

In this subsection we apply the resummation in the context of the UV fixed-point model for MM of \cite{Alexandre:2019iub}, for which \eqref{Zastnew}, 
\eqref{fpgeff} are valid, to the case of the final stage of the formation of \emph{composite} (solitonic) MM-anti-MM pairs (cf.\ Fig.~\ref{fig:compform}). The important point to notice is that the parameter $\varepsilon$, which characterizes the allowed class of boundary conditions \eqref{MepsM0}, is \emph{ phenomenological}, chosen in such a way that the DQC \eqref{fpgeff} is satisfied. 

In the composite MM case, the collective coupling to photons, $e_A$, is extremely suppressed due to the factor \eqref{collcoupl}, as a consequence of the complicated diagram of Fig.~\ref{fig:compform}. As $e_A$ represents the collective coupling to the photon in the diagram, it is natural to assume in the case of composite MM the following suppression:
\be\label{eaorder}
e_A \propto  e^{-2/\alpha} \, \vert \mathcal J \vert \, e\,,
\ee
where $\mathcal J$ is a finite quantity to be determined below.

The effects of the suppression factor $e^{-2/\alpha}$ in \eqref{eaorder}  on the magnetic charge $g_m =g^\star_\text{eff}$ (cf.\ \eqref{fpgeff}) can be compensated by explicitly selecting the parameter $\varepsilon$ as follows:
\be\label{ea2}
\varepsilon \propto \frac{7}{9} \alpha\, e^{-2/\alpha}\,,
\ee
so that the DQC \eqref{schquant2},  \eqref{fpgeff}, with MM number $n=2n^\prime$, is satisfied, provided one chooses\footnote{ This is a natural and perhaps necessary choice, as this avoids the dependence of the parameter $\varepsilon$, entering the boundary-condition \eqref{bc}, on the MM topological number, and thus the MM species. 
We also remind the reader that the case $n^\prime=2$ is the value of the lowest-lying electroweak MM solutions with magnetic charge $2\gd$, that are currently constructed explicitly in the literature~\cite{Mavromatos:2020gwk}. However, we stress once again that in our work here we keep $n^\prime \in \mathbb Z$ general, in order to match experimental searches for magnetic charges which are both odd and even integer multiples of the fundamental Dirac charge \gd.}
\be\label{choiceofJ}
\mathcal J = 2n^\prime, \quad n^\prime \in \mathbb Z\,.
\ee
This determines the order of magnitude of the wave function renormalization \eqref{Zastnew} at the UV fixed point, $Z^\star$, and thus of the quantum resummation effects. On account of \eqref{ea2} we have:
\be\label{zstarea}
Z^\star = \frac{e^{2/\alpha}}{\alpha} \gg 1\,.
\ee
As we have discussed above, in view of \eqref{MMmasseps}, \eqref{massLambda1}, then,  it is straightforward to 
produce phenomenologically realistic MM masses of order $M_W/\alpha = \mathcal O(10)$~TeV, provided the cutoff $\Lambda$ and the argument of the exponential are chosen accordingly. 

Despite the extreme weakness of the MM -photon coupling $e_A$, 
the quantum resummation effects of the well-behaved (at the non-trivial UV-fixed-point \eqref{fpfull} (or \eqref{fp})) EFT of \cite{Alexandre:2019iub} dominate the final (collapse) stage of the formation of the monopole-antimonopole pair in Fig.~\ref{fig:compform}, thereby compensating the initial suppression. Such (purely quantum) effects are expressed by the very large wave-function renormalizastion \eqref{zstarea}, which yields an unsuppressed coupling of the photon to the MM pair, the magnetixc charge $g_m$ \eqref{fpgeff}, satisfying the unrenormalized DQC \eqref{schquant2}.
This is the extra ingredient, which leads to a picture that supports the unsuppressed production of composite MM at colliders, on equal footing to the elementary MM case. 
In our proposed effective-field-theory approach to MM production in \cite{Alexandre:2019iub} and here, it is the strongly-coupled U(1)$^\prime$ gauge field that leads to the MM selection among the several final states that are produced by the collision of standard model particles. This is reminiscent to the r\^ole of the strong magnetic field, leading to the unsuppressed production of the MM from the vacuum in the dual Schwinger mechanism~\cite{Affleck:1981ag,Gould:2017fve,Gould:2017zwi,Gould:2018efv,Gould:2018ovk,Gould:2019myj,Gould:2021bre}. In our case, the huge value of the MM wavefunction renormalization may reduce the size of the MM to its Compton wavelength ($\sim 1/M^\star$), thereby 
rendering the MM a quantum excitation, and thus avoiding the strong cross-section suppression. For which microscopic composite MM this is valid, however, which could thus be described by our effective UV-fixed-point field theory, is a model-dependent, and still wide-open, problem. 

At this stage, a  detailed connection of microscopic models that can accommodate electroweak MMs with our EFT approach is yet to be determined. In the argumentation of \cite{Drukier:1981fq}, the microscopic theories that admit MM as solitons in their spectra, consisting of appropriately charged SM ingredients, provide the ultimate microscopic Lagrangians, and in this sense, the suppressed production of MM at colliders follows from ``entropic'' arguments, based on the rarity of the MM final states (cf.\ Fig.~\ref{fig:compform}), 
among all other potential final states produced in the collision. In our approach \cite{Alexandre:2019iub}, the standard model of particle physics is augmented with a strongly-coupled ``dark photon'' interaction with the properties outlined in Appendix \ref{sec:app}, which we ascribe physical significance to. This interaction plays, as already mentioned, analogous r\^ole to the strong background magnetic field required for the MM production in the dual Schwinger effect. Whether such a proposal is correct physically can be tested by a MM observation in the above-described collider searches of MM at colliders.

\section{Experimentally Constrained Parameters}\label{sec:mass-limits}

In the previous sections we have used the DQC based on the Schwinger's formula, \eqref{schquant}, or better \eqref{schquant2} for MMs, for which the magnetic charge is really an \emph{even} multiple of \gd. This was the case for the Schwinger quantization condition \eqref{schquant}~\cite{Schwinger:1969ib}, which characterizes non-relativistic dyon--dyon scattering~\cite{Schwinger:1976fr}, but also of 
the microscopic field-theory models of electroweak MMs in the current literature~\cite{Mavromatos:2020gwk,Cho:1996qd,Cho:2013vba,Arunasalam:2017eyu,Mavromatos:2018kcd}, 
which are also known to contain structureless (Dirac) MM in their solutions, in certain limits (e.g.\ see~\cite{Mavromatos:2018kcd}). 

Nonetheless, for experimental searches, we follow a more general approach, also considering MMs with magnetic charges \emph{odd} multiples of \gd. This is achieved by modifying the condition \eqref{schquant2}, and the pertinent mass formulae \eqref{MMmassnover2}
through the replacement  
\be\label{nn2}
n \to 
\frac{n^\prime}{2}\,, \quad n^\prime \in \mathbb Z\,, 
\ee
which allows for magnetic charges to be both even and odd multiples of \gd:
\be\label{schquant3}
g_m = \frac{n^\prime}{2\alpha} \, e = n^\prime \, \gd \, , \quad n^\prime \in \mathbb Z\,. 
\ee
This generalization will be used throughout this section for the interpretation of experimental searches using the fixed-point resummation model developed in the current work. 

The $e_B$ coupling does not enter the amplitudes corresponding to the DY or PF processes in the context of the resummed fixed-point theory \eqref{efflag3}. Only the renormalized coupling $Z^\star \, e_A =g_m $ does, which is the magnetic charge in our definition \eqref{fpgeff}, rendering the cross-section values identical to the ones before resummation, as discussed in Section~\ref{sec:modmagcharge}. For this reason, unlike the HECO case~\cite{Alexandre:2023qjo,Alexandre:2024pbs}, the lower mass limits obtained from the experimental upper bounds on cross sections \emph{are not modified} by the applications of one-loop DS resummation schemes. However, they may be altered by higher-order loop-resummation  calculations.    

Again, we stress here that $e_A$ does not represent an electric charge with which MMs interact with SM particles in addition to the magnetic charge, which would make them act as dyons. Instead, $e_A$ is treated from the experimental point of view as a parameter of the resummation process. Therefore, results from  searches for MMs (not dyons) can be used to constrain the effective theory in question.

Under \eqref{nn2} and \eqref{schquant3}, 
the condition \eqref{fpgeff} becomes
\begin{align}\label{fpgeff2}
Z^\star \, e_A = \frac{7}{9\, \varepsilon} e_A = \frac{n^\prime}{2\alpha}\, e\,, \quad n^\prime \in \mathbb Z\,, 
\end{align}
and, thus, \eqref{epsvseA} leads to:
\be\label{epsvseAnover2}
\varepsilon = \frac{7}{9\,n^\prime} \frac{e_A}{\gd}\,, \quad n^\prime \in \mathbb Z\,.
\ee
The reader observes that \eqref{epsvseA} is consistent with $\varepsilon \ll 1 $ provided that:
\be\label{eAless1}
e_A \ll \frac{9\,n^\prime}{7}\, \gd \simeq 88.07 \, n^\prime\, e\,, \qquad n^\prime \in \mathbb Z\,,
\ee
and the corresponding MM mass formula \eqref{MMmasseps} becomes:
\begin{align}\label{MMmassnover2}
M^\star  = \frac{\Lambda}{2}\, \exp\left(-\frac{8\pi^2}{\frac{e_A}{g_m}\, e_B^2} \right) 
= \frac{\Lambda}{2}\, \exp\left(-\frac{8\pi^2}{\frac{e_A}{n^\prime\,\gd}\, e_B^2} \right)  
= \frac{\Lambda}{2}\, \exp\left(-\frac{n^\prime\, \pi}{\left(\frac{e_A}{e}\right)\, \left(\frac{e_B}{e}\right)^2\, \alpha^2} \right) \,, \quad n^\prime \in \mathbb Z\,.
\end{align}

\begin{figure}[htbp!]
  \centering
    \includegraphics[width=0.49\linewidth]{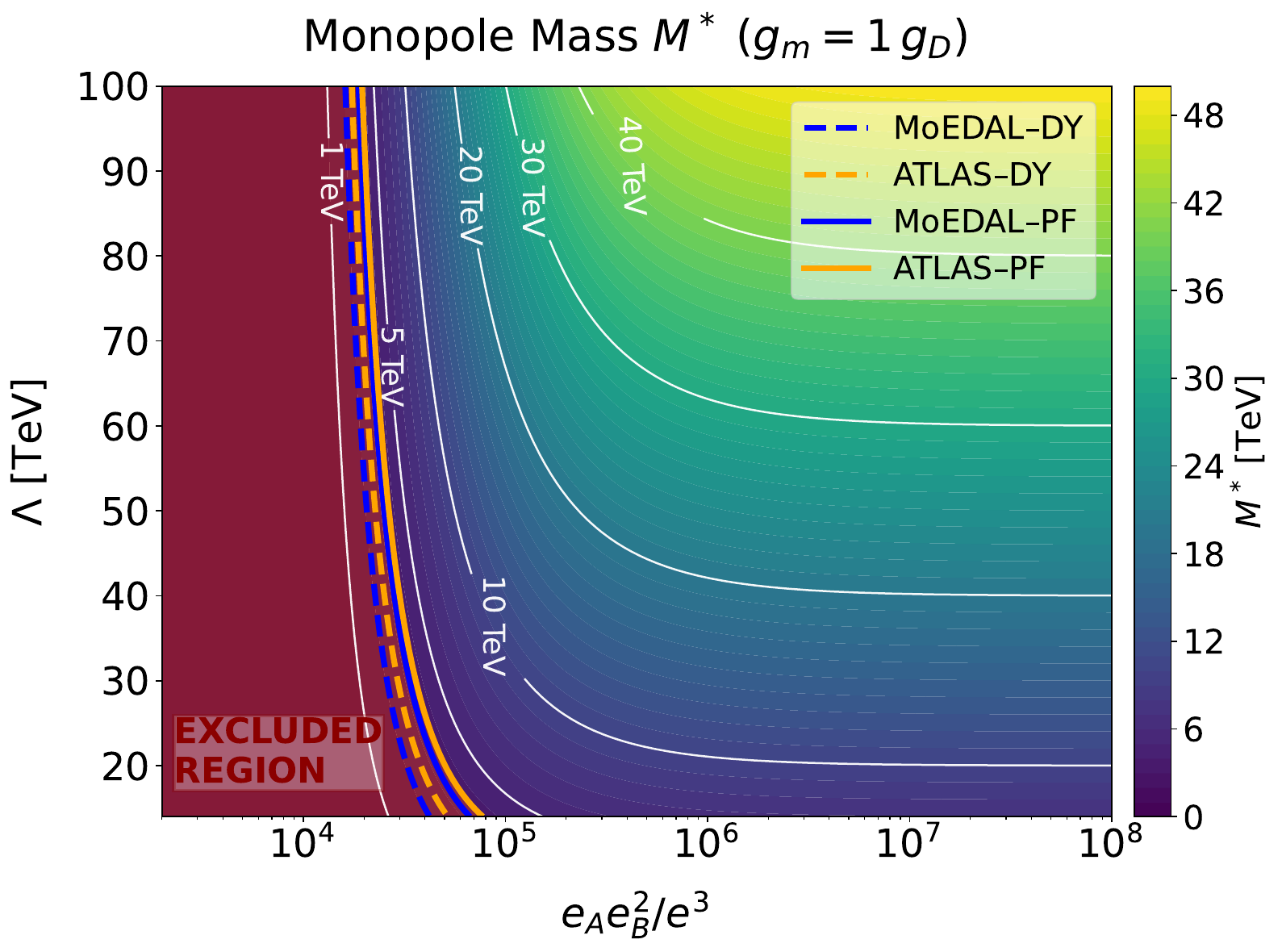}
  \hfill
    \includegraphics[width=0.49\linewidth]{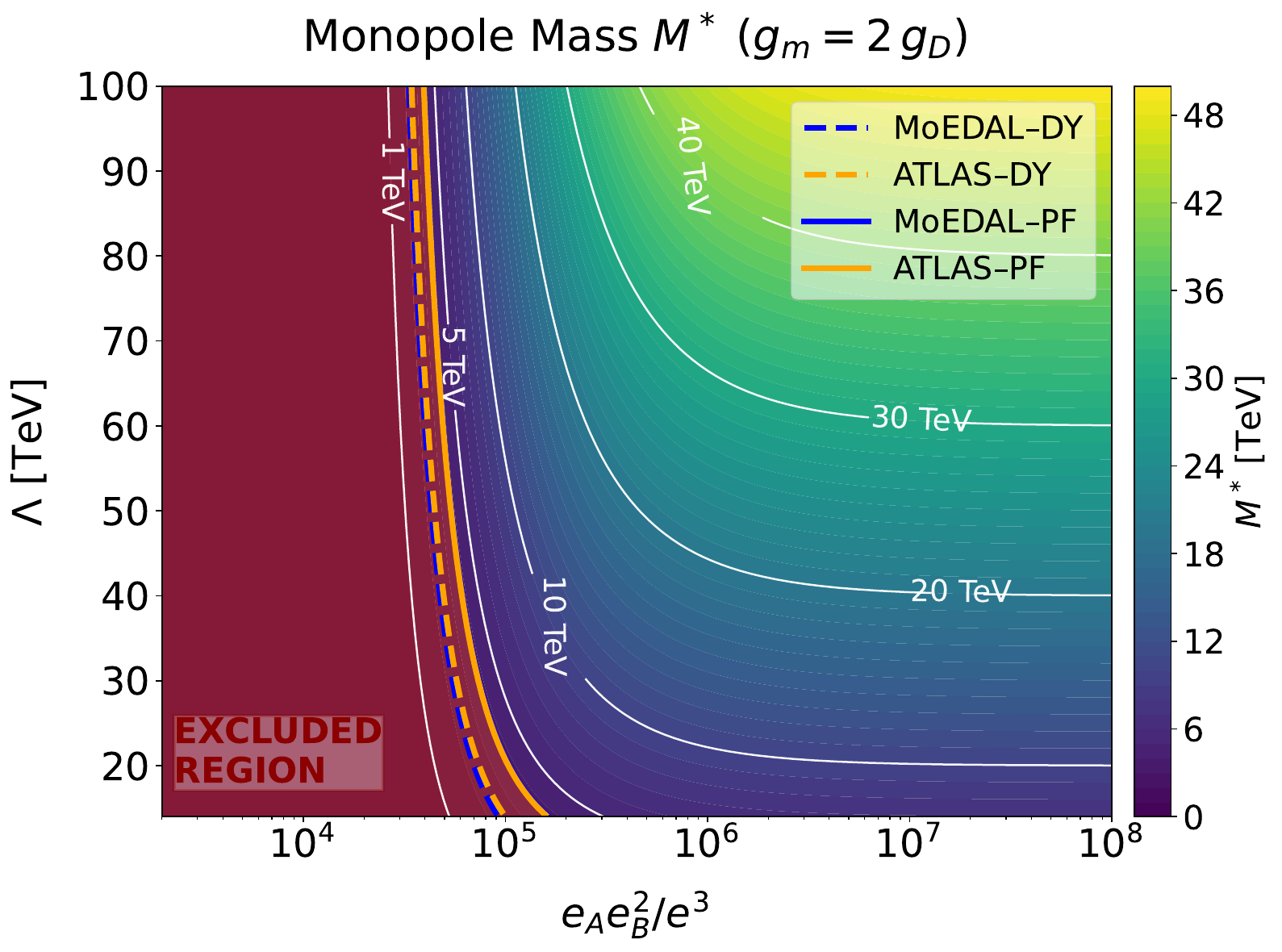}
\hfill\vspace{0.2cm}\hfill
    \includegraphics[width=0.49\linewidth]{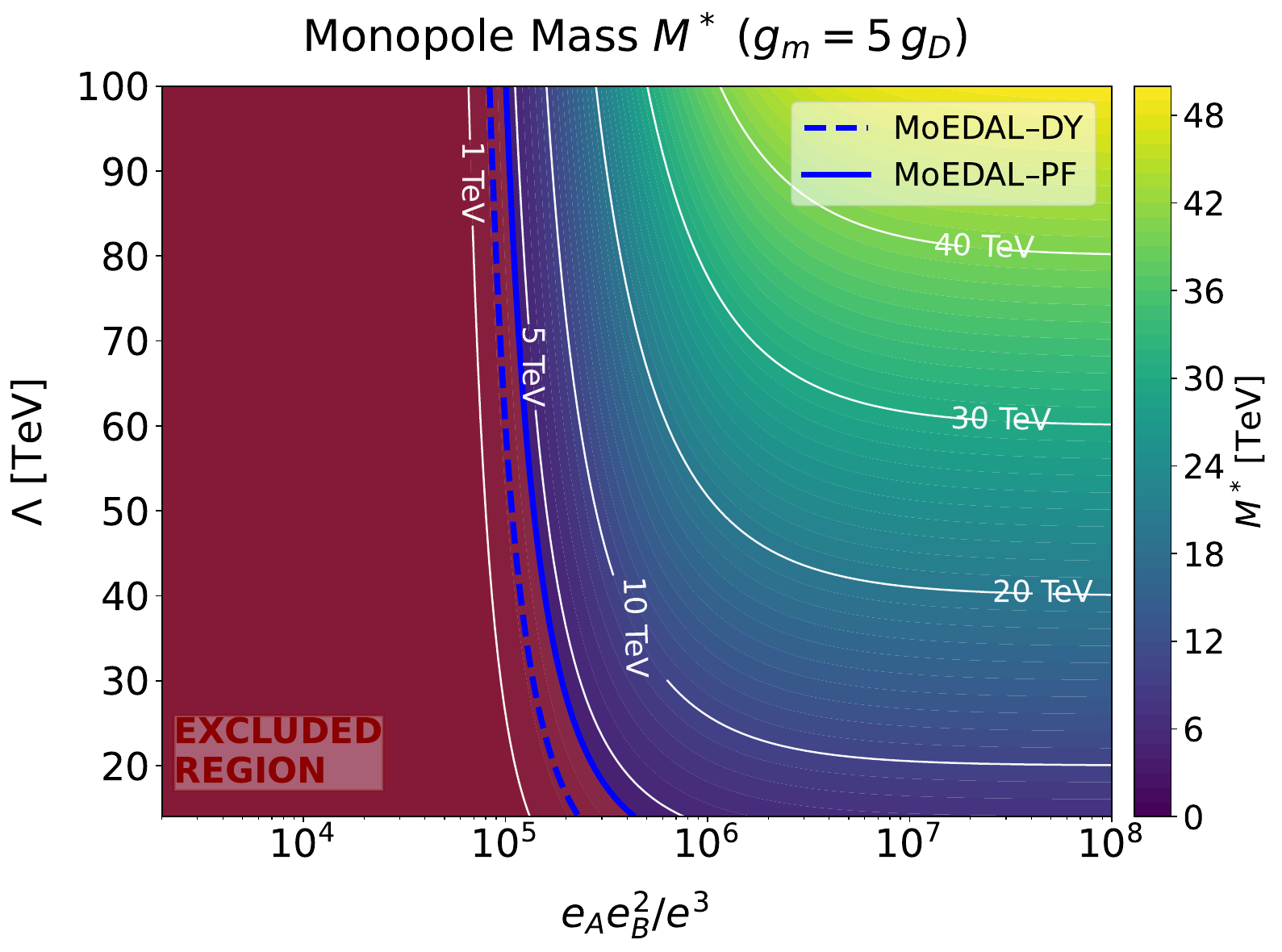}
\hfill
    \includegraphics[width=0.49\linewidth]{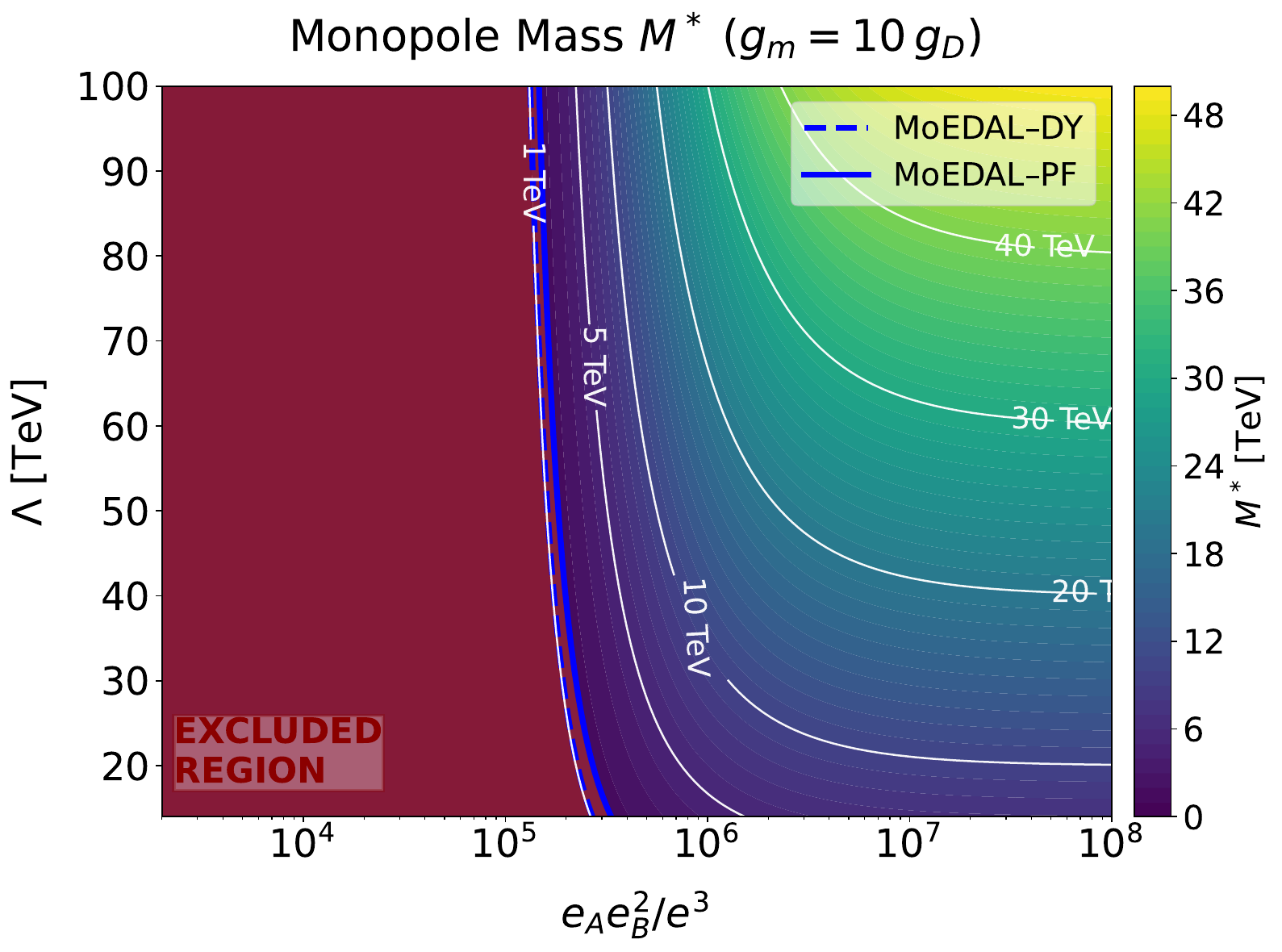}
  \caption{Contour maps of the resummed monopole mass $M^*$ for different magnetic charges $g_m$. The shaded red region indicates the stronger experimentally excluded parameter space from MoEDAL~\cite{MoEDAL:2023ost} and ATLAS~\cite{ATLAS:2023esy} (DY and PF) at $\sqrt{s}=13~\tev$.}
  \label{fig:excl_eaeb_lambda}
\end{figure}

In this analysis, $e_A$ and $e_B$ appear in the explicit expression of the MM mass \eqref{MMmassnover2}, together with the UV cutoff $\Lambda$, as discussed in detail in Section~\ref{sec:massresum}.  
The hierarchy between $e_A, e$ is kept free, provided they are both perturbative, and much smaller than $e_B$ (cf.\ \eqref{ebea}), and such that $\varepsilon \ll 1$. These parameters can be constrained by the experimental mass limits set by the ATLAS~\cite{ATLAS:2023esy} and MoEDAL~\cite{MoEDAL:2023ost} experiments at the LHC. Making use of the $M^\star$ product dependence on $e_A$ and $e_B$, we depict in Fig.~\ref{fig:excl_eaeb_lambda} the two-dimensional mass contours versus the product $e_A e_B^2$ and the mass scale $\Lambda$. For each $g_m$ value, the most stringent bounds released so far at a proton-proton center-of-mass collision energy of $\sqrt{s}=13~\tev$ have been applied for both processes of monopole-antimonopole pair production, DY and PF. Due to the much larger cross section at LHC energies of the PF process compared to the DY, the bounds of the former are the strongest. The lower limits on the product $e_A e_B^2$ slightly depend on $\Lambda$ and saturate for values $\Lambda\gtrsim 50~\tev$. The most stringent limit on the product, $e_A e_B^2 \gtrsim 10^5~e^3$ is obtained for the PF process and for $g_m = 10\gd$, which is valid for any value of $\Lambda$. We point out that as final constraints on the resummation parameters, $\Lambda$, $e_A$ and $e_B$, we consider the strongest ones set experimentally across magnetic charges, production process and experiment. Similar exclusion plots for additional values of magnetic charge are displayed in Fig.~\ref{fig:contour_massvslambdaeaeb2_rest} in Appendix~\ref{sec:exclusions}. 

\begin{figure}[htbp]
  \centering
    \includegraphics[width=0.49\linewidth]{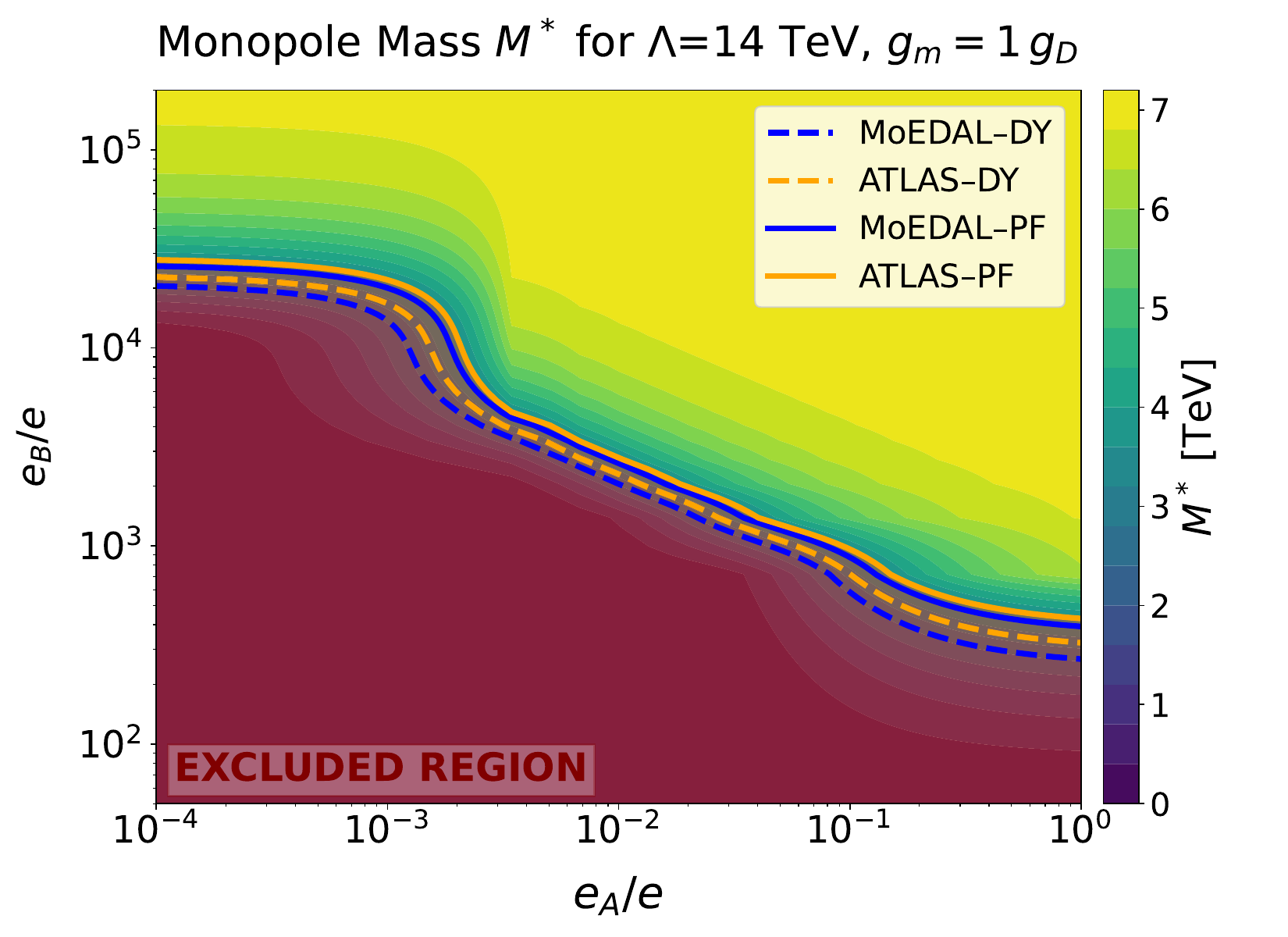}
  \hfill
    \includegraphics[width=0.49\linewidth]{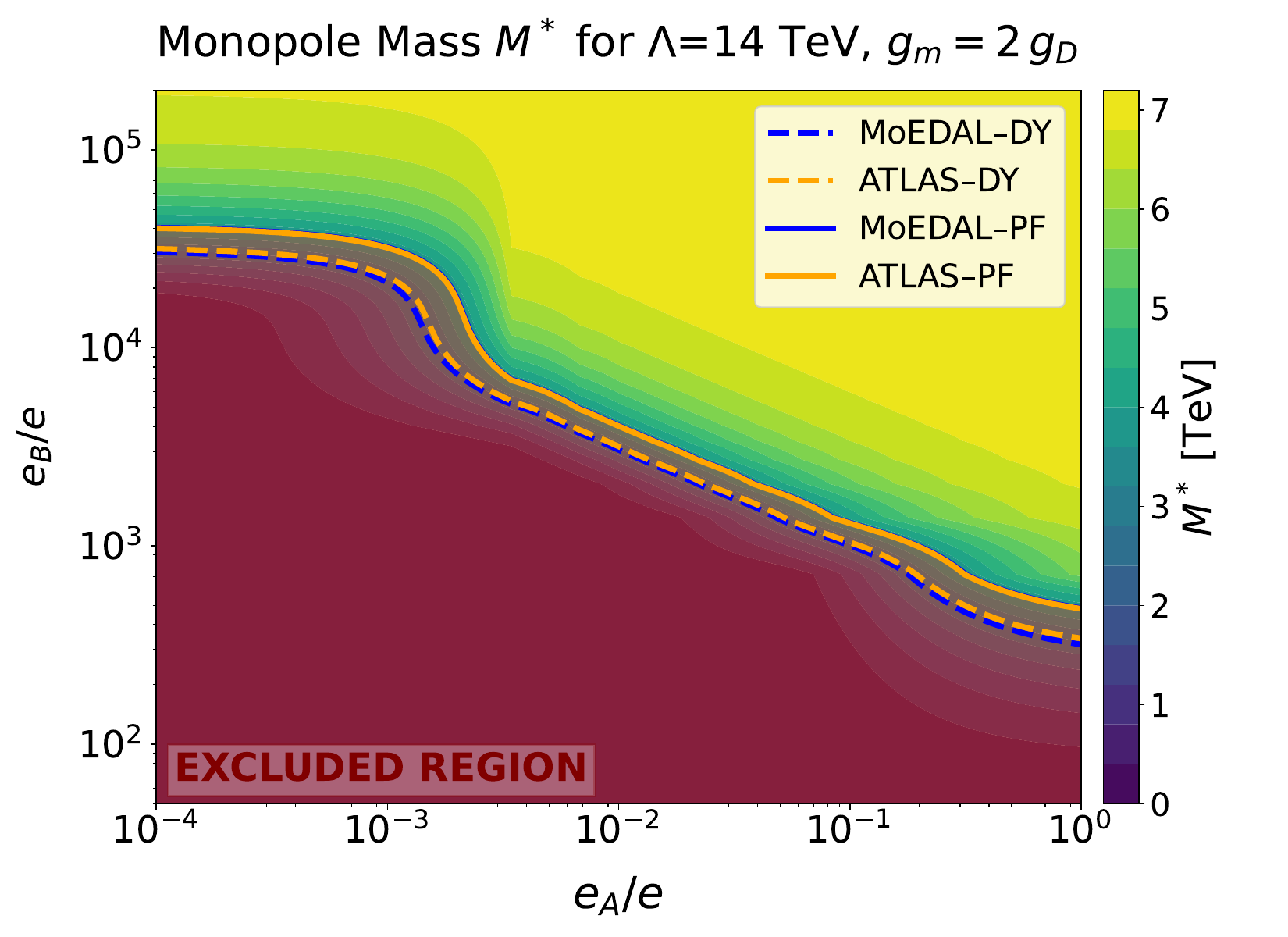}
\hfill\vspace{0.2cm}\hfill
    \includegraphics[width=0.49\linewidth]{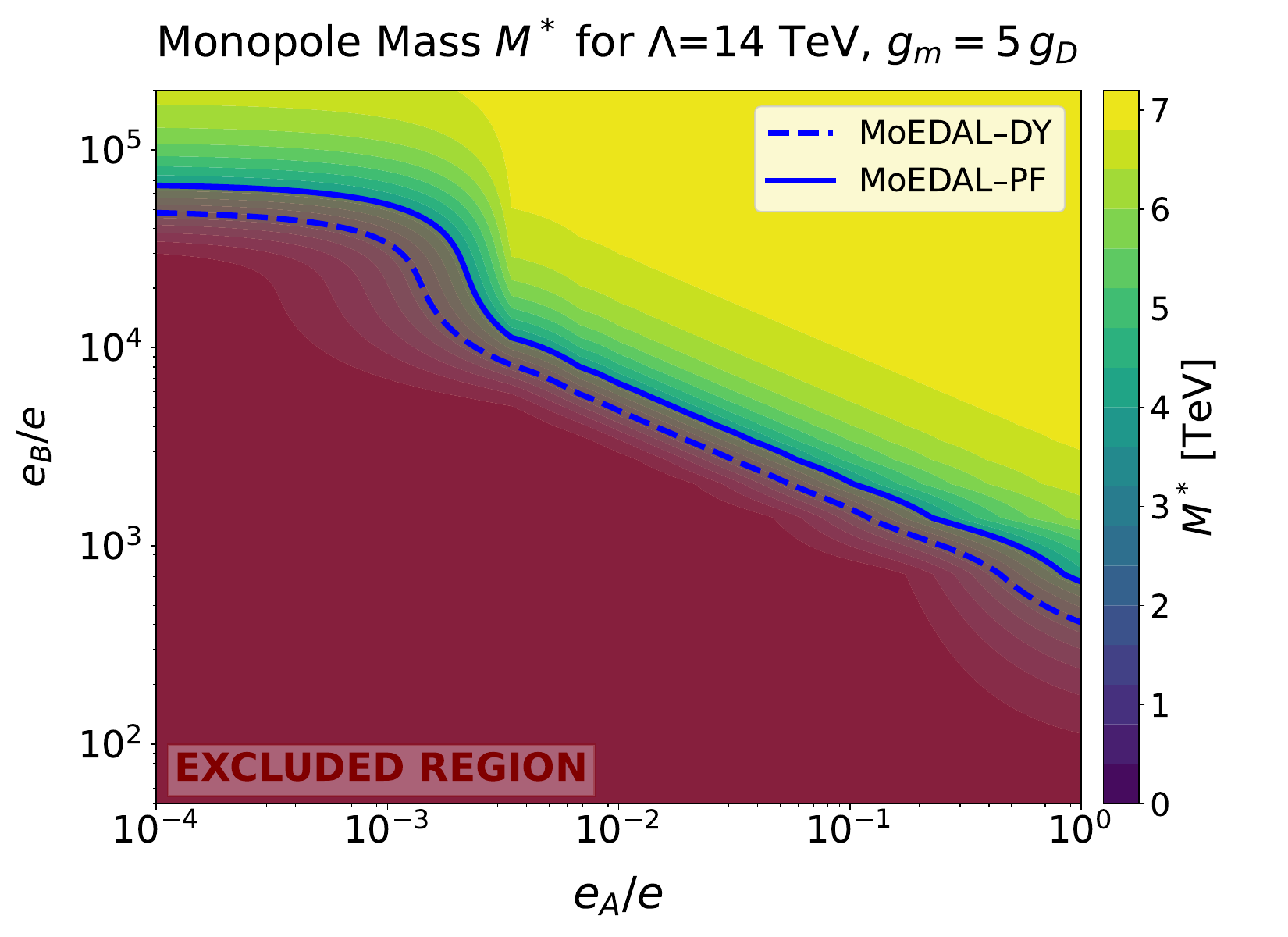}
\hfill
    \includegraphics[width=0.49\linewidth]{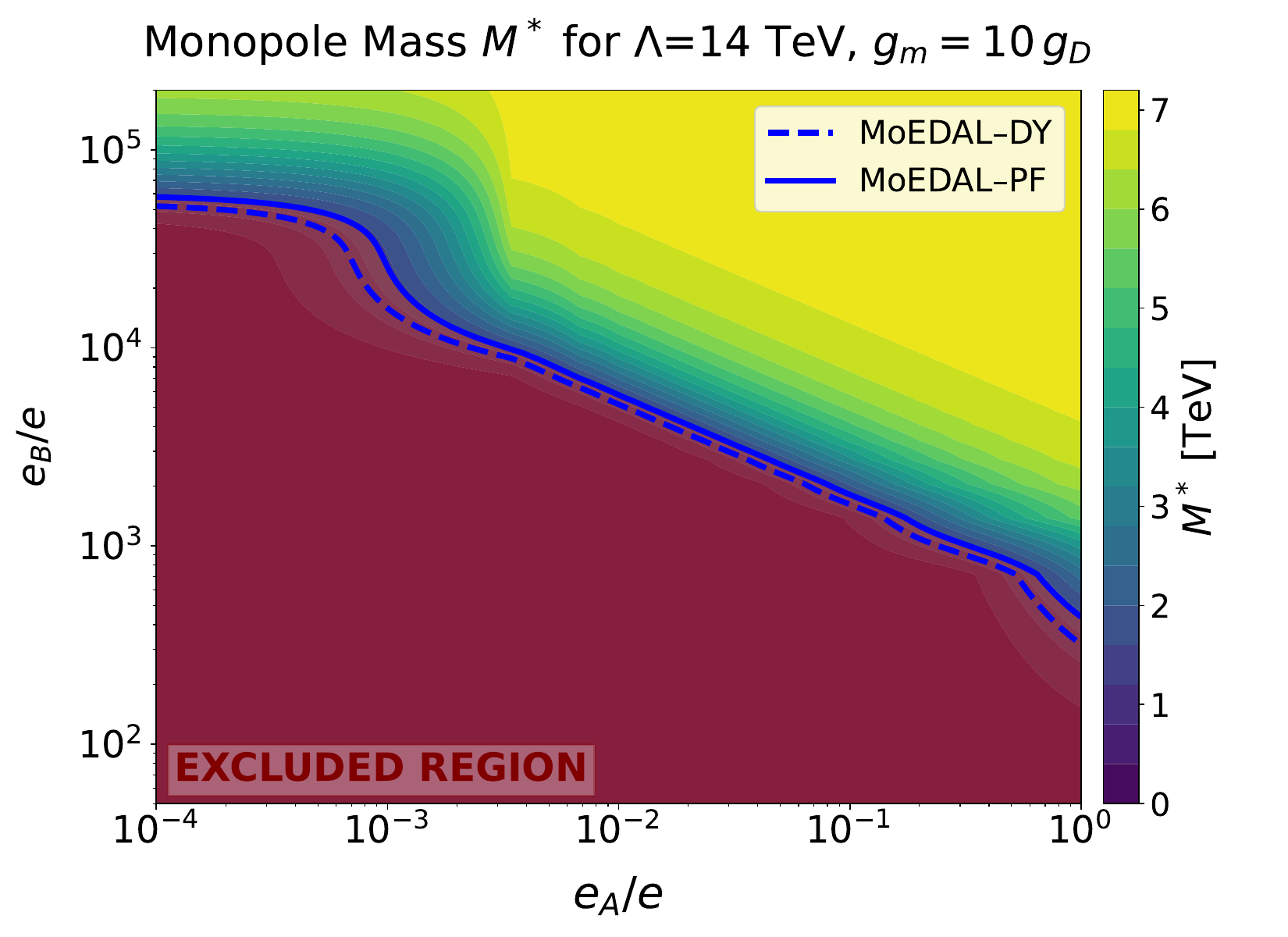}
  \caption{Contour maps of the resummed monopole mass $M^*$ for $\Lambda = 14~\tev$ and different magnetic charges $g_m$.  The shaded red region indicates the experimentally excluded parameter space from MoEDAL~\cite{MoEDAL:2023ost} and ATLAS~\cite{ATLAS:2023esy} (DY and PF) at $\sqrt{s}=13~\tev$.}
\label{fig:excl_lambda14_ea_eb}
\end{figure}

The resummation parameters can be further constrained is we assume the UV cutoff scale equal to the highest energy available in high energy colliders to-date, namely the LHC proton-proton collisions at the LHC, $\Lambda = 14~\tev$. The constraints set by the LHC mass bounds on the $(e_A,e_B)$ plane in this case are depicted in Fig.~\ref{fig:excl_lambda14_ea_eb}. Lower values of $e_A$ and $e_B$ are excluded by ATLAS~\cite{ATLAS:2023esy} and MoEDAL~\cite{MoEDAL:2023ost} mass limits. Similar exclusion plots for additional values of magnetic charge are displayed in Fig.~\ref{fig:monopole_contour_exclusion_lambda14_rest} in Appendix~\ref{sec:exclusions}.

In the specific cases associated with the Zwanziger model~\cite{Zwanziger:1970hk} considered in section \ref{sec:zwanz}, 
in which $e_B^2 = g_m^2$, but with $g_m = n^\prime \gd$ (cf.\ \eqref{schquant3}, 
the analogues of \eqref{massLambda1} and \eqref{massLambda2} (for $e_A \sim e$) read:
\begin{align}\label{massLambda1b}
 M^\star \simeq \frac{\Lambda}{2}\,\exp \left(-\frac{e}{e_A}\,\frac{4\pi}{n^\prime}\right)\,, \quad n^\prime \in \mathbb Z\,, 
\end{align}
and  
\begin{align}\label{massLambda2b} 
 M^\star \simeq \frac{\Lambda}{2}\,  \exp \left(-\frac{4\pi}{n^\prime}\right)\,, \quad n^\prime \in \mathbb Z\,, 
\end{align}
respectively.

\begin{figure}[htbp]
  \centering
   \includegraphics[width=0.49\linewidth]{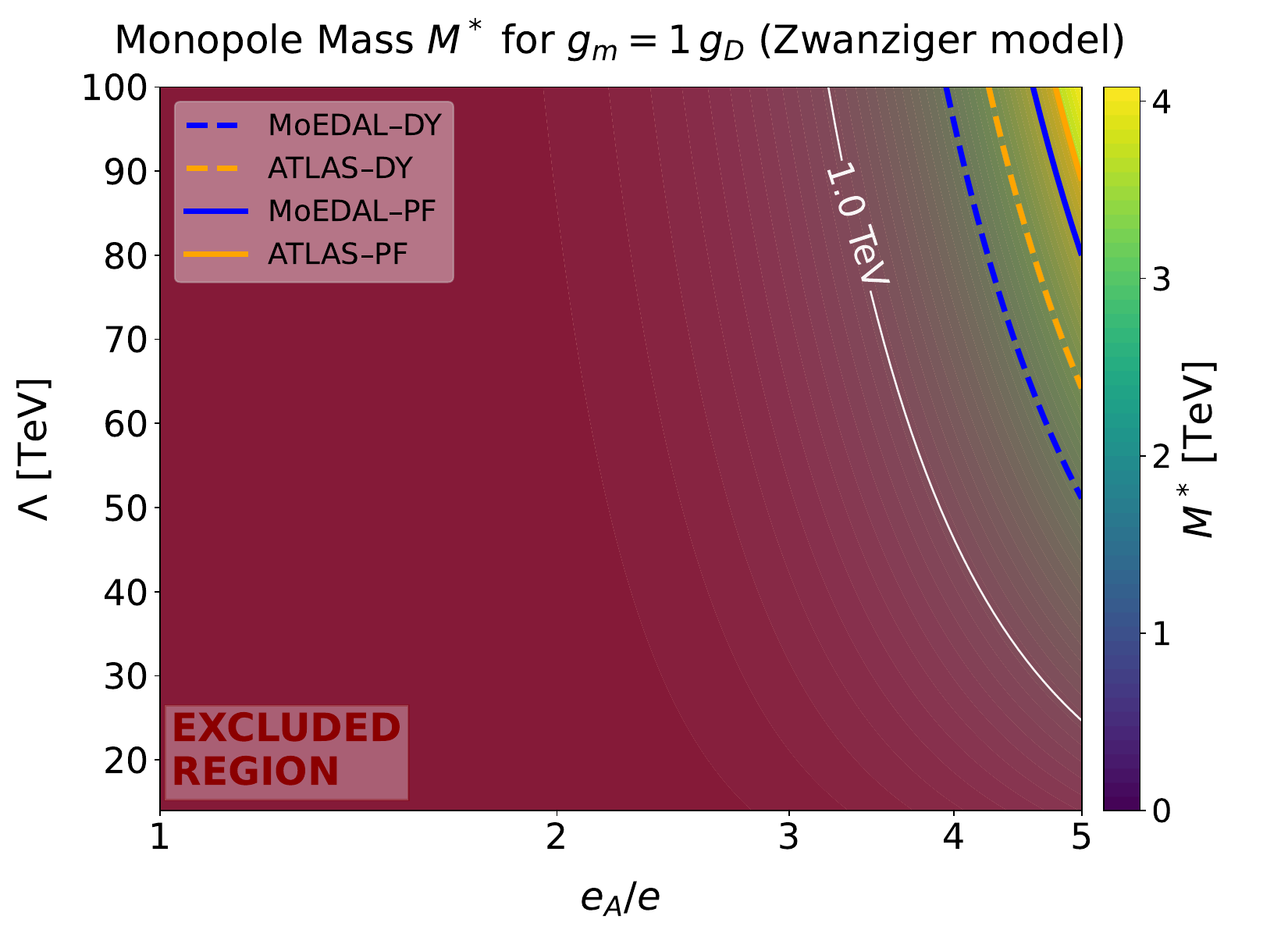}
  \hfill
    \includegraphics[width=0.49\linewidth]{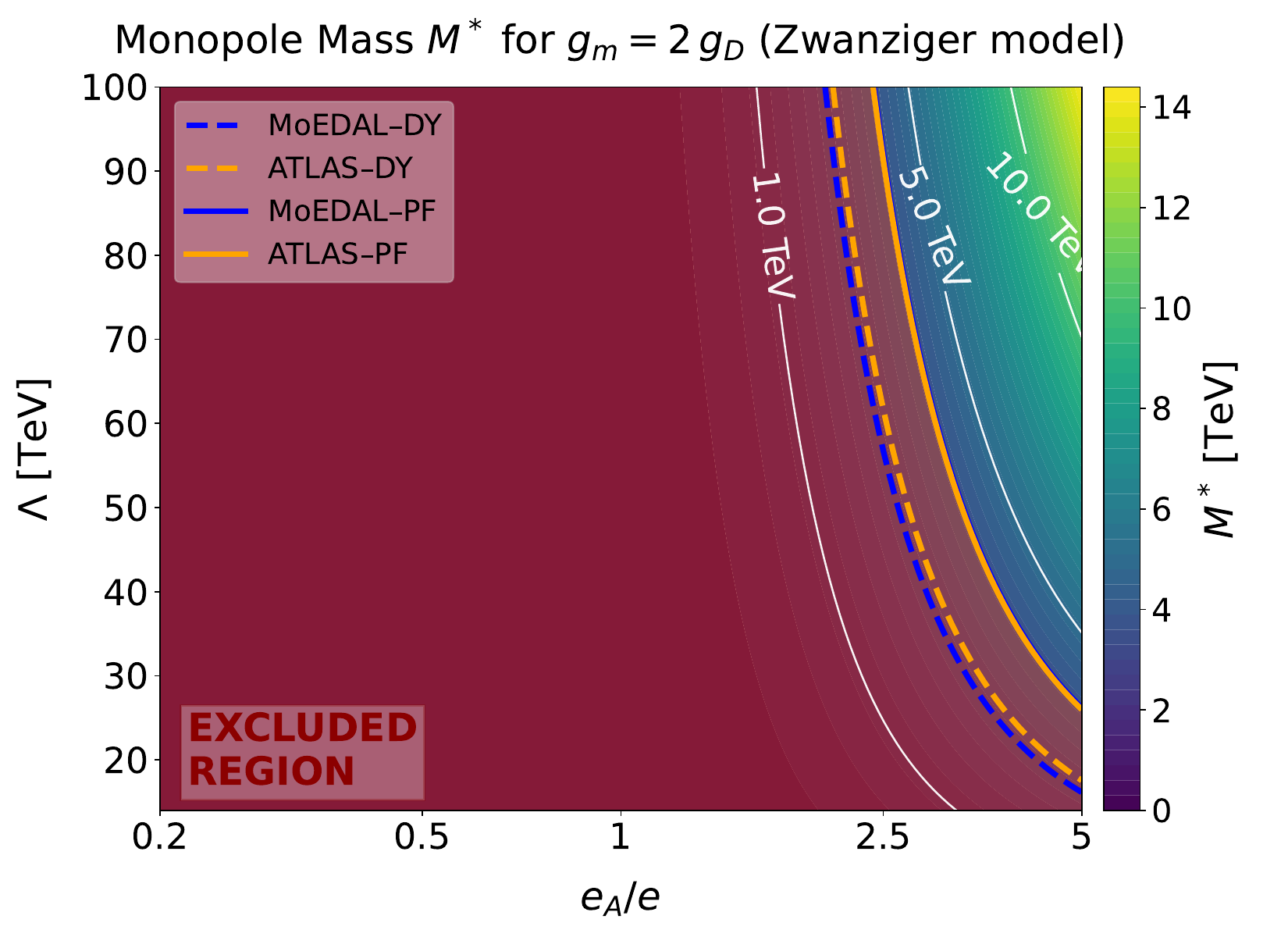}
\hfill\vspace{0.2cm}\hfill
    \includegraphics[width=0.49\linewidth]{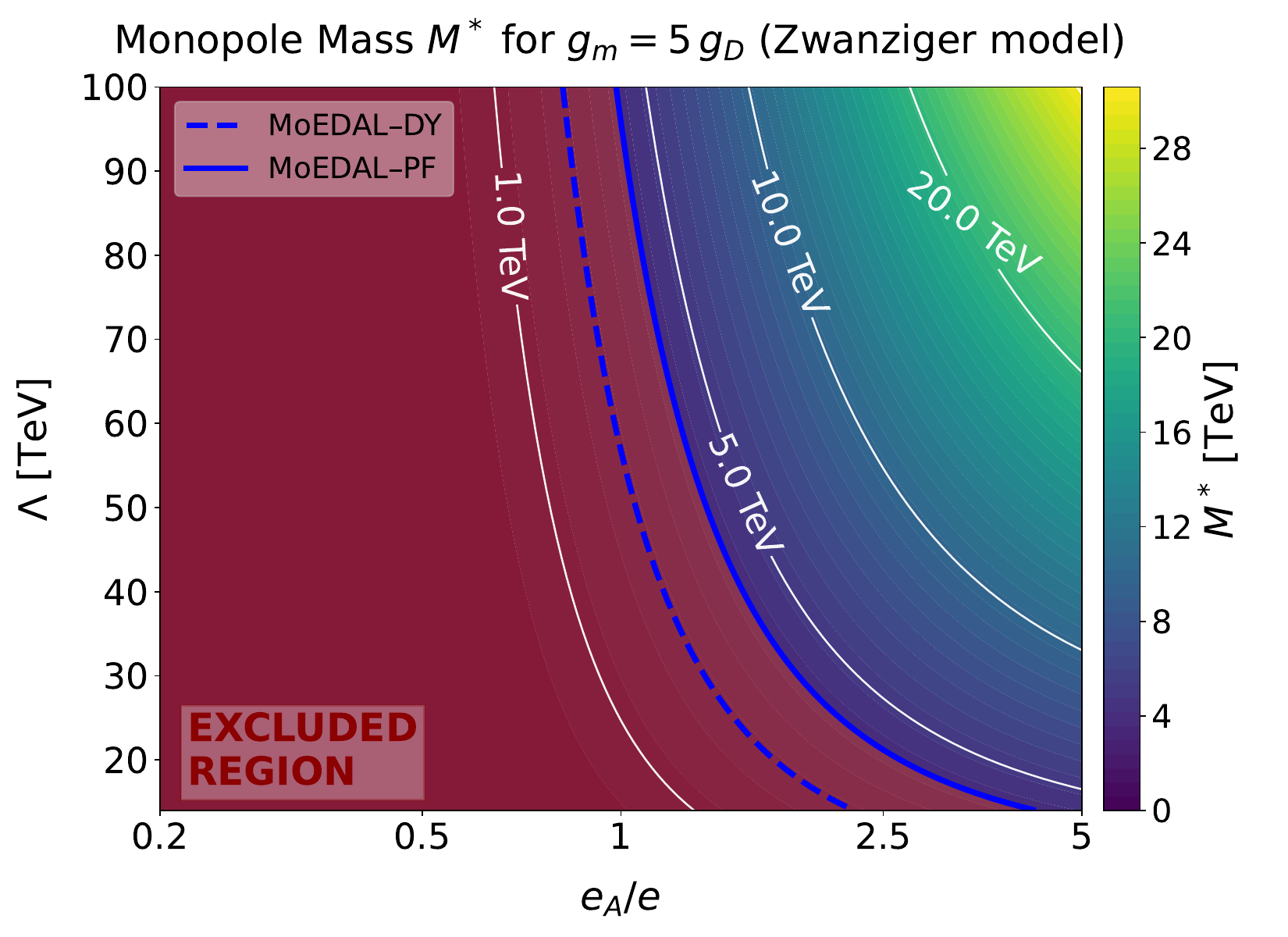}
\hfill
    \includegraphics[width=0.49\linewidth]{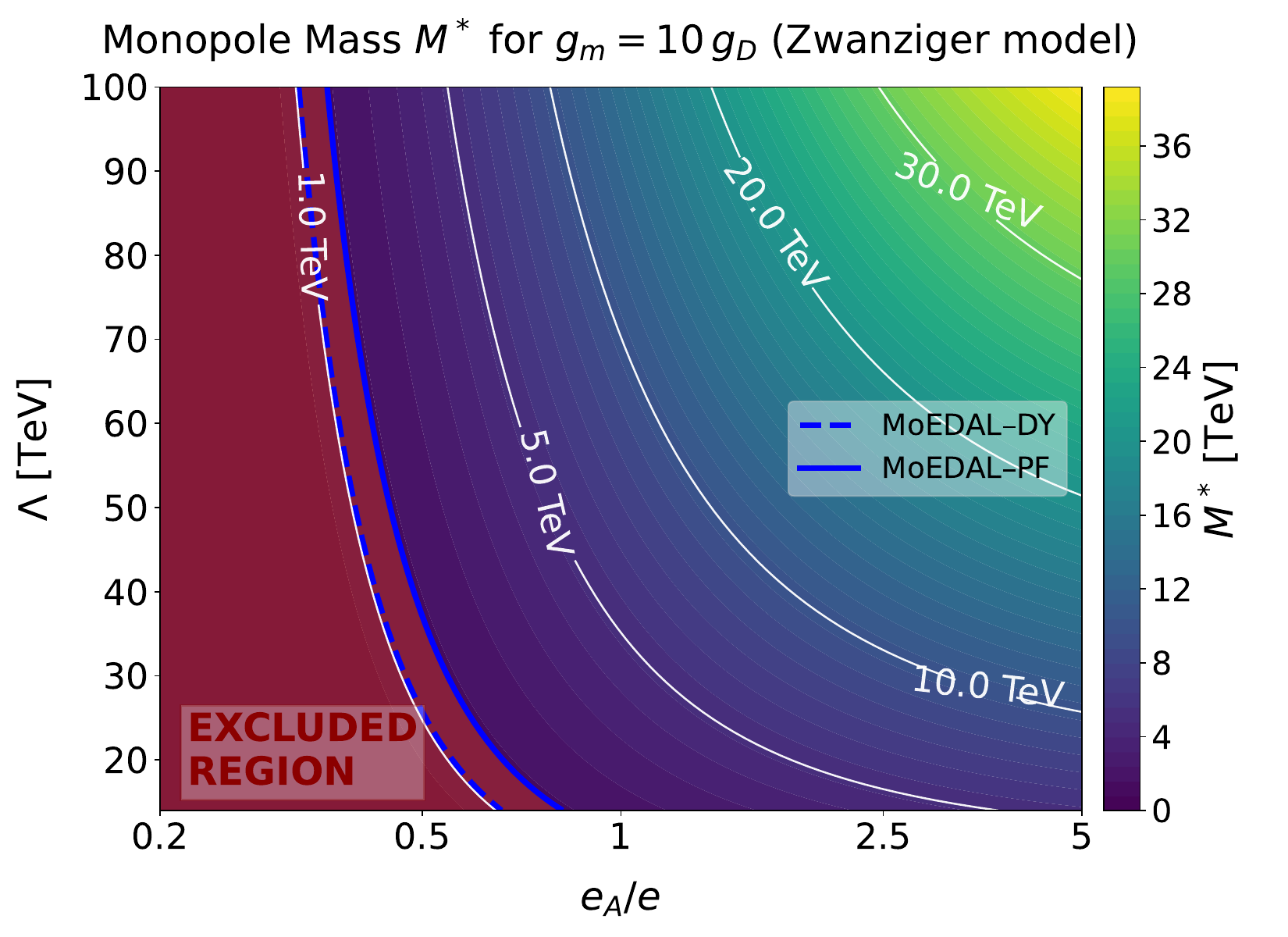}
  \caption{Contour maps of the resummed monopole mass $M^*$ for different magnetic charges $g_m$ assuming the Zwazinger model, that leads to \eqref{massLambda1b}. The shaded red region indicates the experimentally excluded parameter space from MoEDAL~\cite{MoEDAL:2023ost} and ATLAS~\cite{ATLAS:2023esy} (DY and PF) at $\sqrt{s}=13~\tev$.}
  \label{fig:excl_ea_lambda}
\end{figure}

If the Zwanziger model~\cite{Zwanziger:1970hk} is considered assuming \eqref{massLambda1b}, the experimental mass bounds lead to the limits on the $(e_A,\Lambda)$ plane shown in Fig.~\ref{fig:excl_ea_lambda}. We observe that the exclusion region in Fig.~\ref{fig:excl_ea_lambda} varies in a different way versus the magnetic charge compared to  Fig.~\ref{fig:excl_eaeb_lambda}. This is because the $e_B = g_m$ condition of the Zwanziger model practically reverses the $M^\star$ dependence on $g_m$. Hence the excluded region is maximal for the lowest magnetic charge $g_m = 1\gd$. We notice that the entire range $e_A$ for which the coupling hierarchy is respected, including the special case $e_A \sim e$, is excluded for the whole $\Lambda$ range. Similar exclusion plots for additional values of magnetic charge are displayed in Fig.~\ref{fig:excl_ea_lambda_rest} in  Appendix~\ref{sec:exclusions}.

\section{Conclusions and Outlook}
\label{sec:concl}

We have extended the one-loop DS resummation on the EFT model of \cite{Alexandre:2019iub}, describing interactions of MM with ordinary matter, to determine a non-trivial UV-fixed-point structure. The Lagrangian density of the effective theory at the UV fixed point \eqref{efflag3} assumes the form of a QED model in a specific gauge fixing. 

In this effective approach, the MM is viewed as a limiting case of a `dyon' with a tiny (vanishing) electric charge (direct coupling to the electromagnetic photon). The latter is compensated, under non-perturbative renormalization, due to the strongly coupled U(1) gauge group factor, leading to a finite magnetic charge, which obeys the DQC. Moreover, this picture, in conjunction with the broken CP invariance of the effective theory, allows for non-vanishing amplitudes in the Drell--Yan and photon-fusion production processes, used in collider searches of MM. 

The self-consistent definition \eqref{fpgeff} of the magnetic charge as the effective non-perturbatively renormalized coupling of the MM to the electromagnetic photon in the resummed UV-fixed-point theory, justifies the use of tree-level cross sections of the DY and PF production processes~\cite{Baines:2018ltl} in experimental searches for MM at colliders and validates the extracted MM mass bounds~\cite{Mavromatos:2020gwk,Mitsou:2026zcf}. The same holds for processes involving virtual MM such as the light-by-light scattering~\cite{Ellis:2022uxv,Mitsou:2026gvp}.
Consequently, the long-standing issue stemming from the large monopole-photon coupling is resolved. It is reminded, however, that higher-order resummation may result in modifications of the production cross sections and the resulting mass limits.  

Given that the MM mass depends on the EFT parameters, the latter can be constrained from the available experimental bounds in current colliders. In the general case of \eqref{efflag3}, constraints on the UV cutoff $\Lambda$ and $e_A, e_B$ couplings are set. In the specific case where 
our EFT represents quantum aspects of the Zwanziger model~\cite{Zwanziger:1970hk}, i.e. $e_B=g_m$ (cf. \eqref{eAR}), the strongest constraints come from the case $g_m = 1 \gd$ (cf. Fig.~\ref{fig:excl_ea_lambda}). Essentially coupling values $e_A \lesssim 4\, e$ are excluded for \emph{all} $\Lambda$ values attainable in current and future colliders. This exclusion pertains to  the special case $e_A \sim e$ (cf. \eqref{massLambda2b}).  

Although our resummation UV-fixed-point procedure has been developed primarily for elementary (structureless) MMs,  it has the  intriguing feature that it provides the necessary enhancement factors, via the huge values of the wave-function renormalization at the UV fixed point,  which can cancel the large ``entropic'' suppression characterizing the production of composite (solitonic) MM at colliders~\cite{Drukier:1981fq}. Practically, this could be avoided if the huge wave-function renormalization is capable of reducing the radius of the composite monopole down to its Compton wavelength ($\sim 1/M^\star)$, thereby eliminating the classical profile of the MM, making it a quantum excitation. These are important open questions that require further investigation before conclusions are reached, and they are clearly MM-microscopic-model dependent. 
A connection of the effective approach to the microscopic models that incorporate in their solutions solitonic composite MM is currently pending, at present, it is only a conjecture that the dual strongly-coupled U(1)$^\prime$ interaction in our EFT has physical effects, and drives the unsuppressed composite-MM production  at colliders, playing a similar role to the strong magnetic field in the dual Schwinger-effect. 

In the elementary MM case, on the other hand, there is no such a suppression, and the effective resummation theory applies intact. If, therefore, the theory also applies to the composite MM case,  this would imply that the production of both types of monopoles could be treated on equal footing. This remains to be seen.

An important formal aspect of the current work is a clarification of the 
connection of the effective quantum two-gauge-potential field theory of \cite{Alexandre:2019iub} with the classical theory of \cite{Zwanziger:1970hk}, where the gauge potentials satisfy the constraint \eqref{zwanzconstrA}. 
The quantum fluctuations of the potentials in the model of \cite{Alexandre:2019iub},  on the other hand, do not satisfy the constraint, behaving as independent variables. The consistency of this is proved via the background field method, applied to the effective quantum field theory model of \cite{Alexandre:2019iub}, and it relies on the assumption that the presence of background fields does not affect the model's UV fixed-point structure.
In this way, we have argued in favour of the equivalence of our UV fixed-point effective field theory with the Zwanziger~\cite{Zwanziger:1970hk} two-potential formalism, at a quantum level. We have implemented the constraint between the potentials of \cite{Zwanziger:1970hk}, which in that work led to a non-trivial two-point function, yielding the standard Maxwell equations, via a delta-functional constraint in the respective path integral, linking the two quantum U(1) gauge fields of the model~\cite{Alexandre:2019iub}. We derived in this way, the classical Maxwell equations as a stationary point of our path integral, arguing that the constraint does not affect the UV fixed-point structure of the effective theory.

As an outlook of our approach we mention the extension to MM cases of spin zero and one. From the experience we have with scalar HECOs~\cite{Alexandre:2024pbs}, we probably have to include self interactions among scalar MMs, which is a non-trivial task.  Moreover, the extension of the EFT of \cite{Alexandre:2019iub} to vector (spin-one) MMs, is also highly non-trivial, given that unitarity in this case requires a non-trivial magnetic moment for the monopole, in analogy with the neutral gauge boson $Z^0$ of the SM quantum field theory~\cite{Baines:2018ltl}. 

Finally, we mention  that another application of our EFT approach of \cite{Alexandre:2019iub} is in the scattering of ordinary matter with MM~\cite{Schwinger:1976fr,Milton:2006cp}, assuming the latter had been produced in a collider (through either SM particle collisions, or the Schwinger mechanism). In the framework of our UV-fixed point (resummed) gauge theory, such a scattering can be described by cross sections of QED type. 

\section*{Acknowledgments}

NEM thanks the University of Valencia and its Theoretical Physics Department for a visiting Research Professorship supported by the programme  \emph{Atracci\'on de Talento} INV25-01-15.
The research of JA and NEM is supported in part by the UK Science and Technology Facilities research Council (STFC) under the research grants ST/X000753/1. VAM is supported by the Generalitat Valenciana via the Excellence Grant Prometeo CIPROM/2021/073, by the Spanish MICIU / AEI / 10.13039/501100011033 and the European Union / FEDER via the grants PID2021-122134NB-C21 and PID2024-158190NB-C21, by CSIC through grant 2025AEP129, and by the Severo Ochoa project CEX2023-001292-S. EM acknowledges support from the U.S.\ National Science Foundation under grant No.\ 2309505 (FAIN), awarded to the University of Alabama MoEDAL group for the project ``Searching for Magnetic Monopoles and Other Exotics with MoEDAL''. 

\section*{Data Availability Statement}

This manuscript has no associated data. [Authors' comment: Data sharing not applicable to this article as no datasets were generated or analysed during the current study.]

\section*{Code Availability Statement}

This manuscript has no associated code/software. [Authors' comment: Code/Software sharing not applicable to this article as no code/software was generated or analysed during the current study.]


\appendix

\section{Connection of the Lagrangian \eqref{Lagrangian} to Zwanziger's effective Lagrangian}\label{sec:app}

In this appendix we shall justify the use of the quantum version of \eqref{Lagrangian}, as an appropriate quantized version of Zwanziger's effective Lagrangian for MM~\cite{Zwanziger:1970hk}. To this end, we consider a background expansion of the gauge potentials $A_\mu$ and $B_\mu$ (the ``dual'' potential to $A_\mu$ in the notation of \cite{Zwanziger:1970hk}) about a background appropriate for MM:
\begin{align}\label{bckgr}
A_\mu = A^{(0)}_\mu + \mathcal A_\mu \, ,\qquad     B_\mu = B^{(0)}_\mu + \mathcal B_\mu \,,  \qquad \mu =0,1,2,3\,,
\end{align}
where the superscript $(0)$ refers to  
background values,
while ${\mathcal A}_\mu$, ${\mathcal B}_\mu$ denote quantum fluctuations, which we assume to be non-singular, in the sense that the corresponding Bianchi identities 
are valid for the relevant Abelian field strengths:
\begin{align}\label{bianchi}
\varepsilon_{\mu\nu\rho\sigma} \,\partial^\nu F_A^{\rho\sigma}(\mathcal A)  =  0 =
\varepsilon_{\mu\nu\rho\sigma} \,\partial^\nu  F_B^{\rho\sigma} (\mathcal B)\,, \qquad \mu , \nu , \rho, \sigma = 0, 1,2,3\,, 
\end{align}
where $\varepsilon_{\mu\nu\rho\sigma}$ denotes the covariant Levi-Civita tensor in (3+1)-dimensional Minkowski spacetime.
On the other hand, the MM background potentials $A^{(0)}_\mu$, $B^{(0)}_\mu$,
include 
singular parts that require regularization~\cite{Shnir:2005vvi} (such as Dirac-string line singularities~\cite{Dirac:1931kp,Dirac:1948um}).
In this case, 
the Bianchi identities \eqref{bianchi}  may not be satisfied for such parts.

Examples of the singular background potentials 
can be found 
in Dirac's original theory~\cite{Dirac:1931kp}, where the MM is characterized by a vanishing physical electric field, $\vec E_i=0$, $i=1,2,3$, and a non-trivial (but singular) field strength for the $A_i^{(0)}$ vector potential, which is such that 
\be\label{singpart}
\vec B_i \equiv \epsilon_{ijk} \partial_i \vec A_j = \vec B_{\rm MM\,i } + \vec B_{\rm string\, i}\,, \,\quad i,j,k=1,2,3\,, 
\ee
where 
\be\label{BMMsing}\vec B_{\rm MM\, i} = \frac{g_m \hat x_i}{r^2}\,,
\ee
is the physical radial magnetic field, 
sourced by the MM, which is the one coupled to physical electric matter charges, leading to the charge quantization \eqref{schquant2}. The potential $A_i$ is singular along the direction of the Dirac-string line singularity. 
The latter is associated with the part $B_{i\, \rm string}$ of \eqref{singpart} which needs regularization.  Such a regularized Dirac string, say along the negative $x_3$-axis, reads~\cite{Shnir:2005vvi} 
\be\label{regsin}B_{3\,\rm string} = 4\pi\, g_m \delta(x_1)\, \delta(x_2)\, \Theta(-x_3)\,. 
\ee
After regularization, one has $\nabla_i B_i =0$ (i.e.\ a satisfaction of the Bianchi identities \eqref{bianchi}  for the vector potential $\vec A$), since the two parts on the right-hand-side of \eqref{singpart} produce equal in magnitude and opposite in sign contributions~\cite{Shnir:2005vvi}. 
In our context~\cite{Alexandre:2019iub}, the corresponding field strength of the $B^{(0}_\mu$ dual classical background potential is related to the above via the constraint \eqref{zwanzconstrA}, below. 

In his subsequent relativistic approach~\cite{Dirac:1948um}, Dirac has formulated the effect of his regularized string by 
adding to the standard Maxwell-type field strength of the $A_\mu$ potential, which satisfies the Bianchi identity, a string-term $\mathcal C_{\mu\nu}$, which does not:
\be\label{relstringterm}
F_{\mu\nu} = \partial_{\mu} \, A_{\nu]} - \partial_{\nu} \, A_{\mu]}+ \widetilde{\mathcal{C}}_{\mu\nu}\,, 
\ee
where $\partial^\mu \, \mathcal C_{\mu\nu} = J_{m\, \nu}$, with $J_m$ the magnetic current. In this formalism the role of the Dirac-string singular part 
in \eqref{singpart},
is played by ${\mathcal C}_{0i}$,
for which we 
have: $\nabla_i {\mathcal C}_{0i} = g_m \delta (\vec x) - g_m \delta (\vec x - \vec x_f)$  
for a string stretching between a monopole of magnetic charge $g_m$ at, say, $\vec x=0$ 
and an antimonopole of charge $-g_m$ at $\vec x_f$, which in Dirac's model is extended to infinity.

In the effective  MM field theory  of Zwanziger~\cite{Zwanziger:1970hk}, the gauge potentials $A_\mu^{(0)}$, $B_\mu^{(0)}$ do depend also on the LIV configuration of the Dirac string (due to its fixed direction in 3-space), in models where such a string is present, e.g.\ the Dirac (elementary) MM case~\cite{Dirac:1948um}, or the Cho-Maison MM~\cite{Cho:1996qd} and its variants~\cite{Cho:2013vba,Ellis:2016glu,Arunasalam:2017eyu,Mavromatos:2018kcd}. According to the argumentation of \cite{Terning:2018udc}, which has been adopted 
in the phenomenological approach of \cite{Alexandre:2019iub} and here, such LIV Dirac-string-dependent terms appear at most in phases of scattering amplitudes, or are entirely absent if the charge quantization conditions \eqref{schquant}, \eqref{schquant2} are satisfied, and thus they do not affect the physical observables, such as cross sections. Under this assumption, it was argued in \cite{Alexandre:2019iub} that the constraint of \cite{Zwanziger:1970hk} among the classical background potentials $A_\mu^{(0)}, B_\mu^{(0)}$, which involve in general non-local current-source terms depending on the Dirac string, 
can be simplified to the following Lorentz covariant form
\be\label{zwanzconstrA}
F_{A\,\mu\nu} (A^{(0)}) + {\widetilde F}_{B\, \mu\nu} (B^{(0)}) \, ``=" \, 0\,,
\ee
where 
\be\label{dual}
{\widetilde F}_B^{\mu\nu} = \frac{1}{2}\,  \varepsilon^{\mu\nu\rho\sigma} F_{B\, \rho\sigma} 
\ee
denotes the dual field strength, and the symbol $``="$ denotes  equality  only up to Dirac-string (LIV) dependent terms (in this truncated approach to the work of \cite{Zwanziger:1970hk}, one is practically dealing only with the $B_{\rm MM i}$ terms in \eqref{BMMsing}). 

In other type of monopoles, such as the 't Hooft-Polyakov MM~\cite{tHooft:1974kcl,Polyakov:1974ek}, and other Grand-Unified-Theory-inspired ones~\cite{Mavromatos:2020gwk}, which are characterized by the absence of Dirac strings, the background potential configurations incorporate the appropriate singularities in a Lorentz-invariant fashion, following the Wu-Yang approach to MM~\cite{Wu:1975es} (for reviews see~\cite{Shnir:2005vvi,Mavromatos:2020gwk}). They should still satisfy though the constraint \eqref{zwanzconstrA}, with $``="$ representing now a proper equality.

In our approach in this appendix we shall be agnostic to the precise form of the background gauge potentials, as long as they exhibit the feature of being proportional to the magnetic charge $g_m$ in the spirit of the aforementioned example of Dirac MM,  \eqref{regsin}.  
Our basic aim here is to 
justify the 
basic assumption underlying the work of \cite{Alexandre:2019iub}, and adopted here, according to which a constraint among the classical background gauge potentials as in \eqref{zwanzconstrA}
can, under some non-trivial but plausible underlying assumptions about the theory,
be consistently incorporated in the full quantum theory in a path integral formalism, leaving the quantum gauge potentials $\mathcal A_\mu, \mathcal B_\mu$ in \eqref{bckgr} unconstrained. In this way, the EFT of \cite{Zwanziger:1970hk} becomes a classical limit of the full quantum theory \eqref{Lagrangian}, proposed in \cite{Alexandre:2019iub} for resummation, and used in the current work to provide an UV fixed-point Lagrangian in the high energy limit, to be used in phenomenological searches of MM production at colliders.

To this end, following~\cite{Farakos:2024ggp},\footnote{We are making a clarification here concerning the Lattice model of scalar monopoles of \cite{Farakos:2024ggp}. In contrast to that case, in which a MM source (MM-like singular magnetic-field background) is considered only for the $A_\mu$-potential, here we do consider a singular MM background for the dual potential $B_\mu$ as well. In the non-perturbative lattice analysis of \cite{Farhi:1984qu}, it was shown that the presence of the singular source induces singularities in the $A$ and $B$ potentials. Here we take into account such singularities by separating them (in~\eqref{bckgr}) from the corresponding quantum fluctuations about them, which are assumed non singular.}  we consider the Euclidean path integral of the effective Lagrangian \eqref{Lagrangian}, but we implement the Zwanziger constraint~\cite{Zwanziger:1970hk} at a full quantum level among the potentials $A$ and $B$, via a path-integration-$\delta$ functional, represented as follows:
\begin{align}\label{dfunct}
&\prod_x\,\left[ \delta\left( \Big[F^{(E)}_{\mu\nu} (A(x)) + {\widetilde F}^{(E)}_{B\, \mu\nu} (B(x))\Big]\, \Big[F^{(E)\,\mu\nu} (A(x)) + {\widetilde F}_B^{(E)\,\mu\nu} (B(x))\Big]\right) \right]\nonumber \\ = &  \lim_{\xi^2 \to 0^+}
\frac{1}{\sqrt{2\pi}\, \vert \xi \vert}
\, \exp\left(-\int d^4x\, \frac{1}{2\xi^2}\, \Big[F^{(E)}_{\mu\nu} (A(x)) + {\widetilde F}^{(E)}_{B\,\mu\nu} (B(x))\Big]\,\Big[(F^{(E)\,\mu\nu} (A(x)) + {\widetilde F}_B^{(E)\,\mu\nu} (B(x))\Big]\right)\,.
\end{align}
The reader should notice that in view of the Euclidean nature of the path integral, the positivity of the argument of the $\delta$-functional is guaranteed, thereby implying the equivalence of the component constraint \eqref{zwanzconstrA}
with \eqref{dfunct}, when one restricts the latter to the background gauge potentials only. 

The path integral formalism we adopt below is based on the background field method for the splitting \eqref{bckgr}, in which
only the quantum gauge potentials $\mathcal A_\mu, \mathcal B_\mu$ are path integrated over, with the backgrounds being considered as fixed.
When implementing the splitting \eqref{bckgr} in the path integral, we
take into account that the background fields satisfy the classical equations of motion (e.o.m.), stemming from the Lagrangian \eqref{Lagrangian}:\footnote{In what follows, we omit the indication $E$ for Euclidean quantities, for notational brevity.}
\be\label{class}
\partial_\mu F_A^{\mu\nu}(A^{(0)}) = -e_A \left(\overline \chi \, \gamma^\mu \, \chi + \frac{e}{e_A} \overline \psi \, \gamma^\mu \, \psi\right)^{(0)} \equiv J^{e\,(0)}_{\chi} + J^{e\,(0)}_{\psi}\,, \, \quad 
\partial_\mu F_B^{\mu\nu}(B^{(0)}) = -e_B \, \Big(\overline \chi \, \gamma^\mu \, \chi \Big)^{(0)} \equiv J_\chi^{m\,(0)}\,,
\ee
where $e$ ($m$) denotes electric (magnetic) components. 

The corresponding fermion currents $J^\mu_{\psi, \chi}$ in the full quantum field theory undergo a splitting between background (classical) and quantum quantities, consisting of quantum fermionic fields, the latter not satisfying the e.o.m.: 
\begin{align}
J^\mu_{\psi, \chi} = 
J^{(0)\,\mu}_{\psi, \chi} + J^{\mu}_{\psi, \chi}\,. 
\end{align}    
Then, on account of  \eqref{dfunct} and after appropriate partial integrations, assuming the usual boundary conditions for vanishing fields and their derivatives at the boundary at infinity, we easily obtain the following Euclidean path integral for the effective model \eqref{Lagrangian}:
\begin{align}
\label{qmodel}
\mathcal Z \propto \int D{\mathcal A} \, D{\mathcal B} \, D\psi \, D\ovl \psi  \, D\chi  \, D\ovl \chi \, 
\exp \Big(-S_\text{eff} \big[A^{(0)},\, B^{(0)},\, \mathcal A, \, \mathcal B\, ; \, \xi^2 \to 0^+\big] \Big)\,,
\end{align}
where 
\begin{align}\label{effactsplit}
S_\text{eff} \Big[A^{(0)},\, B^{(0)},\, \mathcal A, \, \mathcal B\, ; \, \xi^2 \to 0^+ \Big]\,
 \stackrel {\xi^2 \to 0^+}{=} \, \int d^4x \Big[ &-\frac{1}{4}F_{A\,\mu\nu}(A^{(0)})\, F_A^{\mu\nu}(A^{(0)})
-\frac{1}{4}F_{B\,\mu\nu}(B^{(0)})\, F_{B\,\mu\nu}(B^{(0)}) \nonumber \\ & +  e_A \, \mathcal A_\mu \Big(\overline \chi \, \gamma^\mu \, \chi + \frac{e}{e_A} \overline \psi \, \gamma^\mu \, \psi\Big) + 
 e_B \, \mathcal B_\mu \, \overline \chi \, \gamma^\mu \, \chi 
\nonumber \\ 
& -\frac{1}{4} \Big(1 + \frac{2}{\xi^2}\Big)F_{A\,\mu\nu}(\mathcal A)\, F_A^{\mu\nu}(\mathcal A) 
-\frac{1}{4} \Big(1 + \frac{2}{\xi^2}\Big)\, F_{B\,\mu\nu}(\mathcal A)\, F_B^{\mu\nu}(\mathcal B) \nonumber \\
&-\frac{1}{2\xi^2} \Big(F_{A\,\mu\nu}(A^{(0)}) + \widetilde F_{B\, \mu\nu}(B^{(0)})\Big)\, 
\Big(F_{A\,\mu\nu}(A^{(0)}) + \widetilde F_{B\, \mu\nu}(B^{(0)})\Big) \nonumber \\
&-\frac{1}{\xi^2} \Big(F_{A\,\mu\nu}(A^{(0)}) + \widetilde F_{B\, \mu\nu}(B^{(0)})\Big)\, 
\Big(F_{A\,\mu\nu}(\mathcal A) + \widetilde F_{B\, \mu\nu}(\mathcal B)\Big) \nonumber \\
& + \ovl\psi(i\gamma^\mu [\partial_\mu -i\, e A_\mu^{(0)}] \, -\, m )\psi \nonumber \\ &+ \, \ovl\chi(i\gamma^\mu [\partial_\mu 
-i\, e_A \, A_\mu^{(0)}    -i\, e_B \, B_\mu^{(0)}]\,- \, M )\chi \Big]\,,
\end{align}
and we took into account the Bianchi identities \eqref{bianchi}
for the field strengths of the regular $\mathcal A, \mathcal B$, which imply vanishing  mixed terms 
\begin{align}\label{mixed}
F_{A (B)\, \mu\nu} \tilde F_{A(B)}^{\mu\nu} =0\,, 
\end{align} 
and 
that 
$\widetilde F_{B\, \mu\nu} \, \widetilde F_{B}^{ \mu\nu} = F_{B\, \mu\nu} \,  F_{B}^{ \mu\nu} $, due to the definition 
\eqref{dual}, and properties of the Levi-Civita tensor.

We now make some important remarks regarding the structure of the effective action \eqref{effactsplit}.
The first two  (solely background-dependent) terms
 decouple in the path integral \eqref{qmodel}, as an irrelevant factor not path-integrated over, 
 and hence play no role in our resummation analysis. 
In the limit $\xi \to 0^+$, there are four terms in the expression \eqref{effactsplit} which are apparently divergent (and positive in the Euclidan formalism), and, therefore, if not properly dealt with, will lead to a vanishing partition function in \eqref{qmodel}. Of those, the $\xi^{-2}$ divergences (as $\xi^2 \to 0^+$) in the 
fifth and sixth terms of \eqref{effactsplit}, which are proportional to the squares of the field strengths of the quantum fluctuations of the potentials, $\mathcal A$, $\mathcal B$, can be absorbed in redefinitions of the appropriate gauge couplings. This is achieved by first rescaling the Abelian quantum gauge potentials $\mathcal A$ and $\mathcal B$ as:
\be\label{gpredef}
\mathcal A_\mu \, \to \, \mathcal A^\prime_\mu =  \sqrt{1 + \frac{2}{\xi^2}}\, \mathcal A_\mu\,, \qquad \mathcal B_\mu \, \to \, \mathcal B^\prime_\mu =  \sqrt{1 + \frac{2}{\xi^2}}\, \mathcal B_\mu\,, 
\ee
and, then, redefining the respective couplings:
\be\label{couplredef}
e_{A \, (B)}^\prime = \frac{ e_{A \, (B)} \, \vert \xi \vert}{\sqrt{2 + \xi^2}}\,, \qquad 
e^\prime = \frac{\vert \xi \vert \, e}{\sqrt{2 + \xi^2}}\,,
\ee
where the redefined couplings are assumed finite and non-trivial, as $\vert \xi \vert \to 0^+$. 
On the other hand, the $\xi^{-2}$ divergences (as $\xi^2 \to 0^+$) in the seventh and eighth terms of \eqref{effactsplit}
are eliminated for such backgrounds for which 
\be
\label{zwanconstr}
F_{A\,\mu\nu}(A^{(0)}) + \widetilde F_{B\, \mu\nu}(B^{(0)}) =0\,,
\ee
which is the classical Zwanziger constraint~\cite{Zwanziger:1970hk}, \eqref{zwanzconstrA}.
These are the only types of MM background gauge potentials that survive in the path-integral
\eqref{qmodel}, 
\eqref{effactsplit}.

We now remark that the background-fermion-current couplings in the eleventh and twelfth terms of \eqref{effactsplit} will also become divergent, as $\xi^{2} \to 0^+$, 
after the coupling redefinitions 
\eqref{couplredef}, 
being of the form: 
\be
\label{xisource}
 \lim_{\xi^2 \to 0^+} \int d^4 x \,  \sqrt{1 + \frac{2}{\xi^2}}\,\Big(\ovl\psi\,  e^\prime\,\gamma^\mu \, A_\mu^{(0)}\, \psi + \, \ovl\chi\, \Big[e_A^\prime \, A_\mu^{(0)}\, + \,   e_B^\prime \,  B_\mu^{(0)}\Big]\,\gamma^\mu \, \chi \Big)\,.
\ee
Naively one would think that such terms would force us to set the backgrounds $A_\mu^{(0}=B^{(0)}=0$. But this would not make any sense, given the presence of MM in the theory. However, as already mentioned, both potentials are proportional to the magnetic charge $g_m$ (cf.\ \eqref{singpart}-\eqref{regsin})), 
hence, one can absorb the $\vert \xi \vert^{-1}$ singularity, as $\xi^2 \to 0^+$, by redefining 
\be\label{magcharge}
g_m \, \to \, g_m^{\prime} = \frac{g_m \, \vert \xi \vert}{\sqrt{2 + \xi^2}}\,,
\ee
in analogy with the rest of the couplings (cf.\ \eqref{couplredef}). Notice that in our case the product
$g_m^\prime \, e^\prime $ should satisfy the charge quantization condition \eqref{schquant2}. Hence, in the limit 
$\xi^2 \to 0^+$, one is left with finite, non-trivial background-dependent source terms in the action \eqref{effactsplit}, as appropriate for a MM background:
\be\label{classsource}
S_{\rm bckr~source} \Big[\, A^{(0)\,\prime},\, B^{(0)\,\prime}\,, \,  \chi\,, \,\psi\, \Big] = \int d^4x
\, \Big[\, A_\mu^{(0)\, \prime} \, \Big(  e_A^\prime \,\overline \chi \, \gamma^\mu \, \chi + e^\prime \,  \overline \psi \, \gamma^\mu \, \psi\Big)  +  B_\mu^{(0)\,\prime}\, \Big( e^\prime_B \,\overline \chi \, \gamma^\mu \, \chi \Big) \Big]\,,
\ee
where $A_\mu^{(0)\,\prime}$, $B_\mu^{(0)\,\prime}$ denote the MM background configurations
after the magnetic-charge rescaling \eqref{magcharge}.

The reader might then worry that the presence of the terms \eqref{classsource} would affect the UV fixed point structure of the theory, based on the Lagrangian \eqref{Lagrangian}.
Fortunately this is not the case, because of the argument of 
\emph{background-independence} of the UV fixed-point structure stemming from  the resummation-renormalization procedure adopted here, which makes such background-dependent terms irrelevant. 
Indeed, although their presence could affect formally the Feynman rules of the dressed propagators and vertices, appearing in the resummation procedure, and thus the 
various flows with respect to the transmutation mass scale $k$, nonetheless  the UV fixed-point effective high-energy field theory, 
used in the calculation of physical observables, like cross sections we are concerned with here,
should be in the same equivalence class of UV fixed points as the theory without backgrounds ($A_\mu^{(0)}=B_\mu^{(0)}=0$), provided, of course, that there are no gauge anomalies, which we assume for our purposes. In this sense, gauge-invariant physical observables should be background independent.
This is also the essence of the so-called background field method for treating quantum effective gauge field theories in general.\footnote{It is understood, though, that, while such a property should characterize the full resummed theory, it might be artificially violated under various truncations, like the one-loop resummation discussed here. 
To ensure proper gauge invariance and background independence in such truncated theories is a highly non-trivial issue, especially in singular MM backgrounds we are dealing with. This falls beyond the scope of the present phenomenological work. Our basic assumption here is that the one-loop resummation describes, to a good approximation, the full gauge-invariant theory, and thus background independence is plausible.}.
In this sense, one can add appropriate background terms coupled to the fermionic currents in the effective Lagrangian \eqref{effactsplit}, 
before taking the limit $\xi^2 \to 0$, so as to remove the unwanted divergent terms \eqref{xisource}, thus avoiding 
any restriction on the singular backgrounds $A_\mu^{(0)}, B_\mu^{(0)}$ other than the constraint \eqref{zwanconstr}.

The final result of the above manipulations is therefore a quantum path integral, with respect to the redefined quantum gauge potentials $\mathcal A^{\prime}$, $\mathcal B^{\prime}_\mu$, 
with an effective action of the form:
\begin{align}\label{newaction}
S_\text{eff} \Big[\mathcal A^{\prime}, \, \mathcal B^{\prime}\Big] = \int d^4 x \Big[&-\frac{1}{4}\, F_{A\,\mu\nu}(\mathcal A^{\prime})\,
F_{A}^{\mu\nu}(\mathcal A^{\prime})
-\frac{1}{4}\, F_{B\,\mu\nu}(\mathcal B^{\prime})\,
F_{B}^{\mu\nu}(\mathcal B^{\prime}) \nonumber \\
& + \ovl\psi\, (i\gamma^\mu \partial_\mu -\, m )\psi + \, \ovl\chi\, (i\gamma^\mu \partial_\mu 
\,- \, M )\chi
\nonumber \\
& + \mathcal A_\mu^{\prime} \, \Big( e_A^\prime \overline \chi \, \gamma^\mu \, \chi \, + \, e^\prime \overline \psi \, \gamma^\mu \, \psi \Big) \, + \, \mathcal B_\mu^{\prime} \, e_B^\prime \, \overline \chi \, \gamma^\mu \, \chi \Big]\,.
\end{align}
 We stress once again that the quantum gauge potentials $\mathcal A_\mu^{\prime}$, $\mathcal B_\mu^{\prime}$ in the action \eqref{newaction} are unconstrained, as (correctly) assumed in \cite{Alexandre:2019iub}. 
It is only the background gauge potentials $A^{(0)}, B^{(0)}$ that obey the Zwanziger constraint \eqref{zwanconstr}. 
 Upon dropping the primes from the couplings and the quantum potentials, as these are mere redefinitions, we thus observe that the effective Lagrangian \eqref{newaction} is
of the form \eqref{Lagrangian}, expressed, however, in terms of only the unconstrained, and regular, with no monopole-like singularities, quantum U(1) gauge potentials $\mathcal A_\mu, \mathcal B_\mu$. This Lagrangian, which has no trace of a MM background, constitutes the basis of our resummation procedure, in order to determine the UV fixed point structure of the theory. The latter is then used in the calculation of the relevant cross sections, and thus, the interpretation of MM-collider-search data, which is the main focus of the current work.

\section{Estimating the transmutation mass scale for which the wavefunction renormalization $Z_k$ crosses unity}\label{sec:k1unitarity} 

Here we estimate the scale $k_1 > k_0$, at which $Z(k_1)=1$. To this end, we assume for concreteness couplings  
$e_A, e_B$ in structureless MM  satisfying \eqref{eeA}, i.e.:  
\be\label{epseAequal}
e_A \sim e \, \qquad |e_A| \sim e \ll |e_B|\,, 
\ee
where $e>0$ is the positron charge. 
The second of the relations \eqref{epseAequal} can be satisfied in the model of \cite{Zwanziger:1970hk} by identifying $e_B \sim g_m$ (cf.\ \eqref{eBgm}). For our analysis we concentrate on the concrete example of MM number $n^\prime =2$. This corresponds to the lowest-lying MM sectors in the Schwinger quantization \eqref{schquant}, and also characterizes microscopic models of composite MM in the literature~\cite{Mavromatos:2020gwk}, containing also structureless (Dirac) MM in their solutions at certain limits~\cite{Mavromatos:2018kcd}. 
In such a case, the first relation in \eqref{epseAequal}, when combined with \eqref{fpgeff2},
yields (in order of magnitude)
\be\label{epsval}
\varepsilon = \frac{7 \, e}{18 \, g_D} = 
\frac{7\alpha}{9} 
\sim 0.006 
\ee
for the range of the values of the fine structure constant used in experimental searches $\alpha \in  [1/137 - 1/127]$.  
 Therefore, this case is compatible with the correct boundary condition $\varepsilon \ll 1$.
In this case, the $k$-flow equations for $Z$, \eqref{dZdk}, and $M_r$, \eqref{dMdk}, read:
\begin{align}\label{systemZMflows}
k \, \partial_k Z &\simeq \frac{7}{9\varepsilon} \Big(1 - \frac{k \partial_k M_r}{M_r} \Big) \,,\nonumber \\
\frac{k \, \partial_k M_r}{M_r} &\simeq \frac{1}{1 + \varepsilon \, \frac{M}{M_r}}\,.
\end{align}
At the formal UV fixed point $k \to \infty$ $M_r \to \infty$, and thus the slope of $Z$ vanishes at that point, as can be inferred from \eqref{systemZMflows}:
\be\label{uvslope}
\lim_{k \to +\infty} \partial_k Z \to 0\,.
\ee

\begin{figure}[ht]
 \centering
\includegraphics[clip,width=0.70\linewidth]{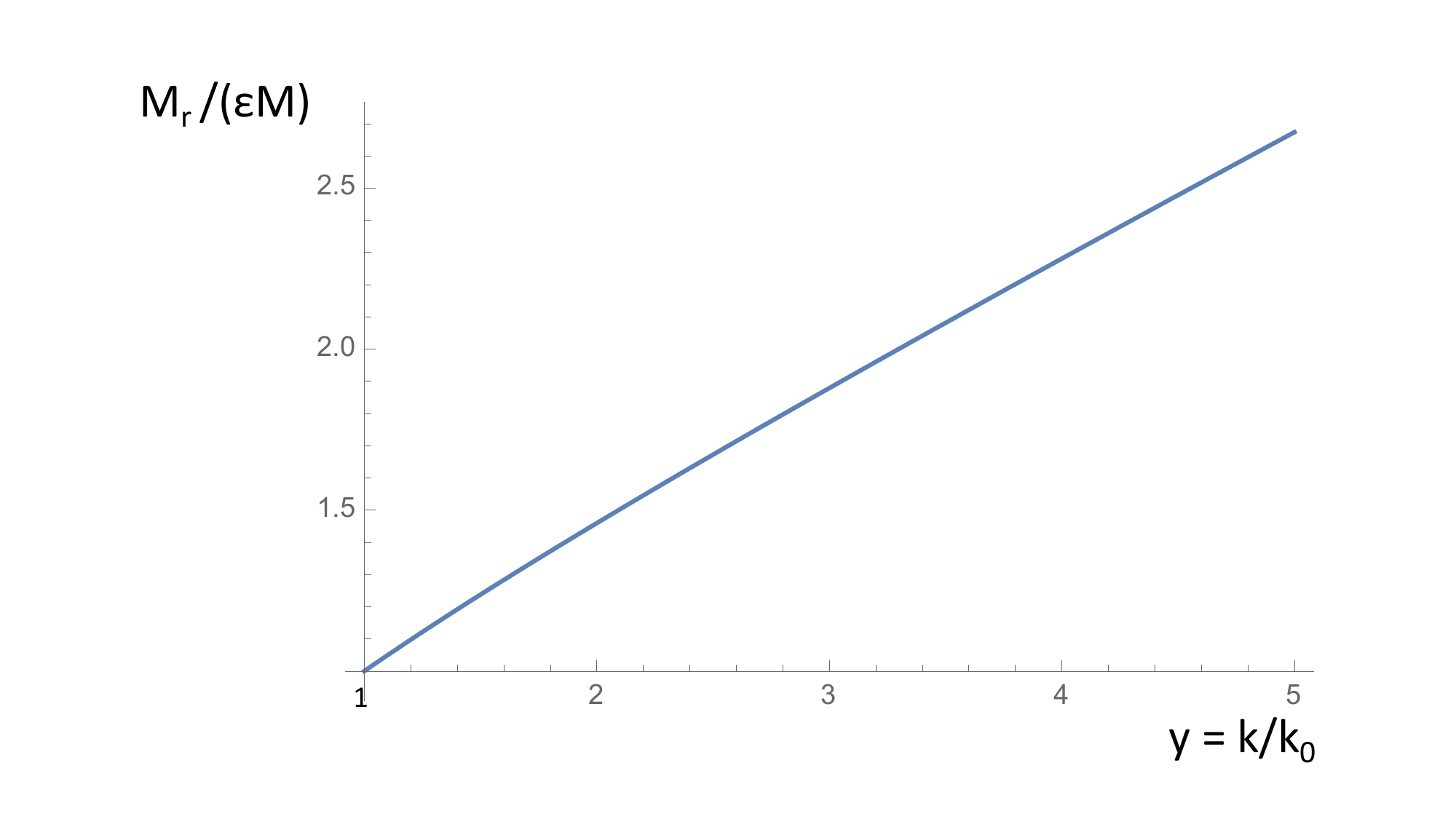} 
\caption{The running MM mass (in units of $\varepsilon M$) \eqref{lambertsol} in our one-loop-based resummation scheme. It is a monotonically increasing function of the transmutation mass scale $k \ge k_0$, proportional to the inverse of the $W_0(ek_0/k)$ branch of the Lambert function, which becomes a linear function of $k$ at the UV region, $ k \to \infty$ \eqref{eta}.}
\label{fig:rmass}
\end{figure}

The second equation can be integrated, taking into account the boundary condition $M_r(k=k_0) \equiv M_0 = \varepsilon M$ (on account of \eqref{MepsM0}):
\be\label{intMeq}
\ln\Big(\frac{M\, \varepsilon}{M_r}\Big) +  \frac{M\, \varepsilon}{M_r} = 1 - \ln \Big(\frac{k}{k_0}\Big)  \,, \qquad k > k_0\,.
\ee
If we denote $x \equiv k/k_0>1$ and $y(x)\equiv M\varepsilon/M_r(x)$, the previous equation can be written
\begin{align}
 y(x) ~ e^{y(x)} = \frac{e}{x} ~,
\end{align}
which can be solved in terms of the so-called Lambert function~\cite{lambert,lambert2,lambert3}, 
\begin{align}\label{Lambertfn}
y(x) = W\left(\frac{e}{x}\right)~,
\end{align}
which, since $x>0$, involves the principal branch $W_0$ of the Lambert function, and finally
\begin{align}\label{lambertsol}
\frac{M_r}{\varepsilon\, M} = \Big(W_0(e k_0/k)\Big)^{-1}~, \qquad k \ge k_0~. 
\end{align}
The running mass, as given by \eqref{lambertsol}, is plotted in Fig.~\ref{fig:rmass}.

On substituting the expression into the flow equation for $Z$ in \eqref{systemZMflows} we obtain:
\begin{align}\label{systemZMflows2}
k \, \partial_k Z &\simeq \frac{7}{9\varepsilon} \Big(1 - \frac{1}{1 + W_0(\frac{ek_0}{k})}\Big)\,,
\end{align}
Integrating over $k$ we can then obtain the $Z(k)$ with the boundary condition that $Z \to Z^\star = \frac{7}{9\varepsilon}$, as $k \to \infty$  (cf.\ \eqref{Zastnew}).
Setting $Z(k_1)=1$ we then determine $k_1$ as a function of $k_0$.

\begin{figure}[ht]
 \centering
\includegraphics[clip,width=0.70\linewidth]{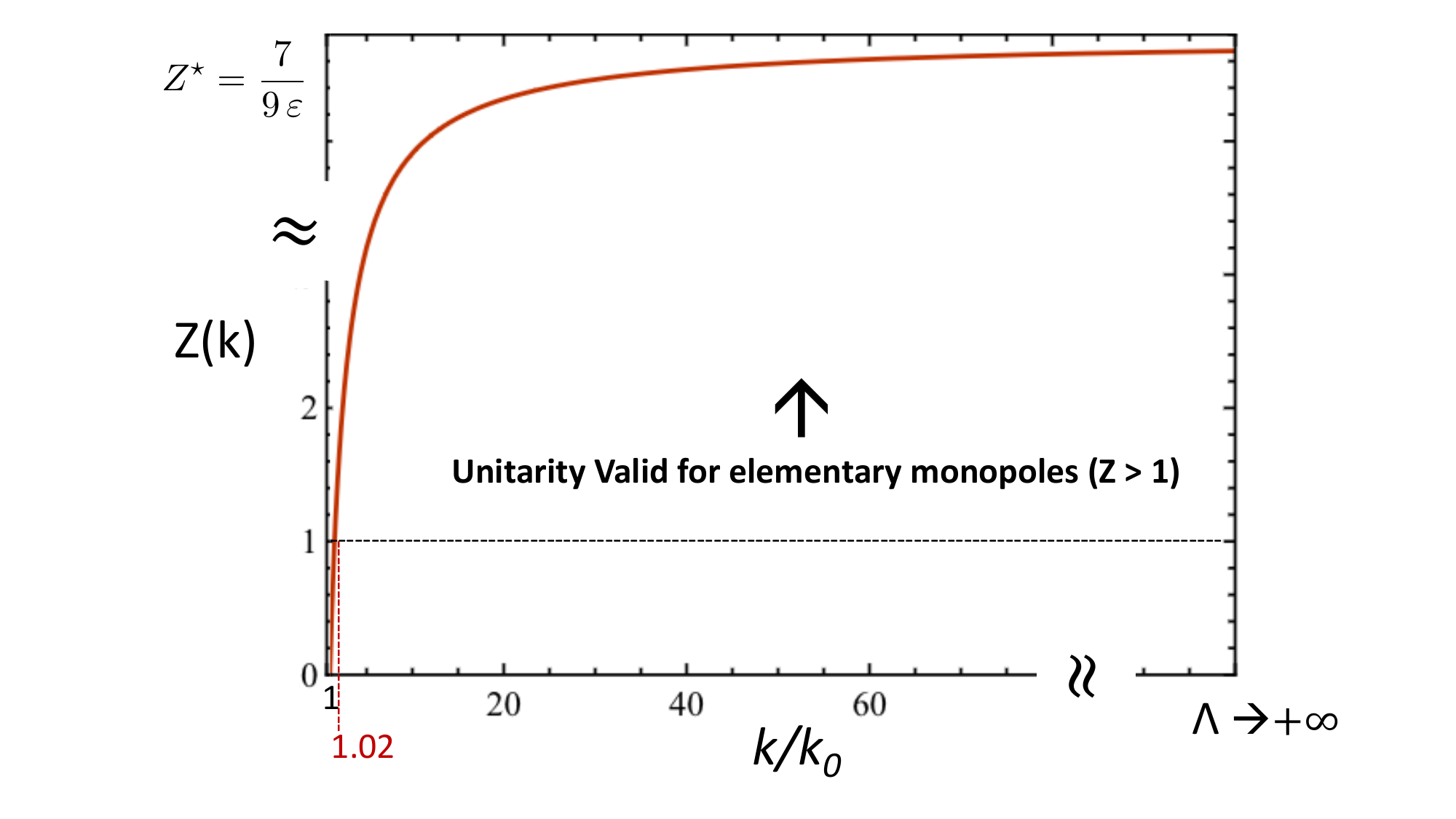} 
\caption{Rough qualitative sketch of the wavefunction  renormalization $Z(k)$, as a monotonically increasing function of the transmutation mass scale $k/k_0$, in the resummed MM effective gauge field theory (not to scale). The function increases abruptly (with a very large slope) from a zero value at $k=k_0=2M$ $(Z(k_0)=0)$, to unity at 
$k_1 = 1.02 \, k_0$ 
$(Z(k_1)=1)$, where it crosses into the unitary regime $Z>1$. From that point onward, it continues to increase with decreasing slope, until it reaches asymptotically the UV fixed point at $k \to +\infty$, where the slope vanishes, and at which the function assumes the large fixed-point value $Z^\star = \frac{7}{9\,\varepsilon} \gg 1$, for $\varepsilon \ll 1$.}
\label{fig:Zfunc}
\end{figure}

Equation~\eqref{systemZMflows2}
cannot be integrated analytically for arbitrary values of $k$. However, we can still exclude the region $k_1 \gg k_0$ analytically, by making use of the asymptotic expansion of the $W(z)$ function for small $0 < z \in \mathbb R$, $z \ll 1, $
$W(z) \simeq z - z^2 + \frac{3}{2} z^3 + {\mathcal O}(z^4)$. 
Indeed, in that case, we can easily see that there is no positive value of $z$ for which $Z=1$, 
for the DQC-compatible values of $\varepsilon \ll 1 $ \eqref{MepsM0}

Therefore the expected regime of $k_1$ is an intermediate one, not far from $k_0$. That is to say, we are seeking to integrate \eqref{systemZMflows2}
in an intermediate regime of $\frac{k}{ek^0} = {\mathcal O}(1)$. 
For such arguments the Lambert function $W_0(z)$ is known to assume values near $W_0(1) = 0.567$. For $ z \lesssim 1$, slightly less than one, the function drops smoothly from the value 0.567 towards a decaying behavior with $x$, as $e/x$. For $z \gtrsim 1$,  slightly larger than one, the function 
is known to rise slowly from the 0.567 value towards a $\ln z - \ln(\ln z)$ behavior (which is the Lambert function's  leading asymptotic behavior for large arguments~\cite{lambert}).  Given that $k > k_0$ in our approach, $k$ is restricted to lie in in this second case in a narrow regime $k \in [k_0, \, e k_0]$. In view of the above, it is a good approximation to assume  a roughly constant value of $W(\frac{ek^0}{k}) \simeq 0.567$ in the region $k^0/k \lesssim 1$. From \eqref{systemZMflows2}, then, for this regime of $k$, we approximately obtain:
\begin{align}\label{systemZMflows3}
k \, \partial_k Z &\simeq \frac{7}{9\varepsilon} \Big(1 - \frac{1}{1 + 0.567}\Big) \simeq \frac{0.28}{\varepsilon} \,, \qquad k \gtrsim k_0\,,
\end{align}
which can be integrated with the boundary condition $Z(k_0)=0$ (cf.\ \eqref{Zto0}), to yield:
\be\label{zflowtotal}
Z(k) \simeq \frac{0.28}{\varepsilon} \ln\Big(\frac{k}{k_0}\Big) \,, \qquad k \gtrsim  k_0\,.
\ee
The scale $k_1$ is then determined by setting $Z(k=k_1) =1$ in \eqref{zflowtotal}, which yields:
\be\label{k0primedet}
k_1 = k_0 \, \exp \Big(\frac{\varepsilon}{0.28}\Big) > k_0 \,, 
\ee
consistently with the above assumptions.

For the concrete DQC-compatible example case,
\eqref{epsval}, we are considering here, one obtains
\be\label{k0primedet2}
k_1 \simeq 1.02 \, k_0 \,,
\ee
which gives an approximate estimate for the transmutation scale at which the wave function renormalization $Z$ crosses the unitarity bound $Z(k_1) = 1$.\footnote{In the $k$-region of interest, $W[z]$ slightly increases from the value $W[1]=0.567$ to $W[e]=1$,
as $k/k_0$ decreases from $k=e\, k_0$ to $k = k_0 $. Even if we replace the $W[1]=0.567$ value, used in the middle side of \eqref{systemZMflows3},
by the value of $W[z]$ at the mean value of the interval $k \in (1, e)\, k_0$, that is 
$W[\frac{1+e}{2}]= 0.715$, we obtain instead of \eqref{k0primedet2}, the value, $k_1 \simeq 1.019 \, k_0$, that is practically  the same as in \eqref{k0primedet2}, to within the accuracy in this work. This justifies our approximate treatment above.}
From \eqref{systemZMflows3}
we observe that
the $Z(k)$ function
has a very large slope (in units of $k_0$), of order $k_0 \,\partial_k Z \Big|_{k=k_0} \simeq \frac{0.28}{\varepsilon}$, with $\varepsilon \ll 1$, and a slightly less, but also very large, slope at $k_1 $ (cf.\ \eqref{k0primedet2}):
$k_0 \,\partial_k Z \Big|_{k= 1.02 \, k_0}\simeq \frac{0.28}{1.02\,\varepsilon} = \frac{0.27}{\varepsilon}$, with $\varepsilon \ll 1$. 
Thus, there is an abrupt monotonic increase of the wavefunction renormalization $Z$ from $Z(k_0)=0$ to $Z(k_1)=1$ 
within the small region of $k \in [k_0,\,  1.02\, k_0]$. After crossing to the unitary regime at $k_1$, the wavefunction continues its monotonic increase until it reaches the very large fixed point value $Z^\star \simeq \frac{7}{9\varepsilon}$ at the UV fixed point $k \to \infty$ (in practice at $k \to \Lambda$ (cf.\ \eqref{kLambda})), where its slope vanishes (cf.\ \eqref{uvslope}).
The qualitative behavior of the wavefunction renormalization, as a function of the transmutation scale $k \ge k_0$, is represented in Fig.~\ref{fig:Zfunc}.

\FloatBarrier
\section{Additional exclusion plots}\label{sec:exclusions} 

\begin{figure}[htbp]
  \centering
    \includegraphics[width=0.49\linewidth]{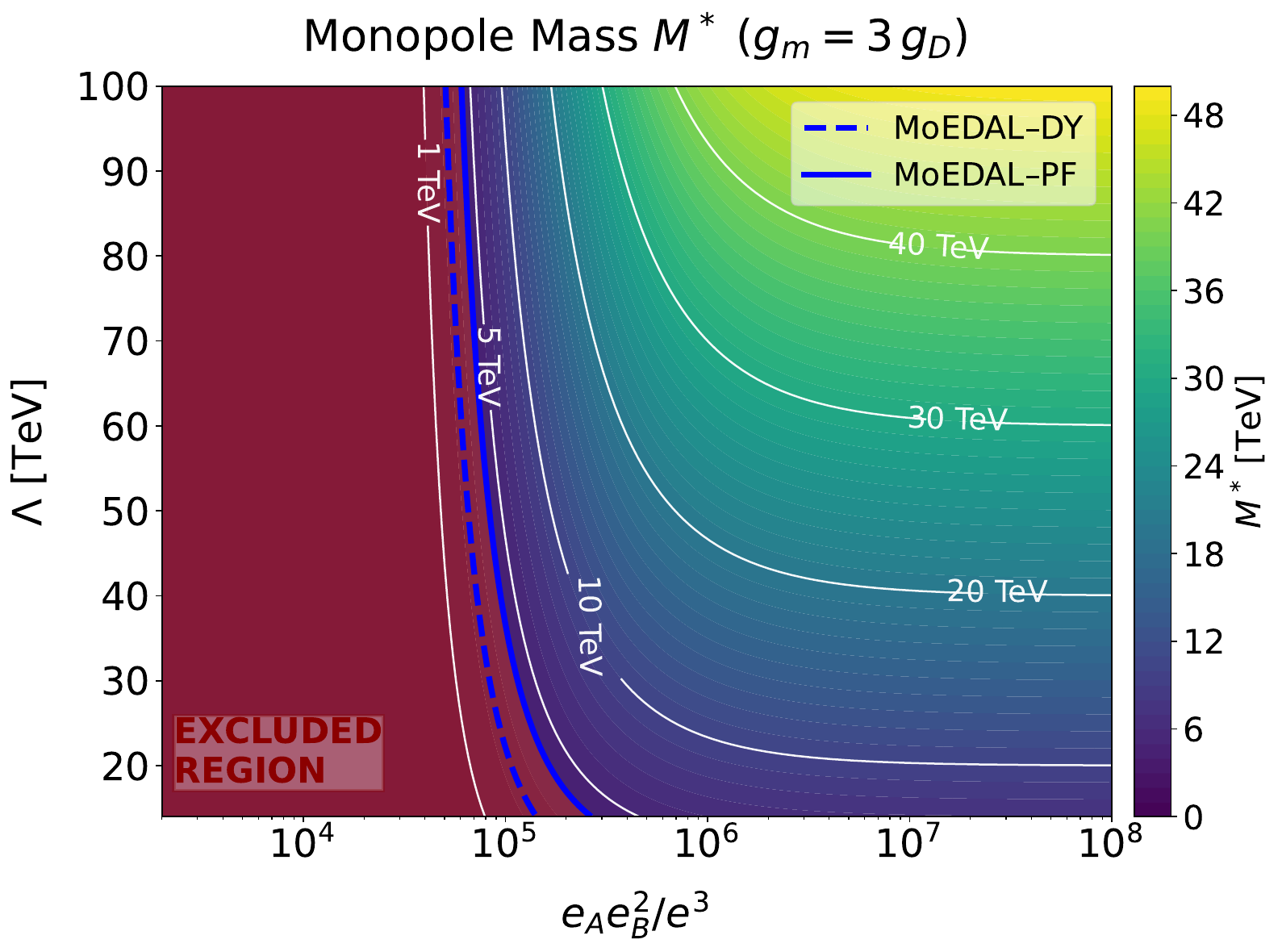}
  \hfill
    \includegraphics[width=0.49\linewidth]{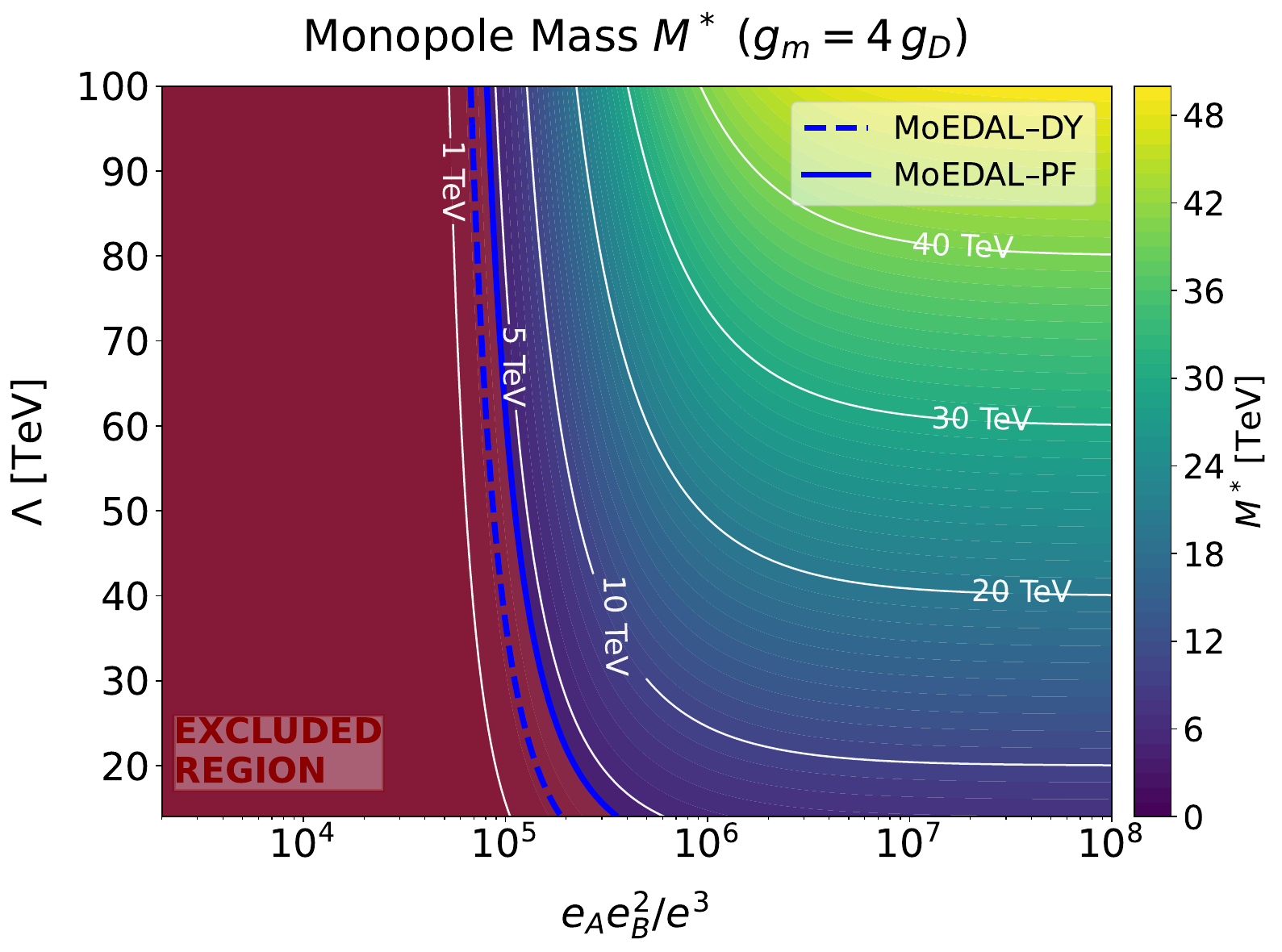}
\hfill\vspace{0.2cm}\hfill
    \includegraphics[width=0.49\linewidth]{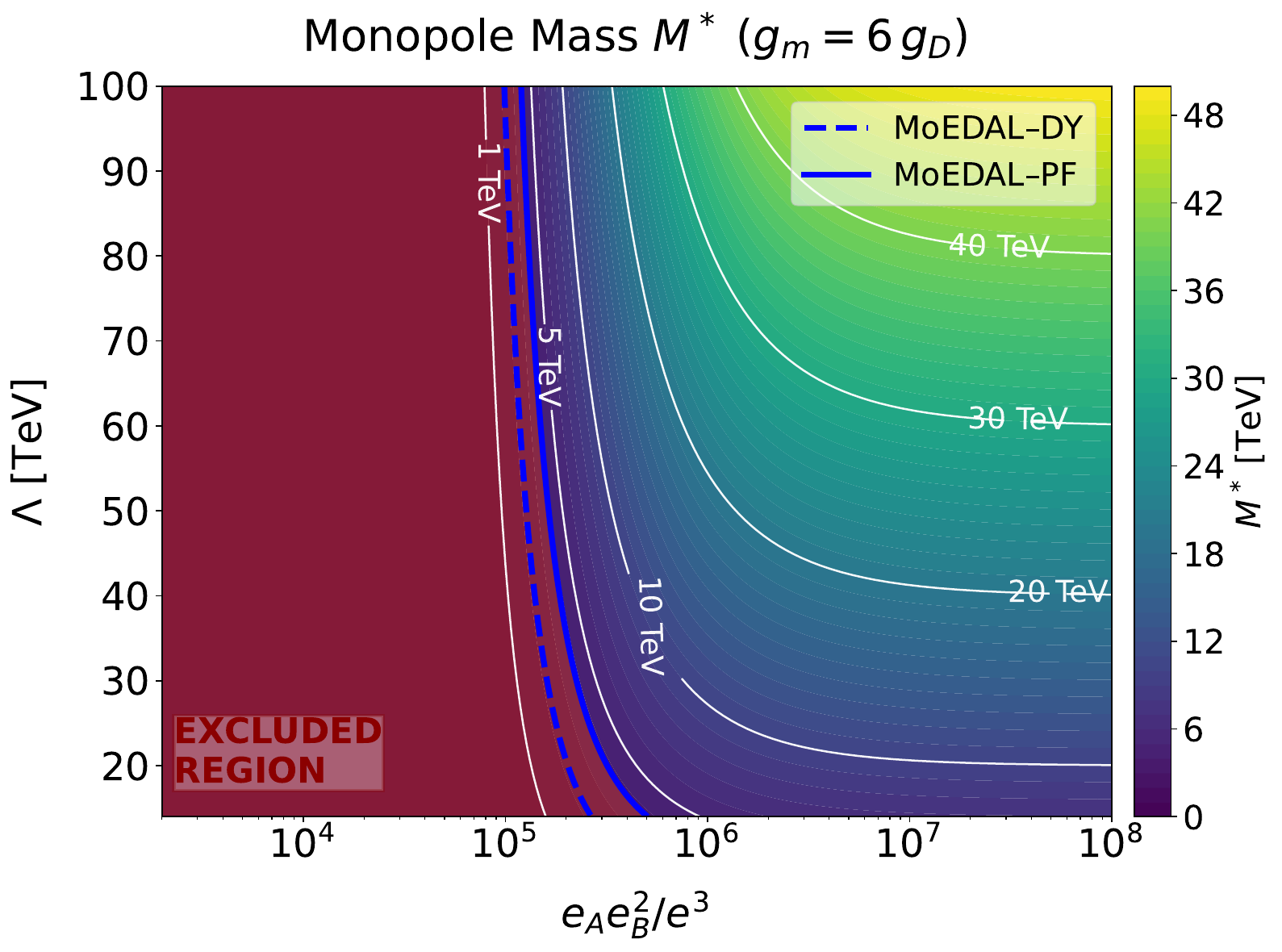}
\hfill
    \includegraphics[width=0.49\linewidth]{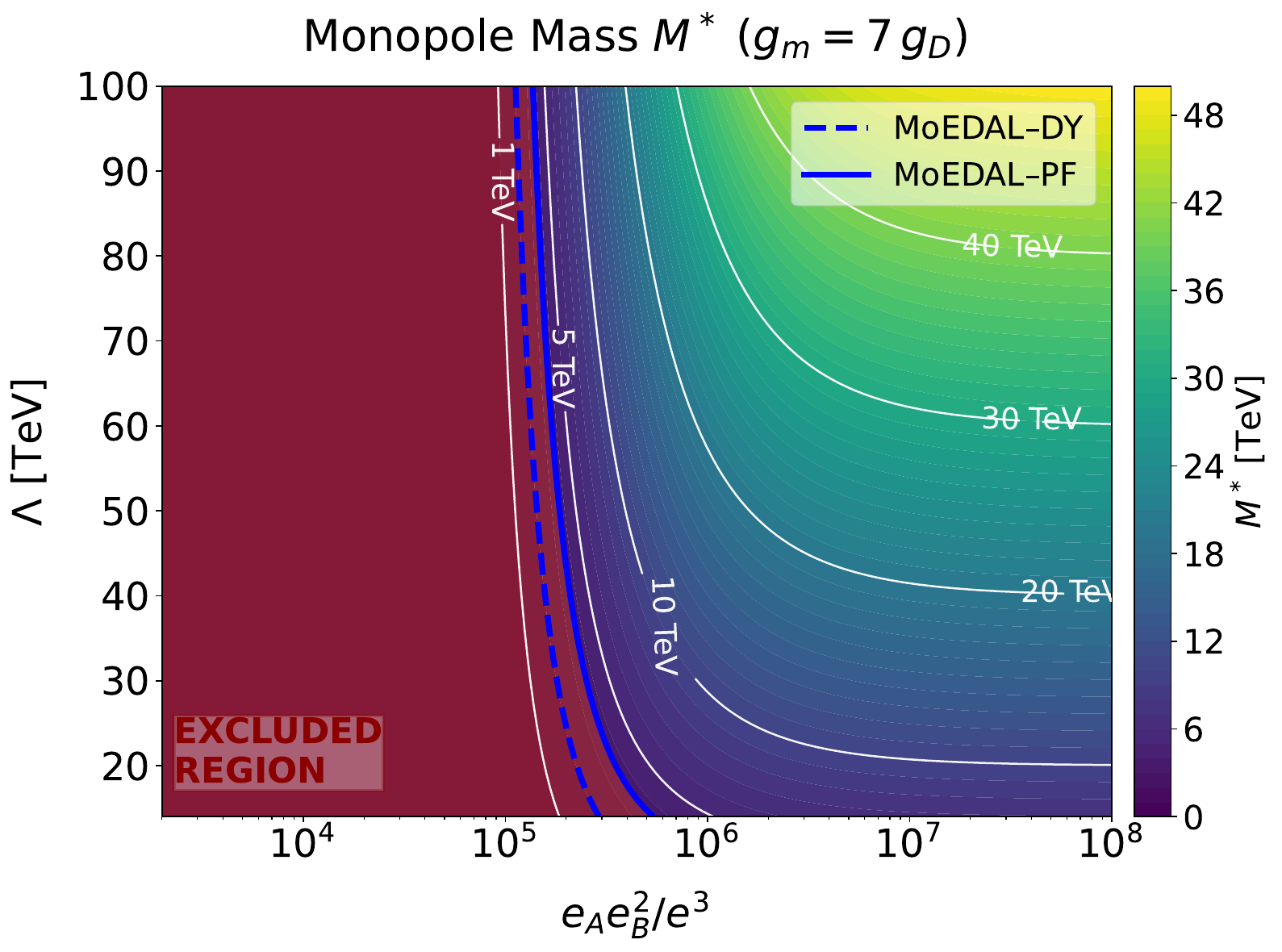}
\hfill\vspace{0.2cm}\hfill
    \includegraphics[width=0.49\linewidth]{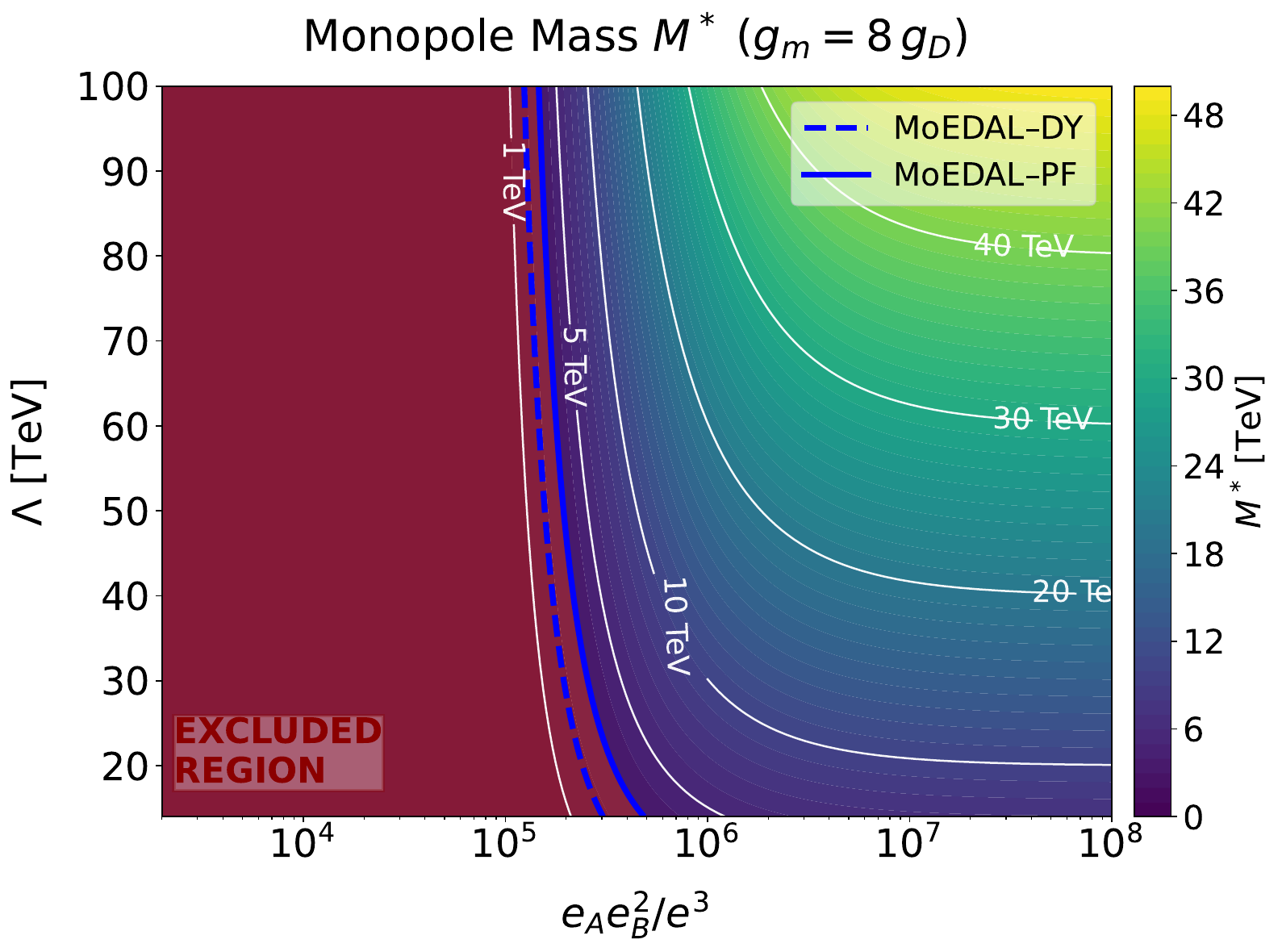}
  \hfill
    \includegraphics[width=0.49\linewidth]{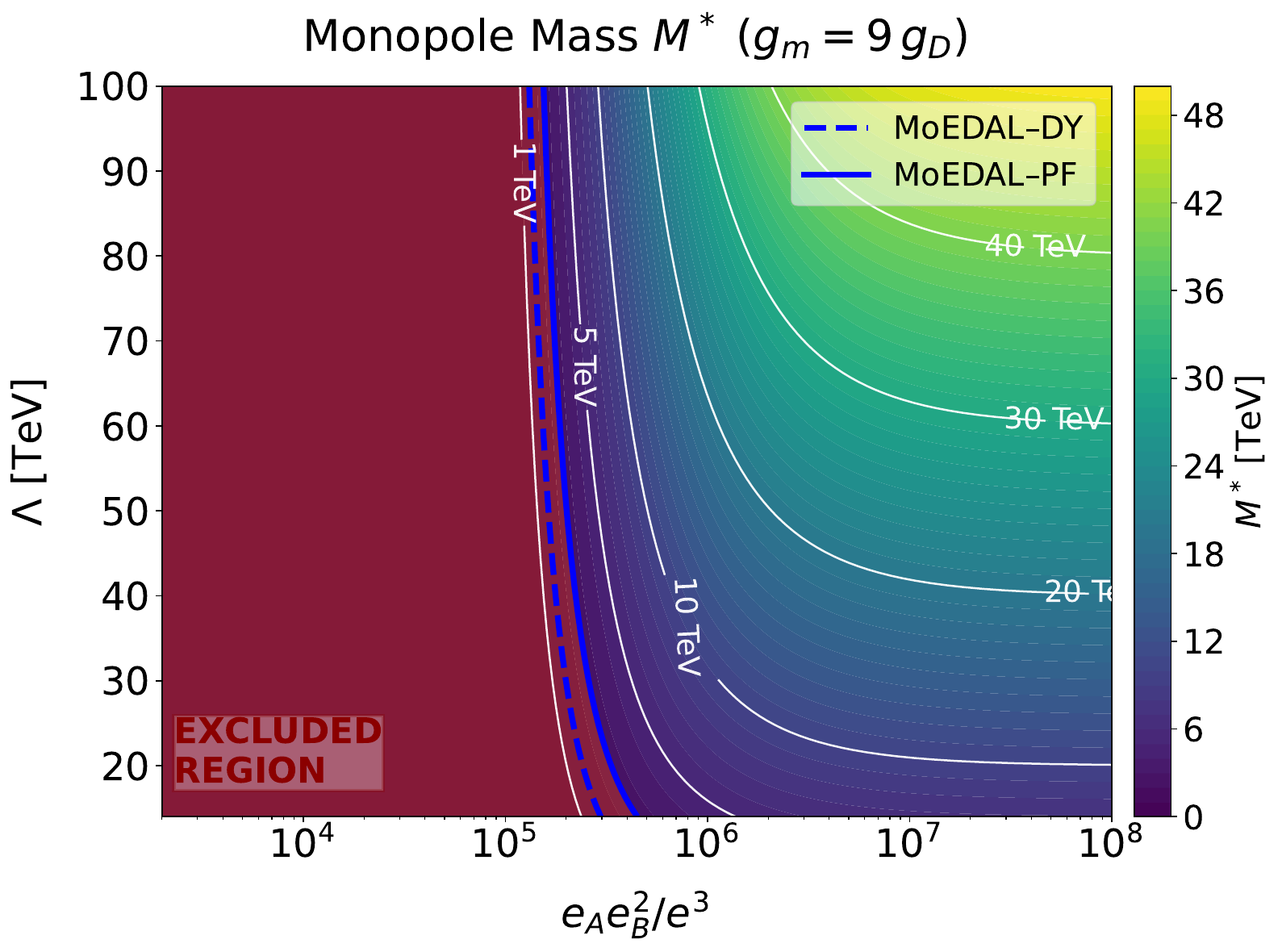}
  \caption{Contour maps of the resummed monopole mass $M^*$ for different magnetic charges $g_m$. The shaded red region indicates the experimentally excluded parameter space from MoEDAL~\cite{MoEDAL:2023ost} and ATLAS~\cite{ATLAS:2023esy} (DY and PF) at $\sqrt{s}=13~\tev$.}
  \label{fig:contour_massvslambdaeaeb2_rest}
\end{figure}

\begin{figure}[htbp]
  \centering
    \includegraphics[width=0.49\linewidth]{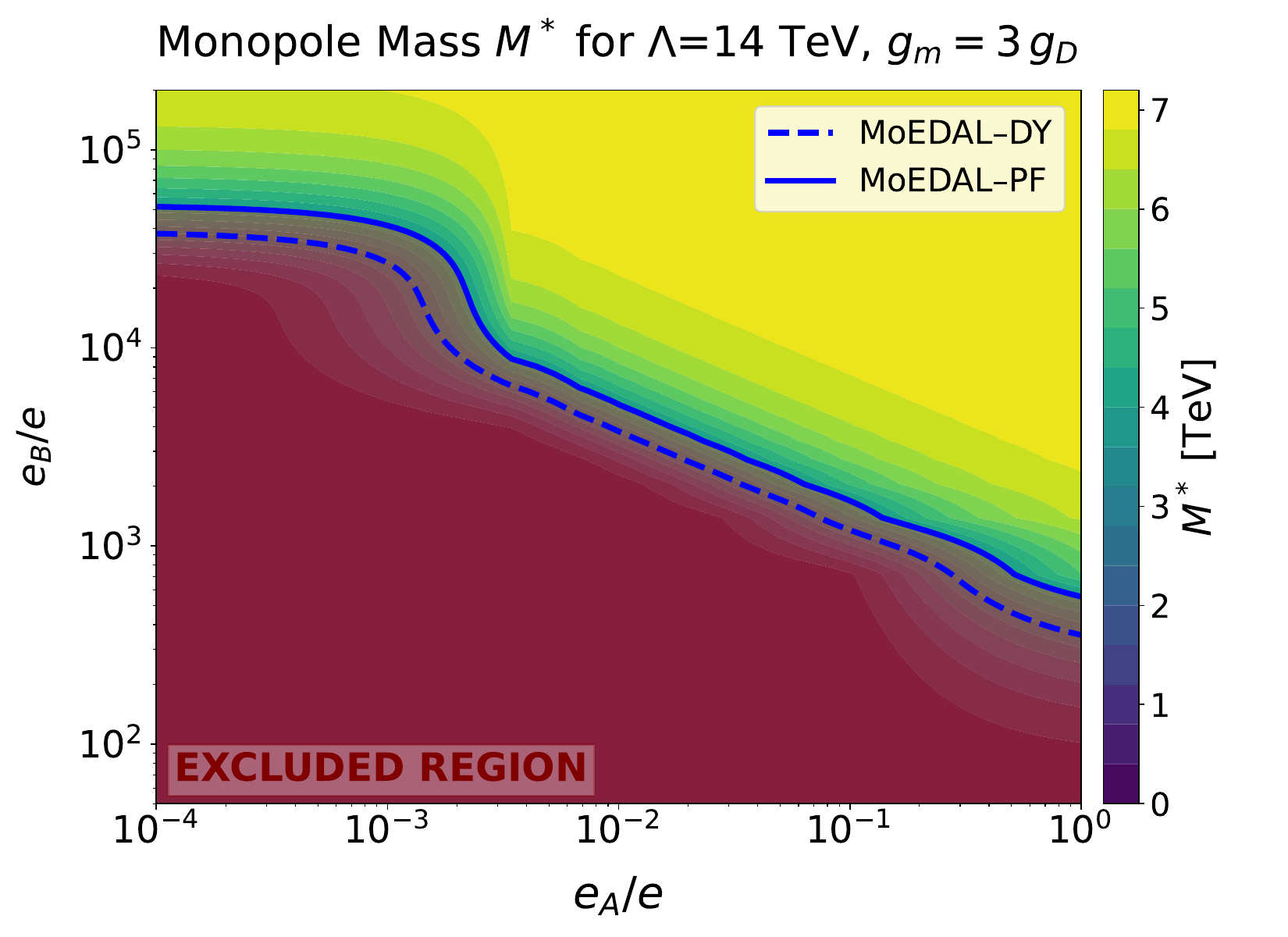}
  \hfill
    \includegraphics[width=0.49\linewidth]{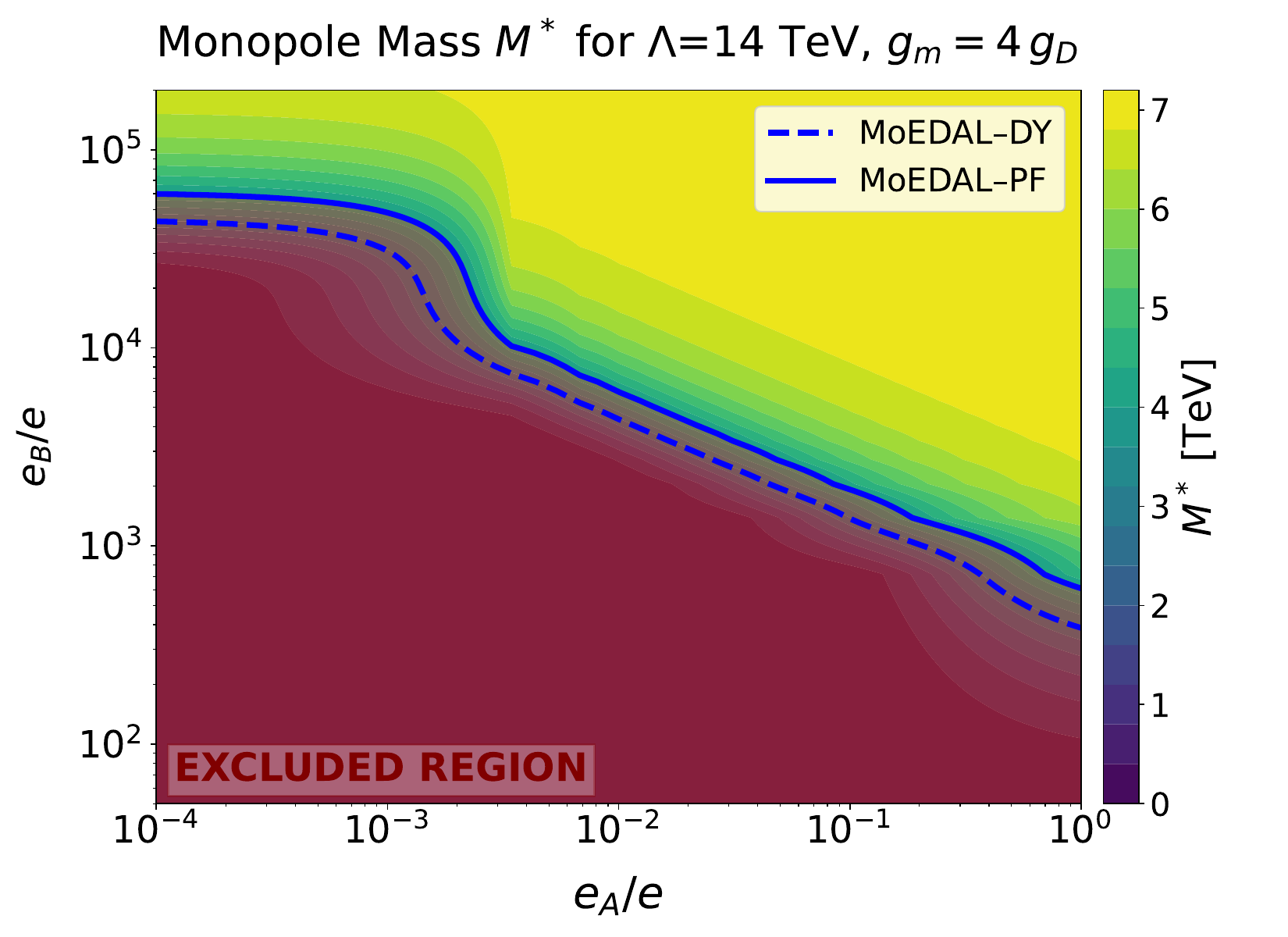}
\hfill\vspace{0.2cm}\hfill
    \includegraphics[width=0.49\linewidth]{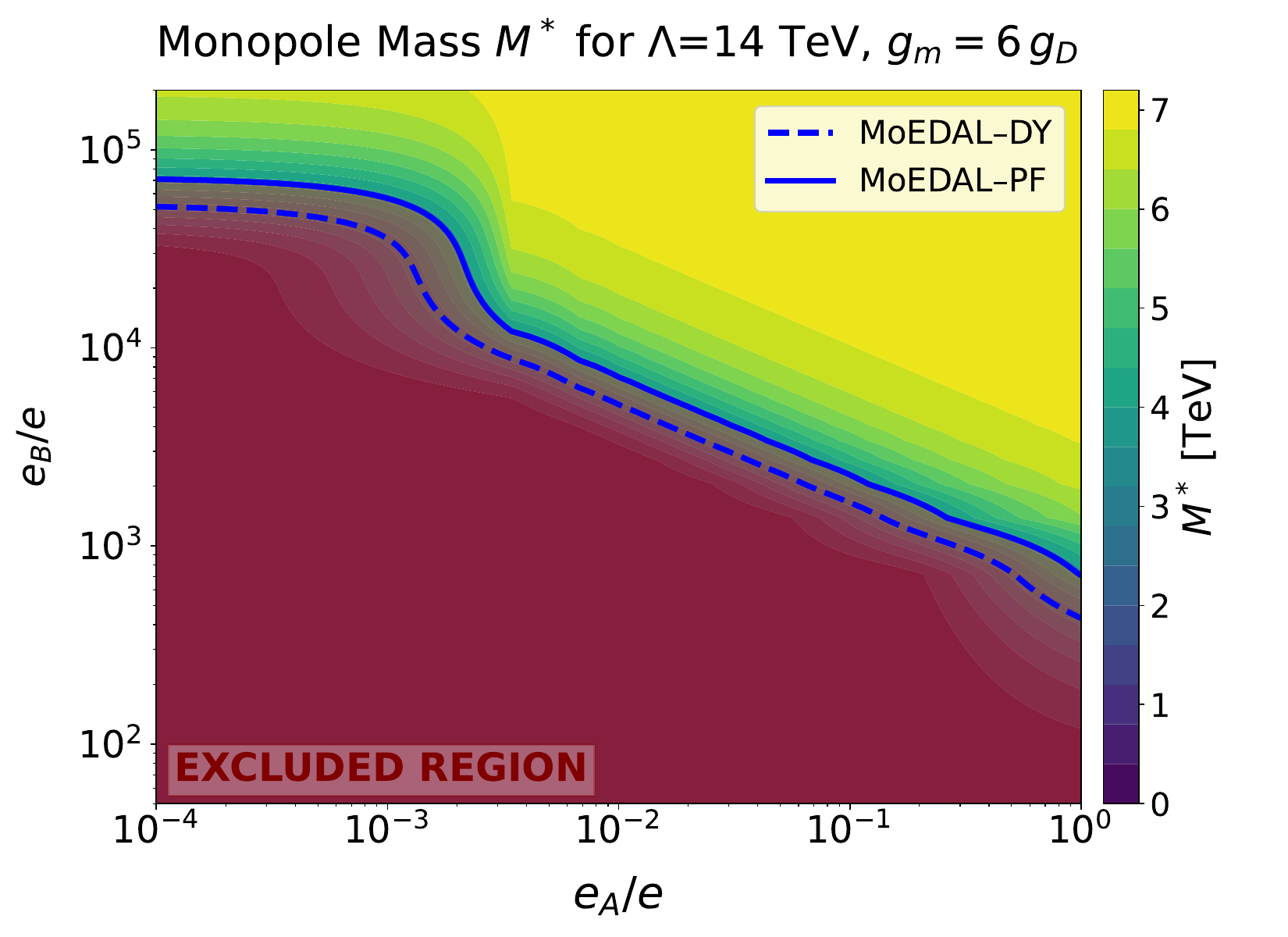}
\hfill
    \includegraphics[width=0.49\linewidth]{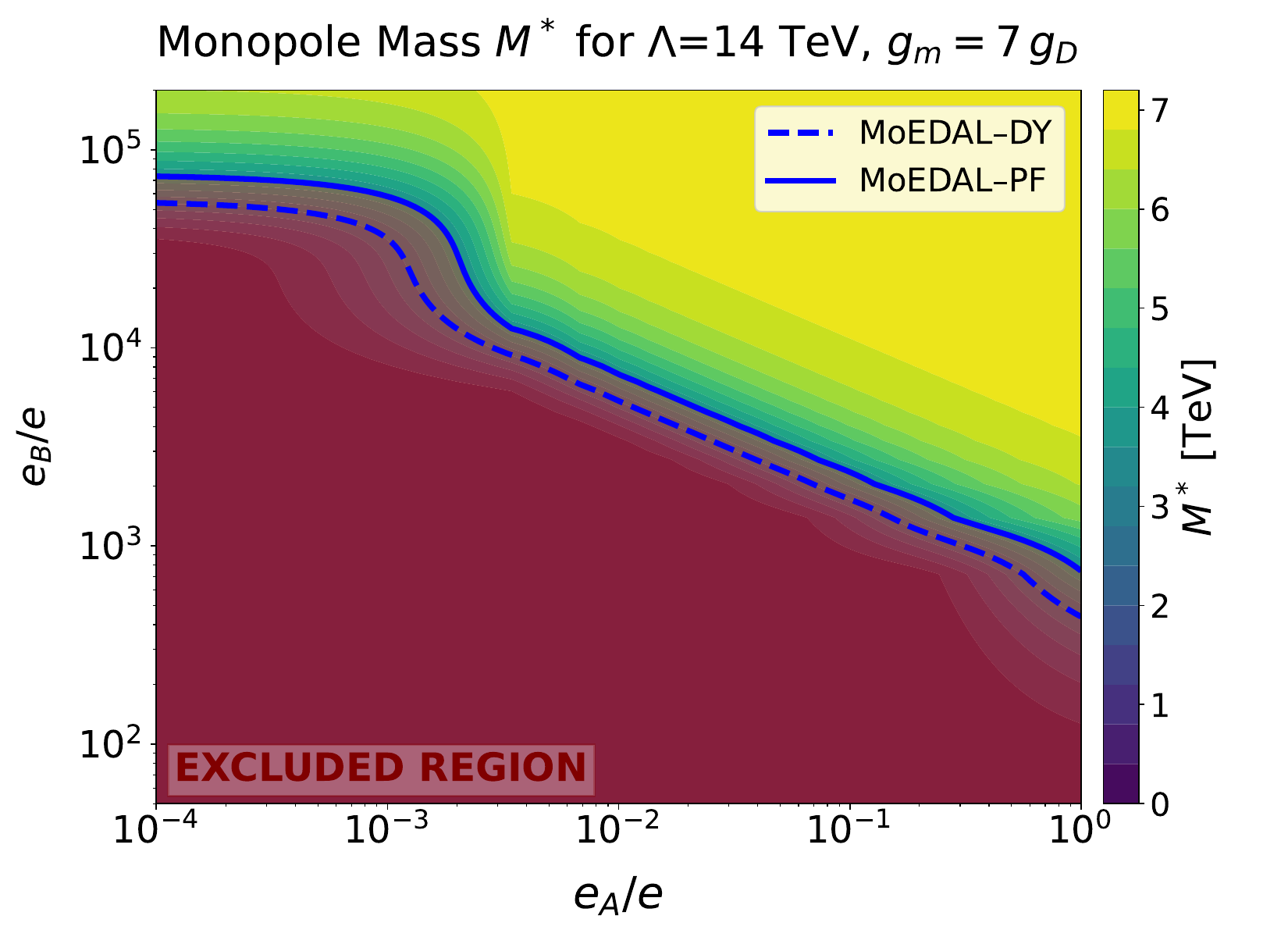}
\hfill\vspace{0.2cm}\hfill
    \includegraphics[width=0.49\linewidth]{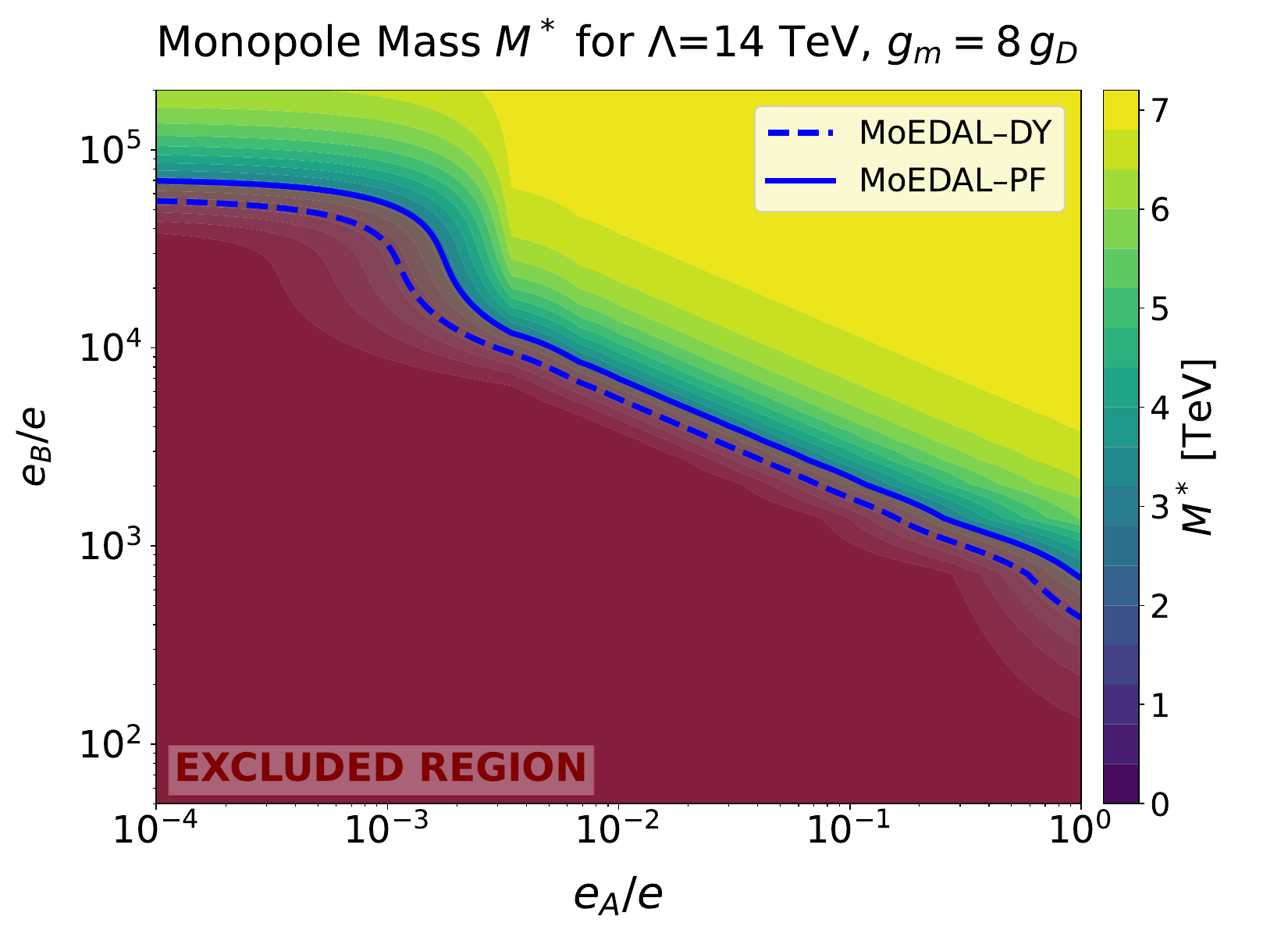}
  \hfill
    \includegraphics[width=0.49\linewidth]{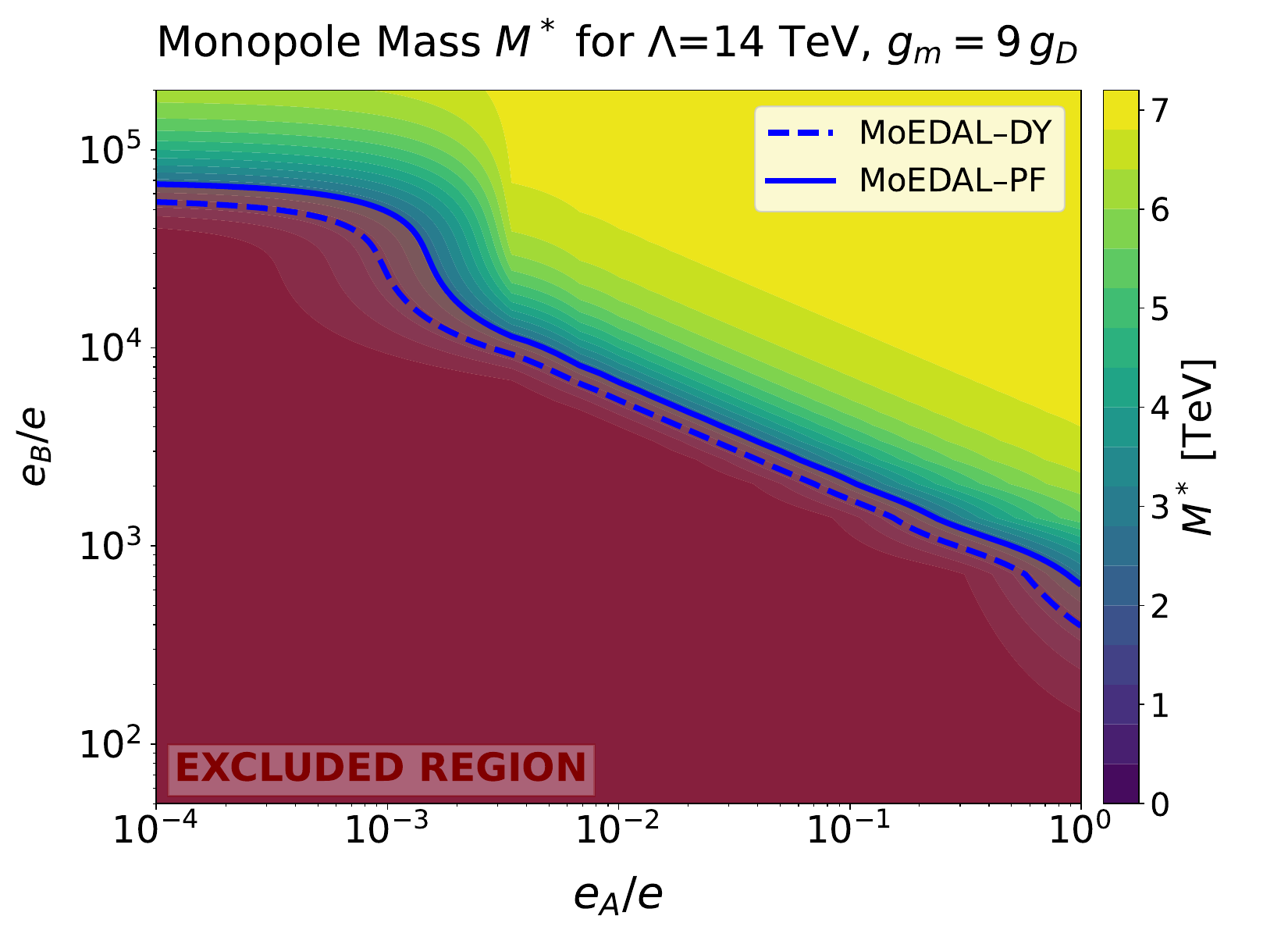}
  \caption{Contour maps of the resummed monopole mass $M^*$ for $\Lambda = 14~\tev$ and different magnetic charges $g_m$.  The shaded red region indicates the experimentally excluded parameter space from MoEDAL~\cite{MoEDAL:2023ost} and ATLAS~\cite{ATLAS:2023esy} (DY and PF) at $\sqrt{s}=13~\tev$.}
\label{fig:monopole_contour_exclusion_lambda14_rest}
\end{figure}

\begin{figure}[htbp]
  \centering
   \includegraphics[width=0.49\linewidth]{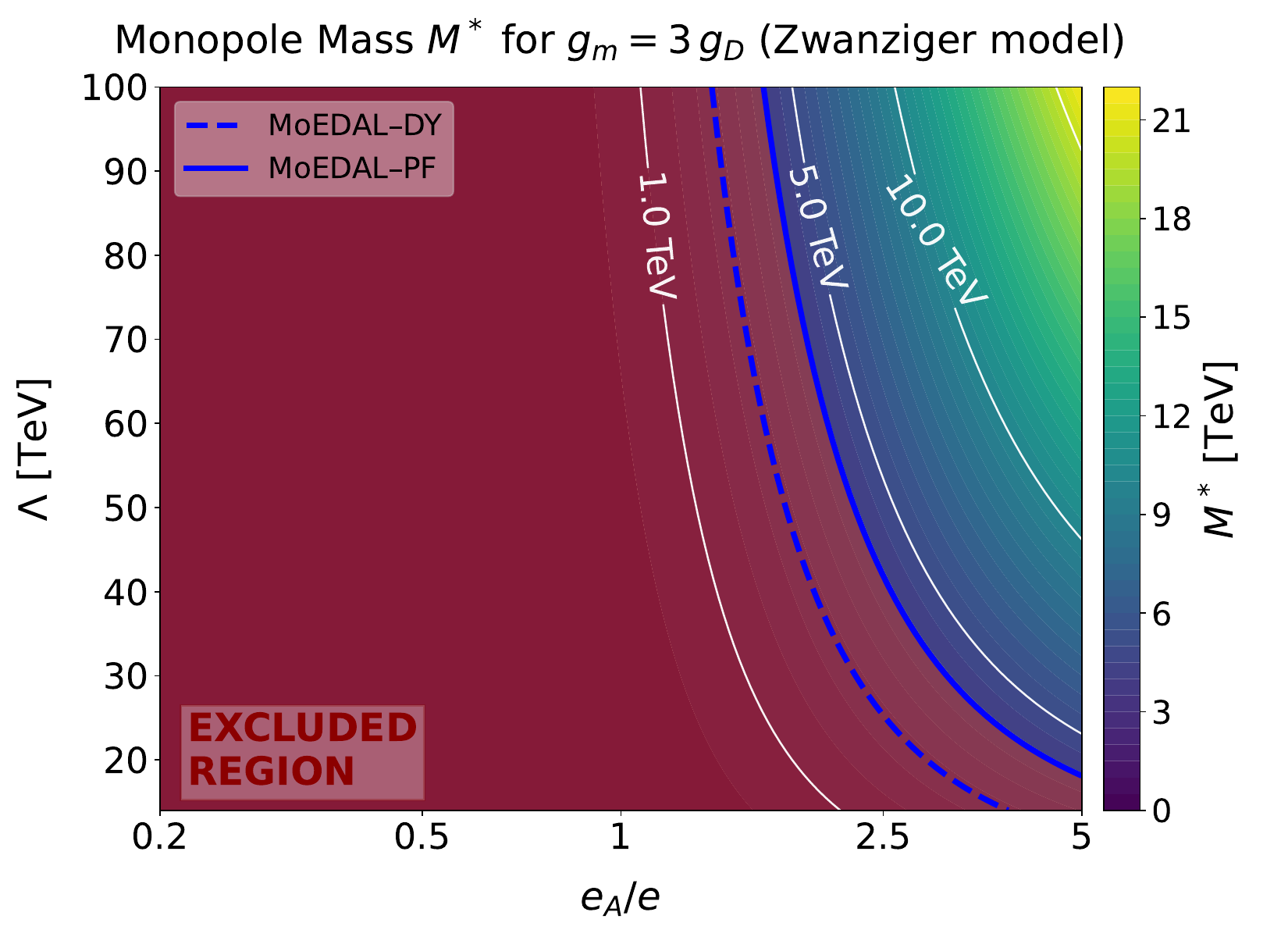}
  \hfill
    \includegraphics[width=0.49\linewidth]{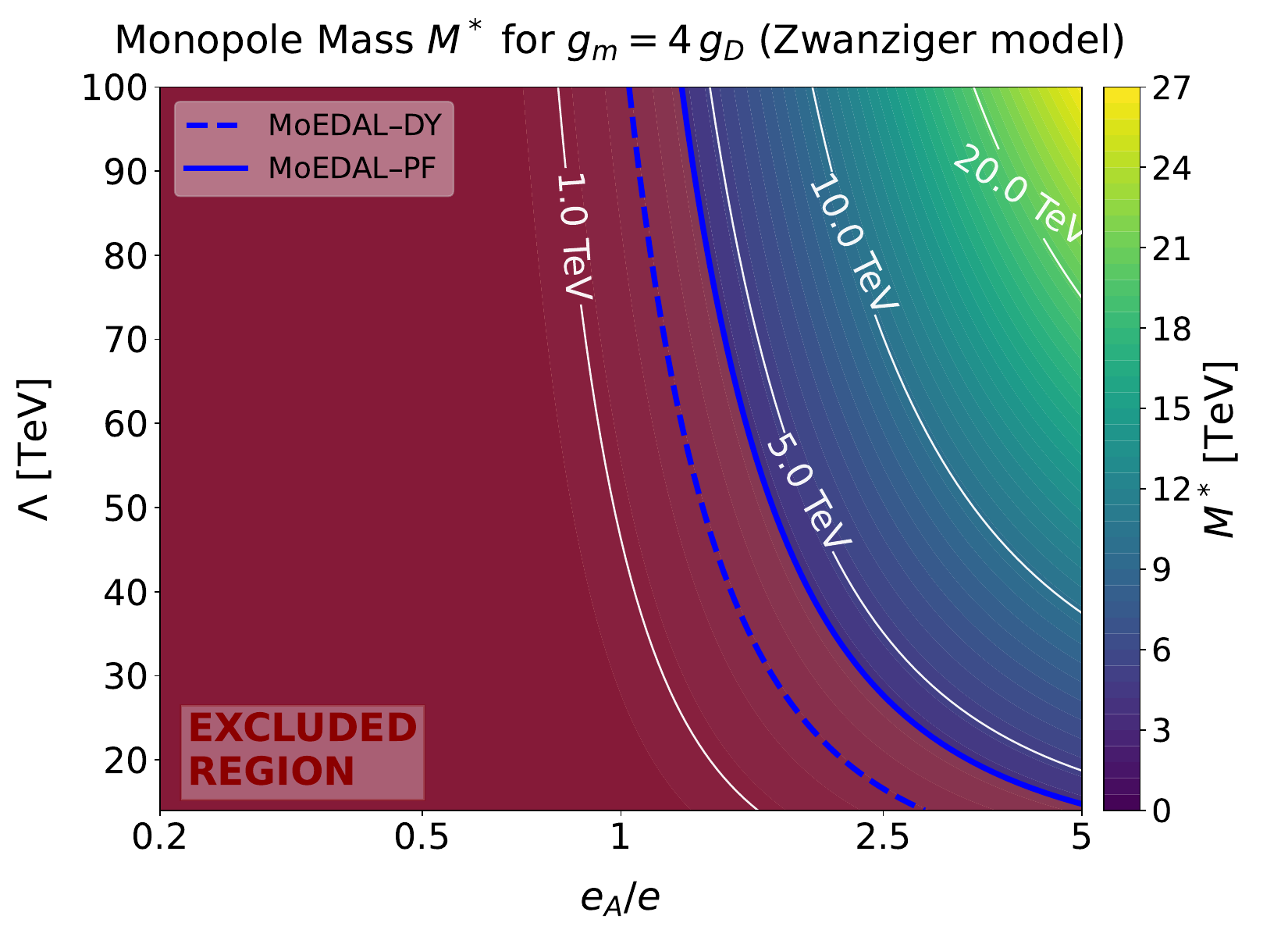}
\hfill\vspace{0.2cm}\hfill
    \includegraphics[width=0.49\linewidth]{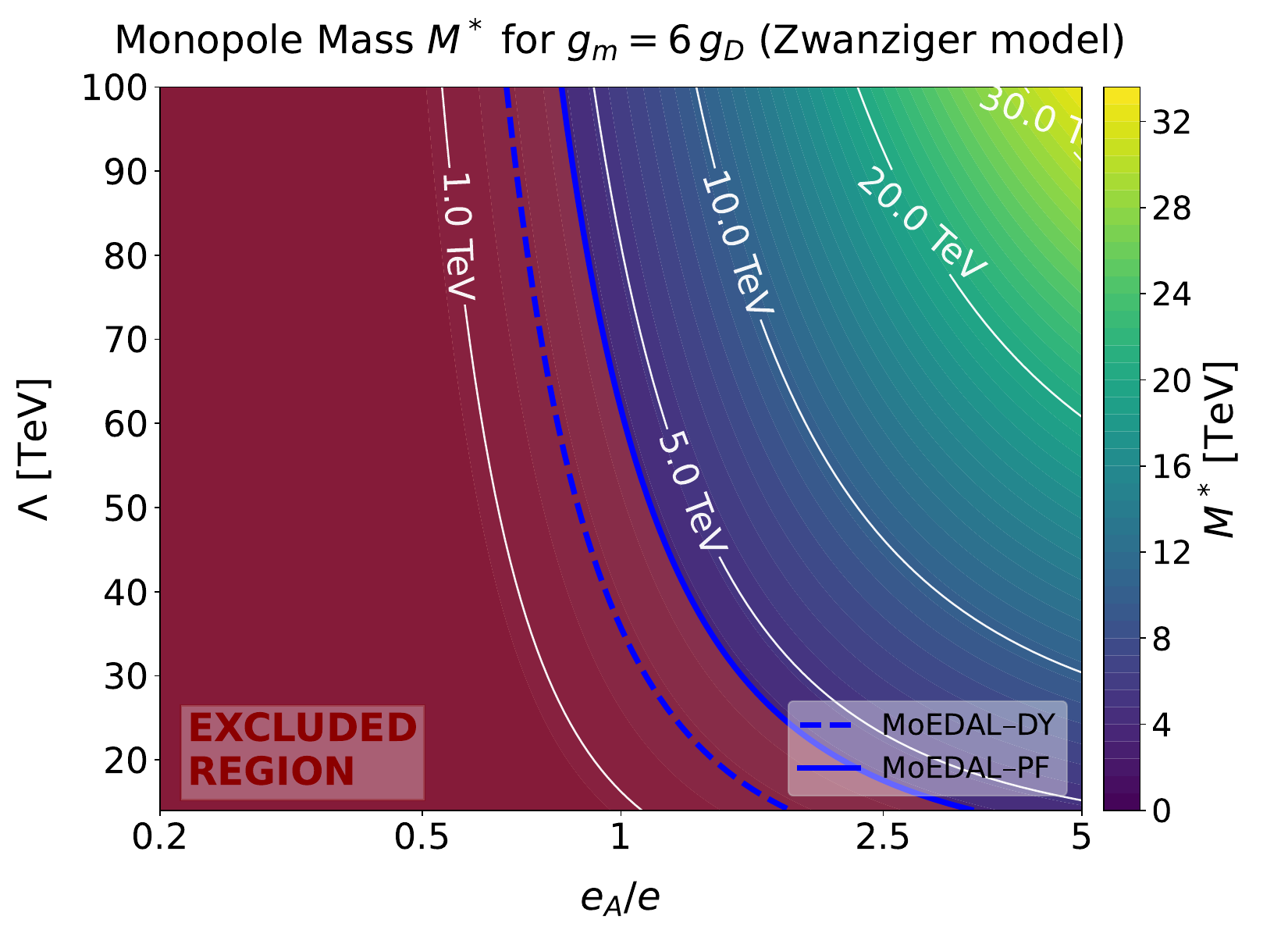}
\hfill
    \includegraphics[width=0.49\linewidth]{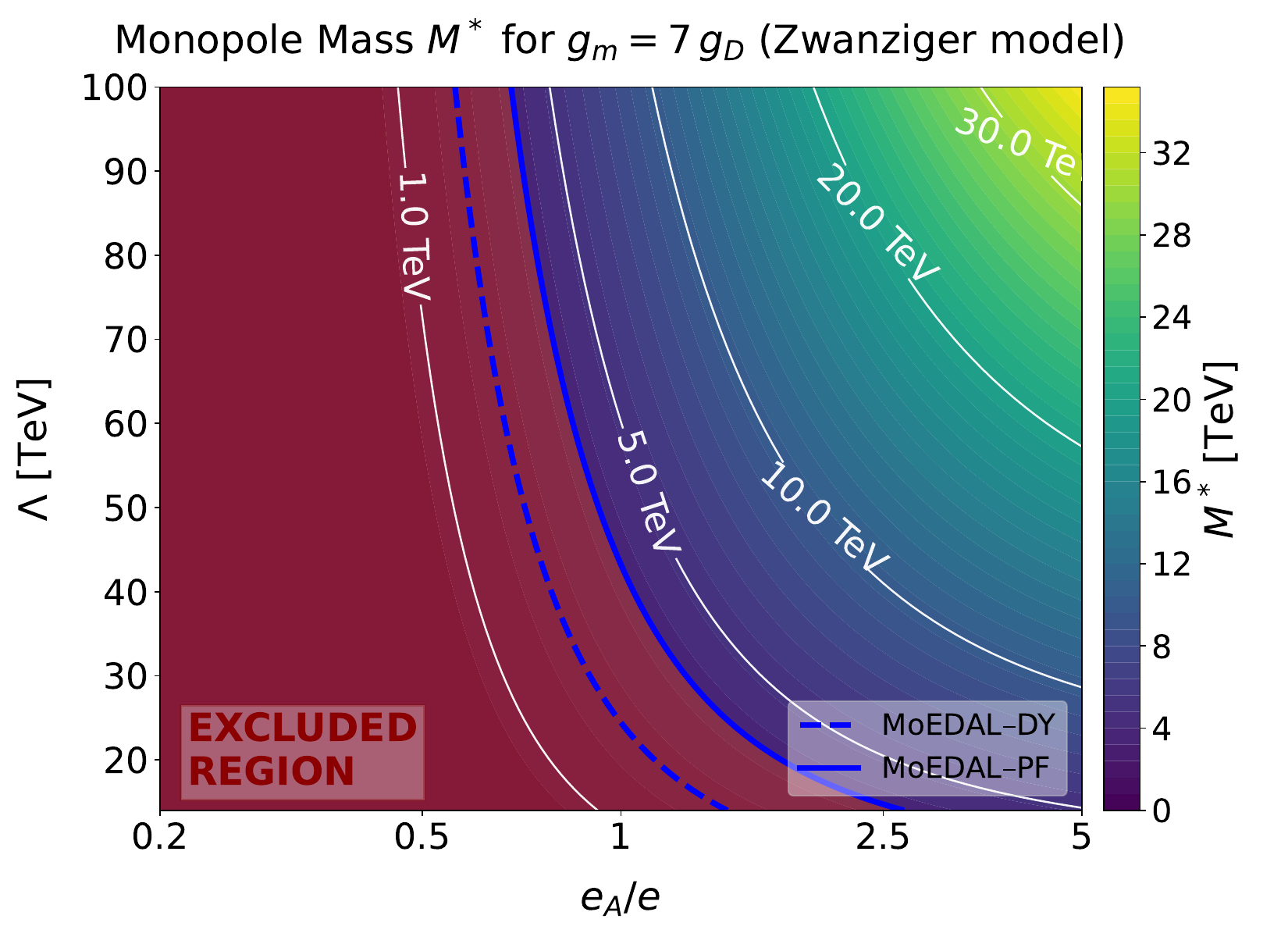}
\hfill\vspace{0.2cm}\hfill
    \includegraphics[width=0.49\linewidth]{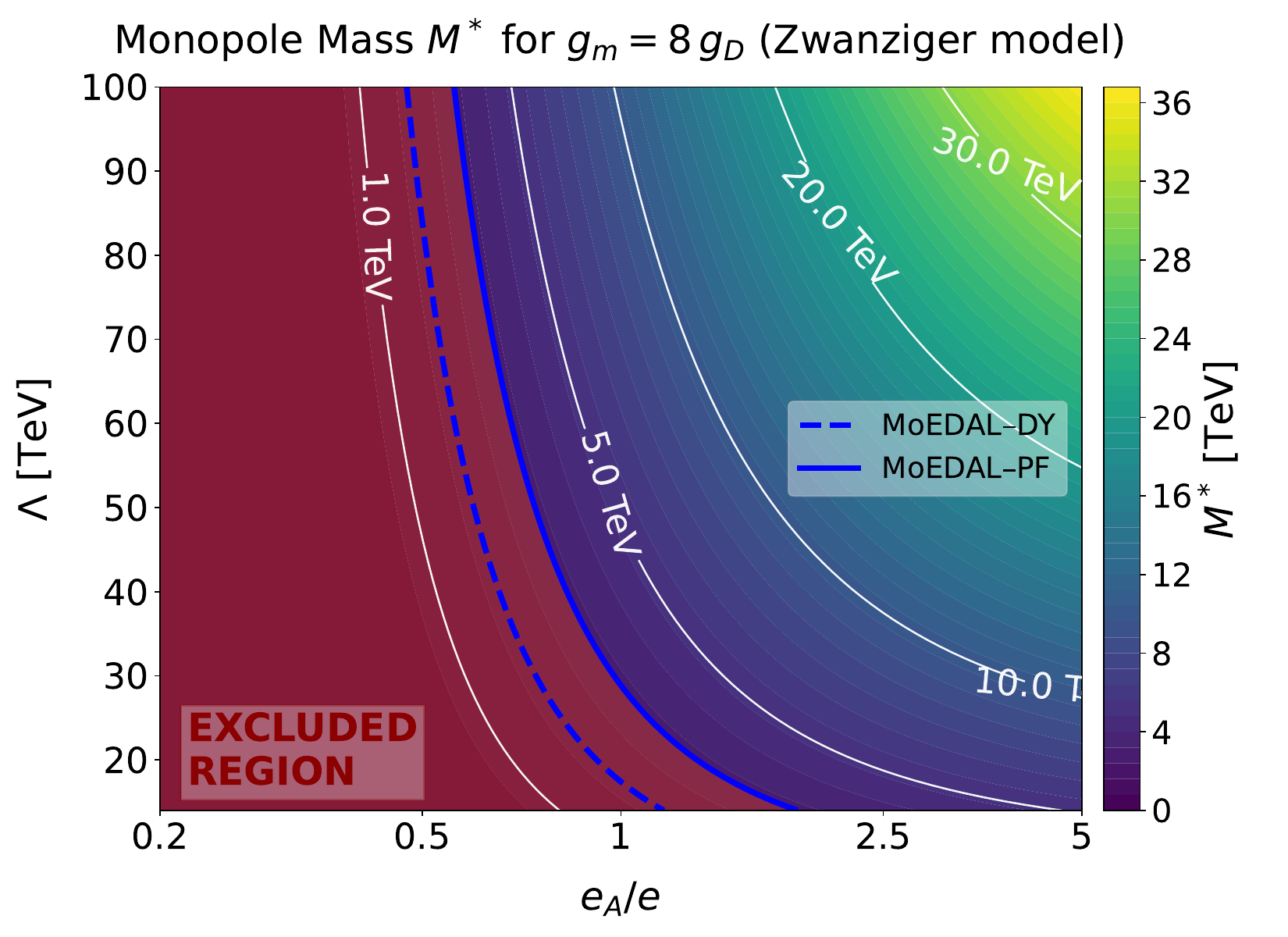}
\hfill
    \includegraphics[width=0.49\linewidth]{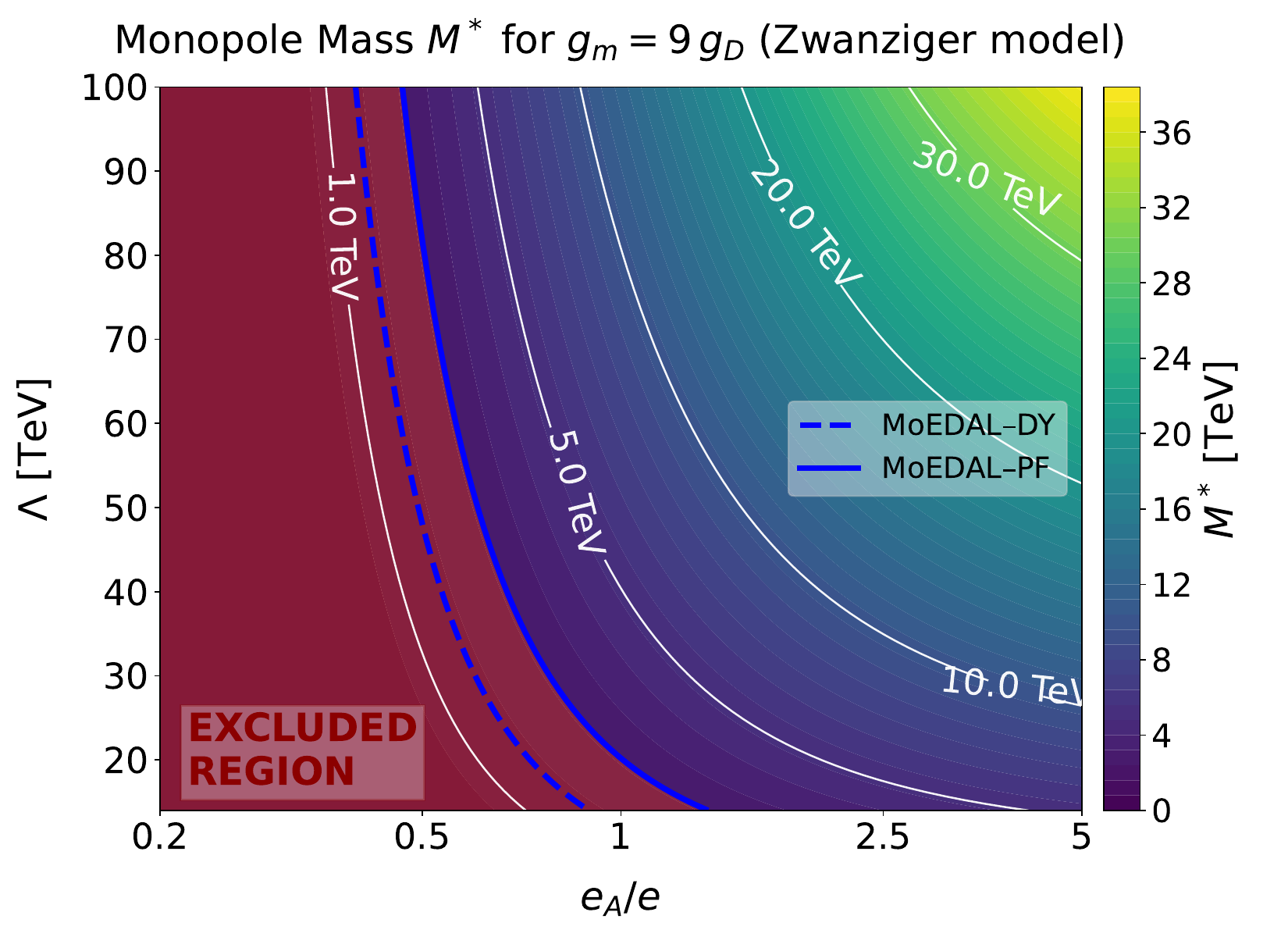}
  \caption{Contour maps of the resummed monopole mass $M^*$ for different magnetic charges $g_m$ assuming the Zwazinger model, that leads to \eqref{massLambda1b}. The shaded red region indicates the experimentally excluded parameter space from MoEDAL~\cite{MoEDAL:2023ost} and ATLAS~\cite{ATLAS:2023esy} (DY and PF) at $\sqrt{s}=13~\tev$.}
  \label{fig:excl_ea_lambda_rest}
\end{figure}

For completeness, in this appendix we provide additional exclusion plots to those in the main text, covering all values of the magnetic charge for which the most stringent mass bounds have been set in current colliders searches, namely by the ATLAS~\cite{ATLAS:2023esy} and the MoEDAL~\cite{MoEDAL:2023ost} experiments at the LHC. The analysis pertains to the EFT \eqref{efflag3}, in both the general, i.e.\ $e_B \ne g_m$ but respecting the hierarchy \eqref{ebea} (cf.\  Figs.~\ref{fig:contour_massvslambdaeaeb2_rest} and \ref{fig:monopole_contour_exclusion_lambda14_rest}), and the Zwanziger, $g_m=e_B$ (cf.\  \eqref{eAR} and Fig.~\ref{fig:excl_ea_lambda_rest}), cases.

\clearpage

\bibliographystyle{apsrev4-2}
\bibliography{Resum_MM}

\end{document}